\documentclass[a4paper,10pt,twoside]{article}
\usepackage{a4}
\usepackage{path}
\usepackage{amsfonts}
\usepackage{amssymb}
\usepackage{amsmath}
\usepackage{fancyhdr}
\usepackage{color}
\usepackage{mathrsfs}
\usepackage[pdftex]{graphicx}

\DeclareMathOperator{\polylog}{Li}

\begin{document}

\title{Analytical Expressions for Fringe Fields in Multipole Magnets}
%\textcolor{red}{\bf DRAFT - version 0.8}}

\author{B.D.~Muratori, J.K.~Jones, ASTeC, STFC Daresbury Laboratory, and Cockcroft Institute, U.K. \\
A.~Wolski, Department of Physics, University of Liverpool, Liverpool, and Cockcroft Institute, U.K.}
\maketitle

\begin{abstract}
Fringe fields in multipole magnets can have a variety of effects on the linear and nonlinear dynamics of particles moving along
an accelerator beamline. An accurate model of an accelerator must include realistic models of the magnet fringe fields. Fringe
fields for dipoles are well understood and can be modelled at an early stage of accelerator design in such codes as MAD8, MADX, GPT
or ELEGANT. However, usually it is not until the final stages of a design project that it is possible to model fringe fields for
quadrupoles or higher order multipoles. Even then, existing techniques rely on the use of a numerical field map, constructed
either from magnetic modelling or from measurement, and which will not be available until the magnet design is well developed.
Substitutes for the full field map exist but these are typically based on expansions about the origin and rely heavily on the
assumption that the beam travels more or less on axis throughout the beam line. In some types of machine (for example, a
non-scaling FFAG such as EMMA) this is not a good assumption. Some tracking codes (such as General Particle Tracer or GPT) use
methods for including space charge effects that require fields to vary smoothly and continuously along a beamline: in such cases,
realistic fringe field models are of significant importance.

In this paper, a method for calculating fringe fields based on analytical expressions is presented, which allows fringe field effects
to be included at the start of an accelerator design project. The magnetostatic Maxwell equations are solved analytically and a
solution that fits all orders of multipoles derived. Quadrupole fringe fields are considered in detail as these are the ones that give
the strongest effects; the simplest specific solution is shown. Two examples of quadrupole fringe fields are also presented. The first
example is a magnet in the LHC inner triplet, which consists of a set of four quadrupoles providing the final focus to the beam, just
before the interaction point. Because of the very strong focus at the collision point, particles in the beam pass through the inner
triplet quadrupoles a considerable distance from the axis and, as a result, fringe field effects are important. Quadrupoles in EMMA
provide the second example.  In both examples, the analytical expressions derived in this paper for quadrupole fringe
fields provide a good approximation to the field maps obtained from a numerical magnet modelling code.
\end{abstract}

\section{Introduction}

Fringe fields represent regions that lie at the edges of a magnet where there is a transition from the nominal field to zero field,
or to the nominal field in an adjacent magnet.  In multipole magnets, the nominal field has no longitudinal component.  However, in
the fringe field region where the fields vary with longitudinal position, Maxwell's equations require the presence of a non-zero longitudinal
field component.  For dipoles, the nominal field has only a single component.  In dipole fringe fields, therefore, the field has only two
components, and (assuming that the fields are independent of the horizontal transverse co-ordinate) analytical expressions for the field
can be obtained by solving the two-dimensional Laplace equation.  For quadrupoles and higher-order multipoles, however, the fields in
the fringe region have three components, and analytical expressions for these fields must be obtained by solving the three-dimensional
Laplace equation.  Fringe fields can impact the motion of particles passing through magnets in a number of ways.  For example, they
can introduce nonlinearities in the equation of motion, or they can make a substantial contribution to the desired effects of the nominal
field.  The latter situation is the case in EMMA, for example, where the large aperture of the quadrupoles compared to their lengths means
that the fringe fields dominate the focusing effects of these magnets. Nonlinearities from fringe fields can become important if the transverse
size of the beam is large, or if the beam traverses a multipole at an angle and some distance from the magnetic axis: this is often the case
in final focus quadrupoles in colliders. The implementation of fringe fields is also important in some tracking codes which include effects
like space charge, as is the case in GPT (General Particle Tracer) \cite{GPT1}, for example. This requires that all fields be continuous so
that there are smooth regions where the fields transit from their maximal value to zero and vice versa.

There are several models available for the study of fringe fields for multipoles, see for example \cite{Etienne1,Etienne2} and references
therein. However, existing models are usually limited to on-axis and mid-plane approximations, meaning that the field in the full space of the
multipole typically has to be computed with elliptic integrals. Therefore, for simple but accurate particle tracking, a significant amount of
effort and computing power goes into just the calculation of the fields that particles see. The results obtained in this paper make possible an
alternative method, based on analytical expressions for the fields as functions of position, that provide exact solutions to the static Maxwell
equations in three dimensions. This allows for arbitrarily smooth fields to be constructed and used in tracking codes as well as the possible
creation of transfer maps. The fringe fields considered here have associated scalar and vector potentials.
From the scalar potential it is possible to inspect the shape of the pole-face for an iron-dominated magnet generating the given field.
Using the vector potential it is possible to perform symplectic integration of the equations of motion for a particle in the field, leading to
the construction of transfer maps for the fringe region. Further, it may be possible to improve the efficiency of the magnet design
process by making some initial assumptions based on the formulae presented in this paper. However, a full three-dimensional
numerical field map will ultimately be needed (obtained from a numerical magnet modelling code) in order to achieve the accuracy that
is needed for validating the design of many accelerators.

This paper develops the mathematical framework that was initially presented in \cite{BM}, gives further details and results, and presents
two examples.  Following the Introduction, fringe fields for dipoles are briefly reviewed.  The formalism used for dipoles is extended to fully
three-dimensional fields in Section 3. A complete solution to the static Maxwell equations, in a form suitable for application to multipole
fringe fields, is derived and presented. Expressions for fringe fields in multipoles of arbitrary order are then given in Section 4. A
particular case of a quadrupole fringe field, together with a of the simple fall-off in the form of an Enge function \cite{engecite}, is then
presented in Section 5. All the salient properties are described in order to demonstrate that the field behaves in the way expected of a
quadrupole, first inside the magnet, then in the fringe field region, and finally at a large distance from the magnet (so that the field
effectively falls to zero). In Section 6, the scalar and vector potentials for fringe fields in the case of a multipole of arbitrary order are
discussed. In Section 7, two quadrupole examples are presented. The first is the HL-LHC inner triplet, where the beam size and trajectory in
the quadrupoles providing strong focusing close to the interaction point make fringe field effects significant. The second example is based on
the quadrupole magnets in the non-scaling FFAG, EMMA.  The large aperture of these magnets compared to their length means that the fringe
fields make a dominant contribution to the focusing effects. Conclusions are given in Section 8.

\section{Fringe fields for dipoles}

The goal is to derive expressions for multipole fringe fields that satisfy Maxwell's equations.  To ensure the validity of the solution and the
corresponding assumptions, it is important to write all equations explicitly.  For static fields in the absence of any electric current, the
equations for the magnetic field $\vec{B}$ are:
\[
\nabla\times\vec{B} = \nabla\cdot\vec{B} = 0.
\]
For dipole magnets, it is sufficient to consider a two dimensional version of the equations.  Taking $B_x = 0$, we are left with:
\begin{equation}
\partial_y B_y + \partial_z B_z = \partial_y B_z - \partial_z B_y = 0,
\label{maxwell}
\end{equation}
together with:
\begin{equation}
\partial_x B_z = \partial_x B_y = 0,
\end{equation}
which excludes all dependence on $x$. Further, we seek fringe fields which have a possible fall-off along the axis of the magnet
given by the six parameter Enge function \cite{engecite,Berz}:
\[
\frac{1}{1 + e^{E(z)}},
\]
with $E(z)$ given by:
\[
E(z) = \sum_{n=0}^5 a_n\left(\frac{z}{D}\right)^{\!\! n}.
\]
All coefficients $a_n$ are constants determined by modelling and/or experiment. $D$ is the full aperture of the dipole.
The main advantage of the Enge
function is that it is analytic and can be made to tend to asymptotic values arbitrarily fast, if required. The main disadvantage of this
function is that, if all coefficients $a_n$ with $n = 0, \ldots, 5$ are included, varying any one coefficient changes the effective length of
the magnet (in this case, a dipole). Other functions which decay sufficiently rapidly may be used instead of the Enge function
\cite{Kato1,Kato2}. For simplicity, we consider in this paper only the case where $a_1 \neq 0$ and all other coefficients are set to
zero. This has the further advantage that varying $a_1$ only changes the fringe field ``hardness'' without altering its length.
Additionally, because we only have one non-zero coefficient, we can normalise it to $1$ without loss of generality.

Maxwell's equations (\ref{maxwell}) imply:
\[
\Delta_{y,z} B_y = \Delta_{y,z} B_z = 0,
\]
where $\Delta_{y,z} = \partial_y^2 + \partial_z^2$. Both wave equations (for $B_y$ and $B_z$) can easily be solved:
\begin{eqnarray}
B_y & = & e(z + iy) + f(z -iy), \nonumber \\
B_z & = & g(z + iy) + h(z - iy). \nonumber
\end{eqnarray}
Requiring that equations (\ref{maxwell}) be solved as well, we end up with:
\begin{eqnarray}
B_y & = & e(z + iy) + f(z -iy), \nonumber \\
B_z &= & -ie(z + iy) + if(z - iy). \nonumber
\end{eqnarray}
If we further restrict ourselves to real magnetic fields, we obtain:
\begin{eqnarray}
B_y & = & e(z + iy) + \bar{e}(z -iy), \label{maxrel1} \\
B_z & = & - ie(z + iy) + i\bar{e}(z - iy). \label{maxrel2}
\end{eqnarray}
$B_y$ and $B_z$ are given by twice the real and imaginary parts of the function $e(z + iy)$, respectively.

Traditionally, dipole fringe fields are described by:
\begin{eqnarray}
B_y & = & \frac{B_0}{2\pi}\left( \pi - \arctan\!\left(\frac{z}{g + y}\right) - \arctan\!\left(\frac{z}{g - y}\right)\right), \nonumber \\
B_z & = & \frac{B_0}{4\pi}\left( \ln(z^2 + (g + y)^2) - \ln(z^2 + (g - y)^2)\right), \nonumber
\end{eqnarray}
where $B_0$ is the nominal strength of the dipole field, and $g$ is a parameter (related to the aperture of the magnet)
affecting the precise shape of the fringe field.
Using the result:
\[
(z + i(g \pm y))(z - i(g \pm y)) = z^2 + (g \pm y)^2,
\]
$B_z$ may be re-written as:
\[
B_z = \frac{B_0}{4\pi}\left[\ln(z + i(g + y)) + \ln(z - i(g + y)) - \ln(z + i(g - y)) - \ln(z - i(g - y))\right],
\]
whence $e(z + iy)$ has the form:
\[
e(z + iy) = \frac{iB_0}{4\pi}\left[\ln(z + i(g + y)) - \ln(z - i(g - y))\right].
\]
Then, using equation (\ref{maxrel1}) we find $B_y$ to be given by:
\[
B_y = \frac{iB_0}{4\pi}\left[\ln(z + i(g + y)) - \ln(z - i(g + y)) + \ln(z + i(g - y)) - \ln(z - i(g - y))\right],
\]
which may be converted into the same form as the $B_y$ component of the magnetic field used in GPT (General Particle Tracer) \cite{GPT2}, for
example. So, in summary, given a complex function $e(z + iy)$, Maxwell's equations may be satisfied automatically by setting
$B_y = 2\mathrm{Re}(e(z + iy))$ and $B_z = 2\mathrm{Im}(e(z + iy))$.

A possibility for having a magnetic field whose $B_y$ component falls off on-axis is given by the six parameter Enge function:
\begin{equation}
B_y = \frac{1}{2(1 + e^{E(z + iy)})} + \frac{1}{2(1 + e^{E(z - iy)})}.
\label{engeby}
\end{equation}
This would force $B_z$ to have the form:
\begin{equation}
B_z = \frac{-i}{2(1 + e^{E(z + iy)})} + \frac{i}{2(1 + e^{E(z - iy)})},
\label{engebz}
\end{equation}
for some complex function $E(z + iy)$. If we consider the simple case $E(z + iy) = z + iy$ (with aperture having unit diameter) then equations
(\ref{engeby}) and (\ref{engebz}) simplify to:
\begin{eqnarray}
B_y & = & \frac{(1 + e^z\cos y)}{1 + 2e^z\cos y + e^{2z}}, \nonumber \\ 
B_z & = & \frac{-e^z\sin y}{1 + 2e^z\cos y + e^{2z}}. \nonumber
\end{eqnarray}
This may be extended to include as many parameters of the Enge function as desired, the only restriction being that $E = E(z + iy)$.
So, if $E(z + iy) = b_1(z + iy) + b_2(z + iy)^2$, with $b_1$ and $b_2$ arbitrary constants, we have:
\begin{eqnarray}
B_y & = & \frac{(1 + e^f\cos h)}{1 + 2e^f\cos h + e^{2f}}, \nonumber \\
B_z & = & \frac{-e^f\sin h}{1 + 2e^f\cos h + e^{2f}}, \nonumber
\end{eqnarray}
where $f = b_1z + b_2(z^2 - y^2)$ and $h = y(b_1 + 2b_2z)$. This can be generalized to:
\[
E(z + iy) = \sum_{n = 1}^N b_n (z + iy)^n,
\]
with $b_j$ constants if an $N$ parameter Enge function is desired. Indeed, any function can be used, provided it is a function of $z + iy$.

\begin{figure}[htb]
\centering
\begin{tabular}{cc}
\includegraphics[trim=0 0 0 8.5cm,clip,width=0.45\textwidth]{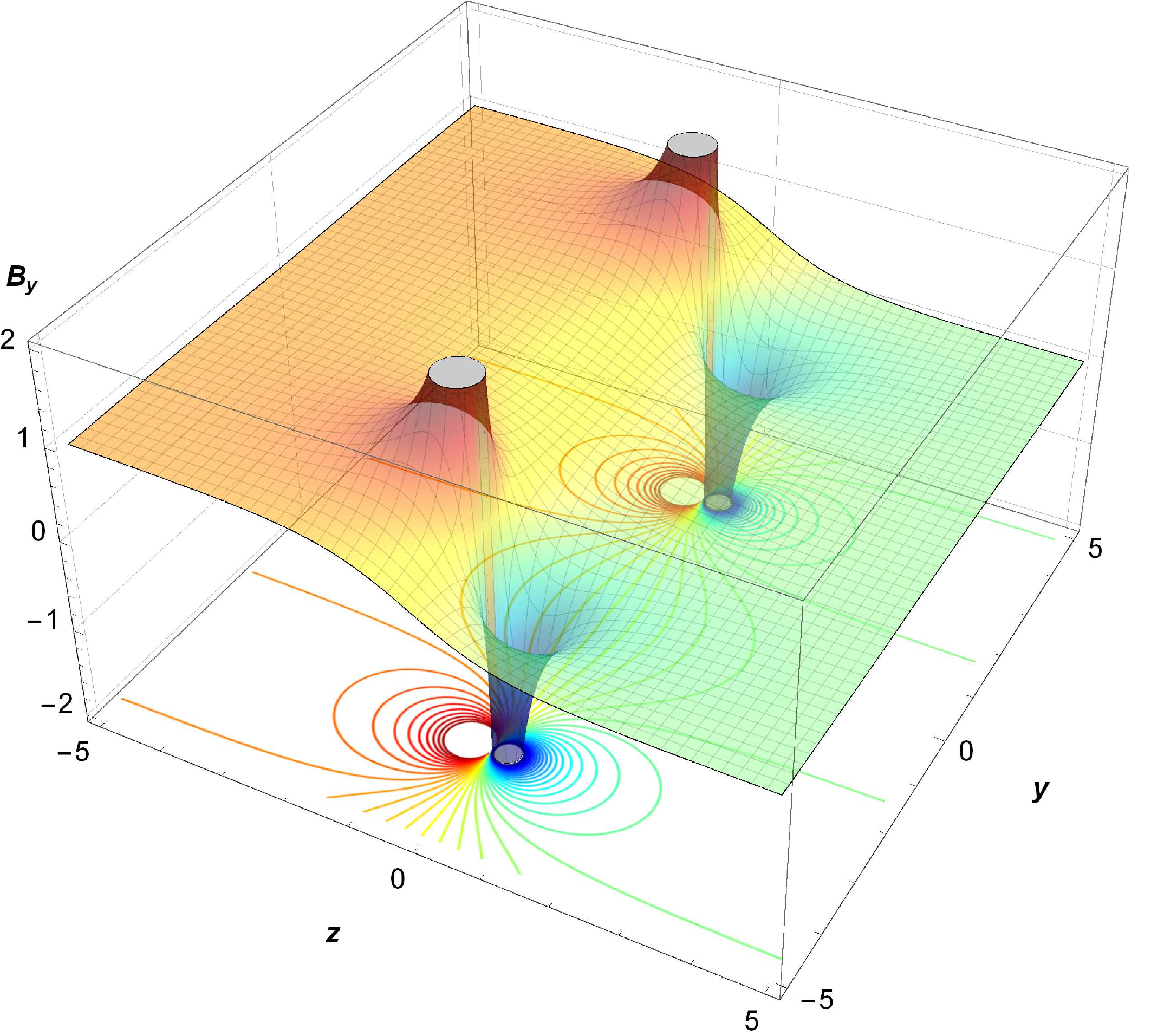} &
\includegraphics[trim=0 0 0 8.5cm,clip,width=0.45\textwidth]{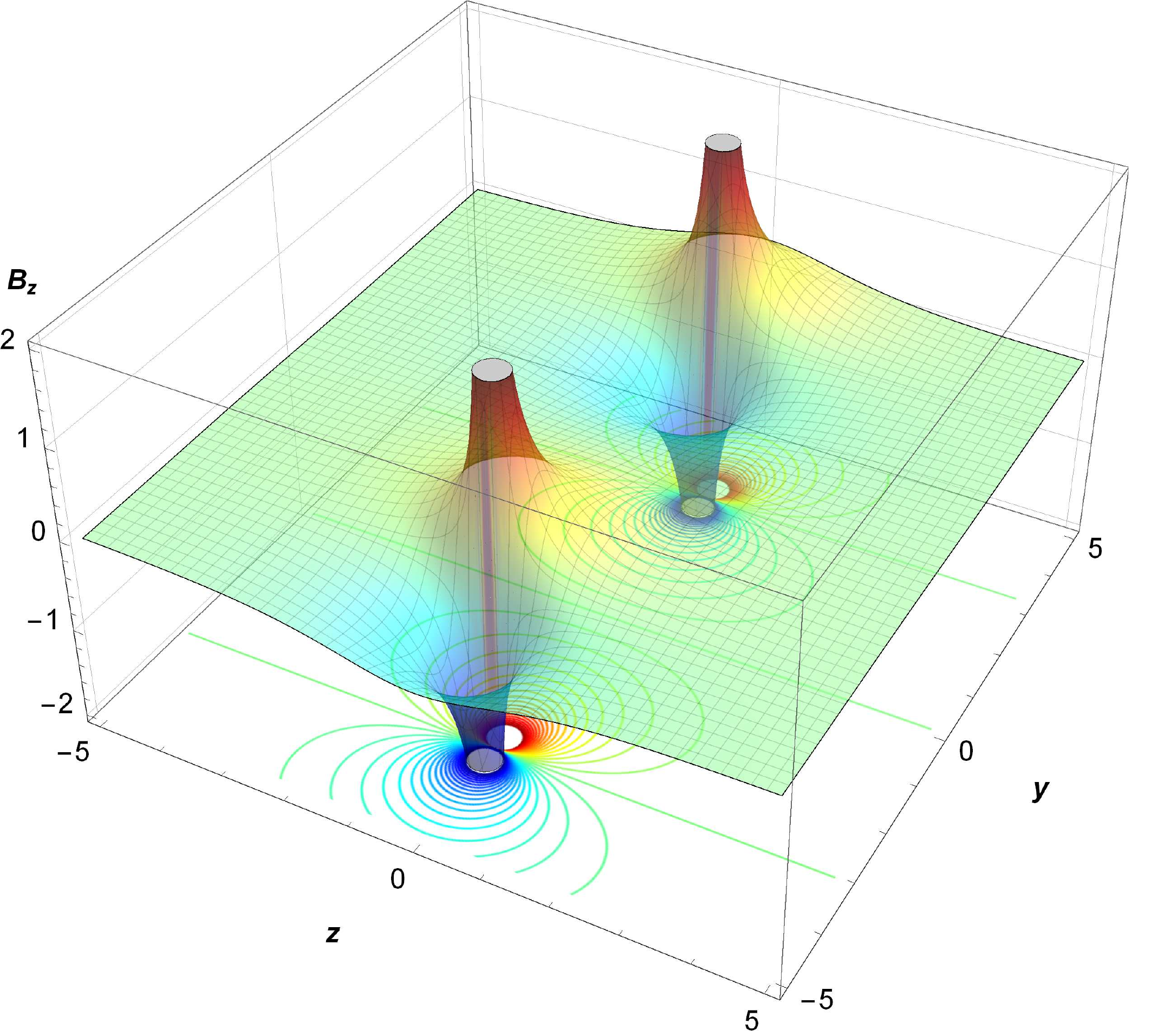}
\end{tabular}
\caption{Dipole field components $B_y$ and $B_z$ with and without fringe field.}
\label{engebybz}
\end{figure}

%\subsection{Analysis of Results for Dipole Fringe Fields}
We can plot the fringe fields (\ref{engeby}) and (\ref{engebz}) in the simplest case ($E(z + iy) = z + iy$), as shown in Fig.~\ref{engebybz}.
Each field component has singularities, of which two are visible in the plots.
From the denominator of (\ref{engeby}) and (\ref{engebz}), it can be seen that the singularities appear when (re-introducing the arbitrary
aperture diameter $D$) $1 + 2e^{z/D}\cos(y/D) + e^{2z/D} = 0$ or $\cos(y/D) = - 1 \rightarrow y/D = \pm n\pi$ where $n$ is an
integer. This does not cause any problems for modelling the field as it is always possible to arrange that the singularities are located
outside the region of interest. It is not possible to avoid the appearance of singularities since any solution to Laplace's equation either
has singularities or is constant. In practical terms, the singularities can be associated with the current in the windings of the dipole.

Note that the singularities have a $\tan$-like behaviour and that, at the exact point where $z = 0$, the value of the field is precisely
half of the maximum and everything is smooth.  This can be verified by setting $z = 0$ in the expressions for the dipole fringe fields above and
using l'Hopital's rule.

\begin{figure}[htb]
\centering
\includegraphics[trim=0 0 0 8.5cm,clip,width=0.45\textwidth]{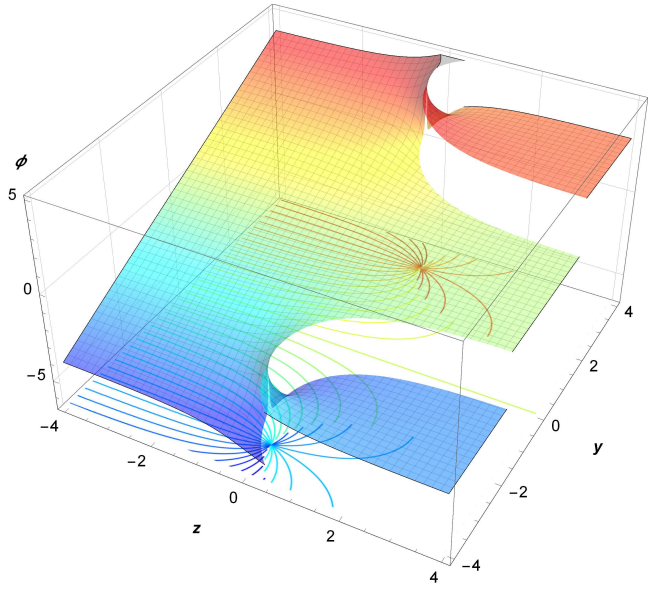}
\caption{Scalar potential $\varphi$ for dipole fringe field together with contours of constant $\varphi$ showing possible pole-face profiles.}
\label{dippf}
\end{figure}

%\subsection{Potentials for dipole fringe fields}
Locally, it is possible to find potentials for the dipole fringe fields. The usual vector potential $\vec{A}$ is related to the field by
$\vec{B} = \nabla\times\vec{A}$.  For magnetostatic fields in the absence of currents, the field can be derived from a scalar potential
$\varphi$, by $\vec{B} = \nabla\varphi$. The existence of the vector potential is ensured by the Maxwell equation $\nabla\cdot\vec{B} = 0$,
while the scalar potential exists because $\nabla\times\vec{B} = 0$. For the simple case discussed above, the scalar potential is given by:
\[
\varphi = y + i\ln\sqrt{\frac{1 + e^{z + iy}}{1 + e^{z - iy}}} + \mathrm{constant.}
\]
with the only gauge freedom being given by the constant.  From the scalar potential, it is possible to obtain a description of the pole-face
geometry, since this is given by surfaces where $\varphi$ is constant. This is shown in Fig.~\ref{dippf} where two profiles of the pole-face
can be seen. It should be remembered that there is a scale invariance in the expressions depending on the dimensions of the dipole.
Figure \ref{dippf} also shows that the pole-face profiles encompass the singularities: this is consistent with the assertion made earlier that
the singularities are associated with the current in the coils of the dipole.

The vector potential has extensive gauge freedom and, in one of its simplest forms, can be given as:
\[
A_x = z - \ln\sqrt{(1+e^{z+iy})(1+e^{z-iy})},
\]
with the other components of $\vec{A}$ set to zero. 

\section{General three dimensional solution in a fringe field}\label{sectiongeneralthreedsolution}

General expressions for the fringe fields in a dipole followed from writing Maxwell's equations in the form (\ref{maxwell}).  In a dipole,
we only needed to consider two field components ($B_y$ and $B_z$) as functions of two co-ordinates ($y$ and $z$).  For higher order
multipoles, it is necessary to consider the dependence of all three field components on all three co-ordinates.  One approach might be to
look for solutions to the three dimensional Laplacian.  A formal solution (due to Whittaker \cite{Piaggio, Whit1, Whit2}) is known, and can
be expressed as:
\[
\varphi(x,y,z) \ = \ \int_0^{2\pi}f(z + ix\cos\vartheta + iy\sin\vartheta)\ {\rm d}\vartheta,
\]
where:
\[
\Delta_{x, y, z}\varphi \equiv \partial_x^2\varphi + \partial_y^2\varphi + \partial_z^2\varphi \ = \ 0.
\]
However, it is difficult to use this result to find expressions that are useful in practice.  The only well-known practical solution is
$f = (z + ix\cos\vartheta + iy\sin\vartheta)^{- 1}$, which gives the standard solution $\varphi = 2\pi/r$ with $r = \sqrt{x^2 + y^2 + z^2}$.

An alternative approach, which we develop here, is to define new variables in terms of which Maxwell's equations can be written for a
general three-dimensional magnetostatic field in a form very similar to (\ref{maxwell}).  This raises the possibility of finding expressions
for fringe fields in higher order multipoles by applying similar techniques to those described in Section 2 for dipoles.

To write Maxwell's equations for three-dimensional fields in a form similar to (\ref{maxwell}), we define new variables:
\begin{eqnarray}
u & = & \frac{1}{\sqrt{2}}(x + iy), \nonumber \\
v & = & \frac{1}{\sqrt{2}}(x - iy), \nonumber \\
\zeta & = & \sqrt{2}z. \nonumber
\end{eqnarray}
We express the magnetic field in terms of components:
\begin{eqnarray}
B_u & = & \frac{1}{\sqrt{2}}(B_x + i B_y), \nonumber \\
B_v & = & \frac{1}{\sqrt{2}}(B_x - i B_y), \nonumber \\
B_\zeta & = & \frac{1}{\sqrt{2}} B_z. \nonumber
\end{eqnarray}
In terms of the new variables, Maxwell's equations can be written:
\begin{eqnarray}
\label{maxnew1}
\partial_u B_u + \partial_z B_z & = & 0, \\
\label{maxnew2}
\partial_v B_v + \partial_z B_z & = & 0, \\
\label{maxnew3}
\partial_z B_u - \partial_v B_z & = & 0, \\
\label{maxnew4}
\partial_z B_v - \partial_u B_z & = & 0.
\end{eqnarray}

From (\ref{maxnew1}) and (\ref{maxnew2}), one can see immediately that, in the absence of any fringe fields, the general solution of Maxwell's
equations for any magnet, acting transversely only and without fringe ($B_\zeta = 0$) is given by:
\begin{eqnarray}
B_u & = & f(v), \nonumber \\
B_v & = & h(u), \nonumber
\end{eqnarray}
for any functions $f(v)$ and $h(u)$. The case of a multipole of order $n$ ($n=1$ for a quadrupole, $n=2$ for a sextupole, and so on) is given
by:
\begin{eqnarray}
B_u & = & iv^n, \nonumber \\
B_v & = & - iu^n, \nonumber \\
B_\zeta & = & 0. \nonumber
\end{eqnarray}
For example, a quadrupole is described by $B_u = iv$, $B_v = - iu$ and $B_\zeta = 0$.

We now assume a very general form that we expect multipole fringe fields should take:
\begin{eqnarray}
\label{ass1}
B_u & = & \frac{f_1(u,v,\zeta) + f_2(u,v,\zeta)e^\zeta}{(1 + 2f_3(u,v)e^\zeta + e^{2\zeta})}, \\
\label{ass2}
B_v & = & \frac{f_4(u,v,\zeta) + f_5(u,v,\zeta)e^\zeta}{(1 + 2f_3(u,v)e^\zeta + e^{2\zeta})}, \\
\label{ass3}
B_\zeta & = & \frac{f_6(u,v,\zeta) + f_7(u,v,\zeta)e^\zeta}{(1 + 2f_3(u,v)e^\zeta + e^{2\zeta})}.
\end{eqnarray}
The form of this field is based on a generalisation of the fringe fields for the dipole case.
Substituting (\ref{ass1})--(\ref{ass3}) into Maxwell's equations (\ref{maxnew1})--(\ref{maxnew4})
gives a set of constraints on the possible forms of the functions $f_1$, $f_2$ etc.
To obtain useful expressions for multipole fringe fields, we need to find solutions satisfying the
various constraints: this is the task that we address in the remainder of this section.

Essentially, there are only two types of derivative that we need to consider.  These are:
\begin{eqnarray}
\partial_u B_u & = & \frac{\partial_u f_1 + \partial_u f_2 e^\zeta}{A} - \frac{2(f_1 + f_2 e^\zeta)e^\zeta \partial_u f_3}{A^2}, \nonumber \\
\partial_\zeta B_u & = & \frac{\partial_\zeta f_1 + \partial_\zeta f_2 e^\zeta + f_2 e^\zeta}{A}
- \frac{2(f_1 + f_2 e^\zeta)(e^\zeta f_3 + e^{2\zeta})}{A^2}, \nonumber
\end{eqnarray}
where $A = 1 + 2f_3 e^\zeta + e^{2\zeta}$. For the remaining derivatives, we simply implement the following changes sequentially:
\begin{eqnarray}
\partial_v B_u \equiv \partial_u B_u & {\rm under} & (u \leftrightarrow v), \nonumber \\
\partial_u B_v \equiv \partial_u B_u & {\rm under} & (f_1 \rightarrow f_4, \ f_2 \rightarrow f_5), \nonumber \\
\partial_v B_v \equiv \partial_u B_v & {\rm under} & (u \leftrightarrow v), \nonumber \\
\partial_\zeta B_v \equiv \partial_\zeta B_u & {\rm under} & (f_1 \rightarrow f_4, \ f_2 \rightarrow f_5), \nonumber \\
\partial_u B_\zeta \equiv \partial_u B_u & {\rm under} & (f_1 \rightarrow f_6, \ f_2 \rightarrow f_7), \nonumber \\
\partial_v B_\zeta \equiv \partial_u B_\zeta & {\rm under} & (u \leftrightarrow v), \nonumber \\
\partial_\zeta B_\zeta \equiv \partial_\zeta B_u & {\rm under} & (f_1 \rightarrow f_6, \ f_2 \rightarrow f_7). \nonumber 
\end{eqnarray}
As all equations (\ref{maxnew1})--(\ref{maxnew4}) are equal to zero, we can take out a factor of $A^2$ to give:
\begin{eqnarray}
(\partial_x f_1 + \partial_x f_2 e^\zeta)(1 + 2f_3 e^\zeta + e^{2\zeta})
+ (\partial_y f_4 + \partial_y f_5 e^\zeta)(1 + 2f_3 e^\zeta + e^{2\zeta}) \quad & & \nonumber \\
+ (\partial_\zeta f_6 + \partial_\zeta f_7 e^\zeta + f_7 e^\zeta)(1 + 2f_3 e^\zeta + e^{2\zeta}) \quad & & \nonumber \\
- 2(f_1 + f_2 e^\zeta)e^\zeta \partial_x f_3 - 2(f_4 + f_5 e^\zeta)e^\zeta \partial_y f_3
- 2(f_6 + f_7 e^\zeta)(f_3 e^\zeta + e^{2\zeta}) & = & 0, \nonumber \\
& & \nonumber \\
(\partial_x f_4 + \partial_x f_5 e^\zeta)(1 + 2f_3 e^\zeta + e^{2\zeta})
- (\partial_y f_1 + \partial_y f_2 e^\zeta)(1 + 2f_3 e^\zeta + e^{2\zeta}) \quad & & \nonumber \\
- 2(f_4 + f_5 e^\zeta)e^\zeta \partial_x f_3 + 2(f_1 + f_2 e^\zeta)e^\zeta \partial_y f_3 & = & 0, \nonumber \\
& & \nonumber \\
(\partial_x f_6 + \partial_x f_7 e^\zeta)(1 + 2f_3 e^\zeta + e^{2\zeta})
- (\partial_\zeta f_1 + \partial_\zeta f_2 e^\zeta + f_2 e^\zeta)(1 + 2f_3 e^\zeta + e^{2\zeta}) \quad & & \nonumber \\
- 2(f_6 + f_7 e^\zeta)e^\zeta \partial_x f_3 + 2(f_1 + f_2 e^\zeta)(f_3 e^\zeta + e^{2\zeta}) & = & 0, \nonumber \\
& & \nonumber \\
(\partial_y f_6 + \partial_y f_7 e^\zeta)(1 + 2f_3 e^\zeta + e^{2\zeta})
- (\partial_\zeta f_4 + \partial_\zeta f_5 e^\zeta + f_5 e^\zeta)(1 + 2f_3 e^\zeta + e^{2\zeta}) \quad & & \nonumber \\
- 2(f_6 + f_7 e^\zeta)e^\zeta \partial_x f_3 + 2(f_4 + f_5 e^\zeta)(f_3 e^\zeta + e^{2\zeta}) & = & 0. \nonumber
\end{eqnarray}
We can now equate coefficients of $e^\zeta$ giving:
\begin{eqnarray}
\label{eq1}
e^{3\zeta}: & \ & \partial_u f_2 + \partial_\zeta f_7 - f_7 \ = \ 0 \\
\label{eq2}
        & \ & \partial_v f_5 + \partial_\zeta f_7 - f_7 \ = \ 0 \\
\label{eq3}
        & \ & \partial_u f_7 - \partial_\zeta f_5 + f_5 \ = \ 0 \\
\label{eq4}
        & \ & \partial_v f_7 - \partial_\zeta f_2 + f_2 \ = \ 0 \\
\nonumber \\
\label{eq5}
e^{2\zeta}: & \ & f_2 \partial_u f_3 + f_6 - f_3 f_7 \ = \ 0 \\
\label{eq6}
        & \ & f_5 \partial_v f_3 + f_6 - f_3 f_7 \ = \ 0 \\
\label{eq7}
        & \ & f_7 \partial_u f_3 - f_4 + f_3 f_5 \ = \ 0 \\
\label{eq8}
        & \ & f_7 \partial_v f_3 - f_1 + f_3 f_2 \ = \ 0 \\
\nonumber \\
\label{eq9}
e^\zeta:    & \ & f_1 \partial_u f_3 + f_3 f_6 - f_7 \ = \ 0 \\
\label{eq10}
        & \ & f_4 \partial_v f_3 + f_3 f_6 - f_7 \ = \ 0 \\
\label{eq11}
        & \ & f_6 \partial_u f_3 + f_5 - f_3 f_4 \ = \ 0 \\
\label{eq12}
        & \ & f_6 \partial_v f_3 + f_2 - f_3 f_1 \ = \ 0 \\
\nonumber \\
\label{eq13}
e^0:    & \ & \partial_u f_1 + \partial_\zeta f_6 \ = \ 0 \\
\label{eq14}
        & \ & \partial_v f_4 + \partial_\zeta f_6 \ = \ 0 \\
\label{eq15}
        & \ & \partial_u f_6 - \partial_\zeta f_4 \ = \ 0 \\
\label{eq16}
        & \ & \partial_v f_6 - \partial_\zeta f_1 \ = \ 0.
\end{eqnarray}
Note that we have not included all the steps and the above equations represent the original set with all possible simplifications taking into
account the set itself. Note also that equations (\ref{eq1})--(\ref{eq4}) and (\ref{eq13})--(\ref{eq16}) may be solved
independently of the rest and can therefore be dealt with later. From equation (\ref{eq8}), using (\ref{eq5}) and (\ref{eq9}), we see that:
\[
f_7(\partial_v f_3 \, \partial_u f_3 + f_3^2 - 1) = 0.
\]
Had we looked at equations (\ref{eq7}) and (\ref{eq11}) instead, using (\ref{eq6}) and (\ref{eq10}), we would have had:
\[
f_6(\partial_v f_3 \, \partial_u f_3 + f_3^2 - 1) = 0,
\]
with the same result from equation (\ref{eq12}). Now, $f_6$ and $f_7$ cannot both be zero as this would mean $B_\zeta = 0$.
Therefore, we must have:
\[
\partial_v f_3 \, \partial_u f_3 + f_3^2 - 1 = 0.
\]
The general solution for $f_3$, using the method of characteristics \cite{Piaggio}, is:
\[
f_3 = \sin h(u,v),
\]
with
\begin{equation}
h(u,v) = \frac{1}{b}u + bv + c, \label{hdefinition}
\end{equation}
where $b$ and $c$ are constant. Without loss of generality and for simplicity, we set $c = 0$. Substituting
this back into (\ref{eq5})--(\ref{eq12}) gives the relations:
\begin{eqnarray}
f_2 & = & b^2 f_5, \nonumber \\
f_1 & = & b^2 f_4. \nonumber
\end{eqnarray}
The equations reduce to just two independent equations that may be written as:
\begin{eqnarray}
\label{eqn6}
\frac{1}{b}f_2 \cos h + f_6 - f_7 \sin h & = & 0, \\
\label{eqn11}
f_6 \cos h + \frac{1}{b}f_2 - \frac{1}{b}f_1 \sin h & = & 0.
\end{eqnarray}
Using $f_1 = b^2 f_4$ and equations (\ref{eq13}) and (\ref{eq14}) we see that we require
\[
b^2 \partial_u f_4 = \partial_v f_4,
\]
which can again be solved by the method of characteristics to give $f_4 \ = \ f_4(h, z)$. Using this with equations (\ref{eq15}) and
(\ref{eq16}) we see that $f_6 \ = \ f_6(h, z)$. Similarly, $f_2 \ = \ b^2 f_5$ applied to (\ref{eq1}) and (\ref{eq2}) and, subsequently
(\ref{eq3}) and (\ref{eq4}) gives $f_5 \ = \ f_5(h, z)$ and $f_7 \ = \ f_7(h, z)$. This leaves six equations to be satisfied from the original
system (\ref{eq1})--(\ref{eq16}), namely (\ref{eqn6}) and (\ref{eqn11}) together with:
\begin{eqnarray}
\label{eqn1}
\partial_u f_2 + \partial_\zeta f_7 - f_7 & = & 0 \\
\label{eqn4}
\partial_v f_7 - \partial_\zeta f_2 + f_2 & = & 0 \\
\label{eqn13}
\partial_u f_1 + \partial_\zeta f_6 & = & 0 \\
\label{eqn16}
\partial_v f_6 - \partial_\zeta f_1 & = & 0.
\end{eqnarray}
After cross-differentiation, equations (\ref{eqn13}) and (\ref{eqn16}) give
\begin{eqnarray}
\partial_{u,v}^2 f_6 + \partial^2_\zeta f_6 & = & 0, \nonumber \\
\partial_{u,v}^2 f_1 + \partial^2_\zeta f_1 & = & 0. \nonumber
\end{eqnarray}

Now, we can re-express the partial derivatives in $u$ and $v$ in terms of $h$ only. The equations simplify to:
\[
\triangle f_1 = \triangle f_6 = 0,
\]
where $\triangle \equiv \partial_h^2 + \partial_\zeta^2$.  We introduce the co-ordinates:
\begin{eqnarray}
w & = & h + i\zeta, \nonumber \\
\bar{w} & = & h - i\zeta. \nonumber
\end{eqnarray}
Note that this operation is equivalent to complex conjugation in the $\zeta$ co-ordinate only and the
function $h$ is untouched. Therefore, we have the solutions:
\begin{eqnarray}
f_1 & = & p_1(w) + q_1(\bar{w}), \nonumber \\
f_6 & = & p_6 (w) + q_6(\bar{w}). \nonumber
\end{eqnarray}
Substituting this back into (\ref{eqn13}) and (\ref{eqn16}), we see that the solutions are further constrained to:
\[
f_1 = - ibp_6 + ibq_6 + k,
\]
with $k$ constant, from which we can get $f_4$ via $f_4 = \frac{1}{b^2}f_1$. Subsequently, we can get
$f_2$ from (\ref{eqn11}) and hence $f_5$ via $f_5 = \frac{1}{b^2}f_2$ and $f_7$ from (\ref{eqn6}).
The general result, in terms of $p_6$ and $q_6$ may be summarised as follows (with $h = \frac{1}{b}u + bv$):
\begin{eqnarray}
\label{res1}
f_1 & = & - ibp_6 + ibq_6 + k, \\
\label{res2}
f_2 & = & (- ibp_6 + ibq_6 + k)\sin h - (bp_6 + bq_6)\cos h, \\
\label{res3}
f_3 & = & \sin h, \\
\label{res4}
f_4 & = & \frac{1}{b}\left(- ip_6 + iq_6 + \frac{k}{b}\right), \\
\label{res5}
f_5 & = & \frac{1}{b}\left(- ip_6 + iq_6 + \frac{k}{b}\right)\sin h - \frac{1}{b}\left(p_6 + q_6\right)\cos h, \\
\label{res6}
f_6 & = & p_6 + q_6, \\
\label{res7}
f_7 & = & \left(p_6 + q_6\right)\sin h + \left(- ip_6 + iq_6 + \frac{k}{b}\right)\cos h.
\end{eqnarray}

We are left with equations (\ref{eqn1}) and (\ref{eqn4}) to be solved. Upon substitution of (\ref{res1})--(\ref{res7}), these are
actually seen to be trivially satisfied with no further constraints on any of the functions $f_i$. In fact, the results constitute a Darboux
transformation, where, given a solution to Maxwell's equations expressed by (\ref{eq13})--(\ref{eq16}), a new solution may be created,
given by (\ref{ass1})--(\ref{ass3}), provided (\ref{eq1})--(\ref{eq12}) are satisfied. Further, the results can be seen to imply
the following solution to Maxwell's equations:
\begin{eqnarray}
\label{resbu}
B_u & = & - ibf(h + i\zeta) + ibg(h - i\zeta), \\
\label{resbv}
B_v & = & - \frac{i}{b}f(h + i\zeta) + \frac{i}{b}g(h - i\zeta), \\
\label{resbzp}
B_\zeta & = & f(h + i\zeta) + g(h - i\zeta).
\end{eqnarray}
with $h$ defined by (\ref{hdefinition}). This solution could have been arrived at by a shorter method, but the above workings show that,
for the kind of functions we are interested in, it is the only solution that fits. We may have been forced to consider more complicated
functions by introducing non-linearities in the coordinate dependance, for example. Note that the constant $b$ is purely a scaling constant of
the coordinates, as well as giving a proportionality between $B_u$ and $B_v$. For the purposes of the remainder of this article, we shall refer
to the solution (\ref{resbu}), (\ref{resbv}) and (\ref{resbzp}) as an \emph{elementary solution} of Maxwell's equations in three dimensions.
In the original (Cartesian) co-ordinates ($x = (u + v)/\sqrt{2}$, $y = (u - v)/\sqrt{2}i$, $z = \zeta/\sqrt{2}$) the elementary solution reads:
\begin{eqnarray}
\label{resbx}
B_x & = & \frac{1}{\sqrt{2}}(B_u + B_v) = - idf(h + i\sqrt{2}z) + idg(h - i\sqrt{2}z), \\
\label{resby}
B_y & = & \frac{1}{\sqrt{2}i}(B_u - B_v) = ef(h + i\sqrt{2}z) - eg(h - i\sqrt{2}z), \\
\label{resbz}
B_z & = & \sqrt{2}B_\zeta = \sqrt{2}f(h + i\sqrt{2}z) + \sqrt{2}g(h - i\sqrt{2}z),
\end{eqnarray}
where:
\begin{eqnarray}
d & = & \frac{1}{\sqrt{2}} \left( \frac{1}{b} + b \right), \nonumber \\
e & = & \frac{1}{\sqrt{2}} \left( \frac{1}{b} - b \right), \nonumber 
\end{eqnarray}
and $h$ is now expressed as:
\[
h = dx + iey.
\]

For some applications, it is useful to have expressions for the scalar potential $\varphi$ and vector potential $\vec{A}$, from which the
field can be derived by:
\begin{equation}
\vec{B} = \nabla \varphi, \label{scalarpotentialdefn}
\end{equation}
or:
\begin{equation}
\vec{B} = \nabla \times \vec{A}. \label{vectorpotentialdefn}
\end{equation}
It can be readily verified by substitution into (\ref{scalarpotentialdefn}) that an expression for the scalar potential for the
elementary solution can be written:
\[
\varphi = -i \tilde{f}(h + i \sqrt{2}z) + i \tilde{g}(h - i \sqrt{2}z),
\]
where:
\begin{eqnarray}
\tilde{f}(\zeta) & = & \int_0^\zeta f(\zeta^\prime) \, \mathrm{d}\zeta^\prime, \nonumber \\
\tilde{g}(\zeta) & = & \int_0^\zeta g(\zeta^\prime) \, \mathrm{d}\zeta^\prime. \nonumber
\end{eqnarray}
Setting the lower limit of each integral to zero fixes the gauge so that $\varphi = 0$ at the origin; a change of gauge can
be made simply by adding a constant to $\varphi$.
It can be verified by substitution into (\ref{vectorpotentialdefn}) that the vector potential for the elementary solution
may be written:
\begin{eqnarray}
A_x & = & 0, \nonumber \\
A_y & = & \frac{\sqrt{2}}{d} \left( \tilde{f}(h + i \sqrt{2}z) + \tilde{g}(h - i \sqrt{2}z) \right), \nonumber \\
A_z & = & -\frac{e}{d} \left( \tilde{f}(h + i \sqrt{2}z) - \tilde{g}(h - i \sqrt{2}z) \right). \nonumber
\end{eqnarray}
Here, we have chosen a gauge in which $A_x = 0$.

Note that any function of $h \pm i\zeta$ may also be expressed as a function of $\zeta \mp ih$: we shall use both forms in what
follows in order to simplify the expressions as appropriate.

\section{Multipole magnets with general fringe fields}

The relations $f_2 = b^2 f_5$ and $f_1 = b^2 f_4$ found earlier imply that $B_u \propto B_v$: this means that no physical
magnetic fields can be represented this way. However, because of the linearity of Maxwell's equations, any linear combination of
elementary solutions gives a solution to Maxwell's equations. Physical solutions corresponding to multipole fields can be constructed
by taking appropriate combinations of elementary solutions. Further (physical) constraints are that the field decays as
$z \rightarrow \infty$ and that the field matches the nominal multipole field inside the magnet. In this section, we describe in detail
the procedure for constructing an expression for the fringe field in a quadrupole, and then generalise our results to multipoles
of any order.  To obtain the correct behaviour of the field as a function of $z$, the functions $f(h + i\zeta)$ and $g(h - i\zeta)$
are each written as a product of two factors.  The first factor, $(h + i\zeta)^n$ and $(h - i\zeta)^n$ respectively, represents the
(nominal) multipole for some constant $n$.  The second factor, $F(h + i\zeta)$ and $G(h - i\zeta)$ respectively, are multiplicative
functions which are chosen to give the desired decay as $z$ becomes large ($z \rightarrow \infty$), and to give the correct field
in the body of the magnet. Thus, we write:
\begin{eqnarray}
f(h + i\zeta) & = & (h + i\zeta)^nF(h + i\zeta), \label{multipolefexpression} \\
g(h - i\zeta) & = & (h - i\zeta)^nG(h - i\zeta). \label{multipolegexpression}
\end{eqnarray}
Following the form of (\ref{resbu}), (\ref{resbv}) and (\ref{resbzp}), let $B_u$, $B_v$ and $B_\zeta$ be given by linear combinations
of the elementary solution:
\begin{eqnarray}
B_u & = & \sum_{j = 1}^{m}c_j \left( - ib_j(h_j + i\zeta)^nF_j(h_j + i\zeta) + ib_j(h_j - i\zeta)^nG_j(h_j - i\zeta) \right), \label{busummation} \\
B_v & = & \sum_{j = 1}^{m}c_j \left( - \frac{i}{b_j}(h_j + i\zeta)^nF_j(h_j + i\zeta) + \frac{i}{b_j}(h_j - i\zeta)^nG_j(h_j - i\zeta) \right),
\label{bvsummation} \\
B_\zeta & = & \sum_{j = 1}^{m}c_j \left( (h_j + i\zeta)^nF_j(h_j + i\zeta) + (h_j - i\zeta)^nG_j(h_j - i\zeta) \right). \label{bzetasummation}
\end{eqnarray}
where $h_j = \frac{1}{b_j}u + b_jv$, and $m$ is a constant determined by the number of copies of the elementary solution needed
to construct a physical multipole of the desired order. The constant $n$ determines the order of the multipole.  Within the body of
the magnet, $B_u = iv^n$, $B_v = -iu^n$, and $B_\zeta = 0$. We also expect to have to satisfy several relations between
the constants $b_j$ and $c_j$.

\subsection{Quadrupole magnets}
\label{quads}

A quadrupole field is obtained by putting $n = 1$ in equations (\ref{multipolefexpression}) and (\ref{multipolegexpression}).
Therefore, within the body of the magnet (far from any fringe field) $F(h_j + i\zeta) = G(h_j - i\zeta) = \pm 1$ for any $j$.
The relative signs of $F(h_j + i\zeta)$ and $G(h_j - i\zeta)$ are to be determined by the required gradient inside the quadrupole
and the fact that there should be no dependence on $z$ at the centre of the magnet.  The required behaviour of the fields
in the body of the magnet can be obtained by taking $m = 2$ (larger values of $m$ can be used, but are not required).
Then, putting $B_u = iv$, $B_v = -iu$ and $B_\zeta = 0$ gives:
\begin{eqnarray}
- ic_1b_1\bigg(\frac{u}{b_1} + b_1v + iz\bigg) - ic_1b_1\bigg(\frac{u}{b_1} + b_1v - iz\bigg)
- ic_2b_2\bigg(\frac{u}{b_2} + b_2v + iz\bigg) - ic_2b_2\bigg(\frac{u}{b_2} + b_2v - iz\bigg) & = & iv, \nonumber \\
- i\frac{c_1}{b_1}\bigg(\frac{u}{b_1} + b_1v + iz\bigg) - i\frac{c_1}{b_1}\bigg(\frac{u}{b_1} + b_1v - iz\bigg)
- i\frac{c_2}{b_2}\bigg(\frac{u}{b_2} + b_2v + iz\bigg) - i\frac{c_2}{b_2}\bigg(\frac{u}{b_2} + b_2v - iz\bigg) & = & - iu, \nonumber \\
- ic_1\bigg(\frac{u}{b_1} + b_1v + iz\bigg) - ic_1\bigg(\frac{u}{b_1} + b_1v - iz\bigg)
- ic_2\bigg(\frac{u}{b_2} + b_2v + iz\bigg) - ic_2\bigg(\frac{u}{b_2} + b_2v - iz\bigg) & = & 0. \nonumber
\end{eqnarray}
Thus, we have three equations to satisfy:
\begin{eqnarray}
2c_1b_1^2 + 2c_2b_2^2 & = & - 1, \nonumber \\
2\frac{c_1}{b_1^2} + 2\frac{c_2}{b_2^2} & = & 1, \nonumber \\
c_1 + c_2 & = & 0. \nonumber
\end{eqnarray}
Therefore, inside the magnet, the coefficients $b_j$ and $c_j$ must satisfy the following constraints:
\begin{eqnarray}
\label{bq}
b_1 & = & \pm \frac{1}{b_2}, \\
\label{cq}
c_1 & = & - c_2 = \frac{1}{2} \left(b_2^2 - \frac{1}{b_2^2}\right)^{-1}.
\end{eqnarray}
We are left with the freedom of choosing one constant, which we take to be $b_2$. We shall consider the significance of this constant
in more detail later, but for now we note that to avoid $b_1$, $c_1$ or $c_2$ becoming singular, $b_2$ must not be equal to $0$ or $1$.
Also, the field is unchanged if we replace $b_2$ by $1/b_2$. Therefore, before implementing any simplifications, we have the following:
%Therefore, dropping the multiplicative constant $c_1 = -c_2$
%(which appears as a factor in every term in each field component) the components of the field in a quadrupole can be written as:
\begin{eqnarray*}
B_x & = & c_1d_1f(\zeta - ih_1) - c_1d_1g(\zeta + ih_1) + c_2d_2f(\zeta - ih_2) - c_2d_2g(\zeta + ih_2), \\
B_y & = & ic_1e_1f(\zeta - ih_1) - ic_1e_1g(\zeta + ih_1) + ic_2e_2f(\zeta - ih_2) - ic_2e_2g(\zeta + ih_2), \\
B_z & = & ic_1f(\zeta - ih_1) + ic_1g(\zeta + ih_1) + ic_2f(\zeta + ih_2) + ic_2g(\zeta + ih_2).
\end{eqnarray*}
Physically, it is necessary that our choices of constants should give a real field in the end. Now, since the two co-ordinates are related
via $u = \bar{v}$ (where the bar denotes complex conjugation) $h_2 = \bar{h}_1$ to within a factor $\pm 1$. In detail, the first constraint
above implies the following two cases:
\begin{eqnarray*}
b_1 = + \frac{1}{b_2}: & \Rightarrow & d_1 = d_2, e_1 = - e_2, h_1 = \bar{h}_2, \\
b_1 = - \frac{1}{b_2}: & \Rightarrow & d_1 = - d_2, e_1 = e_2, h_1 = - \bar{h}_2.
\end{eqnarray*}
Now, in either case, the second constraint ($c_1 = - c_2$) is the same and we factor the resulting constant out for clarity so, in terms of
magnetic field components we are left with (for $b_1 = + 1/b_2$):
\begin{eqnarray*}
B_x & = & df(\zeta - ih) - dg(\zeta + ih) - df(\zeta - i\bar{h}) + dg(\zeta + i\bar{h}), \\
B_y & = & ief(\zeta - ih) - ieg(\zeta + ih) + ief(\zeta - i\bar{h}) - ieg(\zeta + i\bar{h}), \\
B_z & = & if(\zeta - ih) + ig(\zeta + ih) - if(\zeta - i\bar{h}) - ig(\zeta + i\bar{h}),
\end{eqnarray*}
where we have put $h_1 = h$. On the other hand, if we take $b_1 =  - 1/b_2$, then the field components read:
%from which we see that all components of the magnetic field are real as long as the functions $f$ and $g$ are real and $f = -g$.
\begin{eqnarray*}
B_x & = & df(\zeta - ih) - dg(\zeta + ih) + df(\zeta + i\bar{h}) - dg(\zeta - i\bar{h}), \\
B_y & = & ief(\zeta - ih) - ieg(\zeta + ih) - ief(\zeta + i\bar{h}) + ieg(\zeta - i\bar{h}), \\
B_z & = & if(\zeta - ih) + ig(\zeta + ih) - if(\zeta + i\bar{h}) - ig(\zeta - i\bar{h}).
\end{eqnarray*}
So we can see that, for both choices, $B_z$ is real, however, this is not the case for the other two components. In the first case
($b_1 = + 1/b_2$), it is not possible to say anything about $B_x$ or $B_y$ unless $f = g$, in which case both are real. However, in the
second case ($b_1 = - 1/b_2$), all components are automatically real, irrespective of the choices of $f$ and $g$. Note that allowing
the constant $b_1$ to take complex values only increases the complexity of the equations without bringing in any useful additional parameters;
therefore, without loss of generality, we consider only cases where $b_1$ takes real values. Therefore, and so as to be able to keep $f$ and
$g$ independent of each other, we choose $b_1 = - 1/b_2$.

The results for fields in quadrupole magnets can be extended to higher order multipoles in a straightforward way. However, the higher the order,
the more copies of the elementary solution are needed to construct fields with the required properties. In the following subsections, we shall
consider explicitly the cases of sextupole and octupole magnets, to establish a pattern from which the results for a general multipole magnet
can be written down.

\subsection{Sextupole magnets}

A sextupole field is obtained by putting $n = 2$ in equations (\ref{multipolefexpression}) and (\ref{multipolegexpression}).
To obtain the constraints on the various coefficients $c_j$ and $b_j$, we can proceed in close analogy with the case of the quadrupole magnet.
In particular, within the body of the magnet (far from any fringe field) $F(h_j + i\zeta) = G(h_j - i\zeta) = \pm 1$ for any $j$, and the
field is given by:
\begin{eqnarray}
B_u & = & iv^2, \nonumber \\
B_v & = & -iu^2, \nonumber \\
B_\zeta & = & 0. \nonumber
\end{eqnarray}
The corresponding constraints are:
\begin{eqnarray}
- ic_1b_1\bigg(\frac{u}{b_1} + b_1v + iz\bigg)^2 - ic_1b_1\bigg(\frac{u}{b_1} + b_1v - iz\bigg)^2
- ic_2b_2\bigg(\frac{u}{b_2} + b_2v + iz\bigg)^2 - ic_2b_2\bigg(\frac{u}{b_2} + b_2v - iz\bigg)^2 \quad & & \nonumber \\
- ic_3b_3\bigg(\frac{u}{b_3} + b_3v + iz\bigg)^2 - ic_3b_3\bigg(\frac{u}{b_3} + b_3v - iz\bigg)^2 & = & iv^2, \nonumber \\
- i\frac{c_1}{b_1}\bigg(\frac{u}{b_1} + b_1v + iz\bigg)^2 - i\frac{c_1}{b_1}\bigg(\frac{u}{b_1} + b_1v - iz\bigg)^2
- i\frac{c_2}{b_2}\bigg(\frac{u}{b_2} + b_2v + iz\bigg)^2 - i\frac{c_2}{b_2}\bigg(\frac{u}{b_2} + b_2v - iz\bigg)^2 \quad & & \nonumber \\
- i\frac{c_3}{b_3}\bigg(\frac{u}{b_3} + b_3v + iz\bigg)^2 - i\frac{c_3}{b_3}\bigg(\frac{u}{b_3} + b_3v - iz\bigg)^2 & = & - iu^2, \nonumber \\
- ic_1\bigg(\frac{u}{b_1} + b_1v + iz\bigg)^2 - ic_1\bigg(\frac{u}{b_1} + b_1v - iz\bigg)^2
- ic_2\bigg(\frac{u}{b_2} + b_2v + iz\bigg)^2 - ic_2\bigg(\frac{u}{b_2} + b_2v - iz\bigg)^2 \quad & & \nonumber \\
- ic_3\bigg(\frac{u}{b_3} + b_3v + iz\bigg)^2 - ic_3\bigg(\frac{u}{b_3} + b_3v - iz\bigg)^2 & = & 0. \nonumber
\end{eqnarray}
There are four equations to satisfy, namely:
\begin{eqnarray}
c_1b_1 + c_2b_2 + c_3b_3 & = & 0, \nonumber \\
\frac{c_1}{b_1} + \frac{c_2}{b_2} + \frac{c_3}{b_3} & = & 0, \nonumber \\
2c_1b_1^3 + 2c_2b_2^3 + 2c_3b_3^3 & = & - 1, \nonumber \\
2\frac{c_1}{b_1^3} + 2\frac{c_2}{b_2^3} + 2\frac{c_3}{b_3^3} & = & 1. \nonumber
\end{eqnarray}
Finally, the constraints on the coefficients $b_j$ and $c_j$ may be written:
\begin{eqnarray}
c_1 & = & - \frac{b_1}{2\left(b_1^2 - b_2^2\right)\left(b_1^2 - b_3^2\right)}, \nonumber \\
c_2 & = & - \frac{b_2}{2\left(b_2^2 - b_1^2\right)\left(b_2^2 - b_3^2\right)}, \nonumber \\
c_3 & = & - \frac{b_3}{2\left(b_3^2 - b_1^2\right)\left(b_3^2 - b_2^2\right)}, \nonumber \\
b_1 & = & \pm \frac{i}{b_2b_3}. \nonumber
\end{eqnarray}
Note that, for sextupoles and higher order multipoles, it is unlikely that a particular choice of the associated constants results in the field
components being automatically real. This is because of the additional number of constants involved. However, this should not result in any
problem because the real part can just be taken at the end of the explicit construction given that any complex solution to Maxwell's equations
implies that, both the real and imaginary parts, are also solutions of the same equations.

\subsection{Octupole magnets}

An octupole field is obtained by putting $n = 3$ in equations (\ref{multipolefexpression}) and (\ref{multipolegexpression}).  The field is
given by:
\begin{eqnarray}
B_u & = & iv^3, \nonumber \\
B_v & = & -iu^3, \nonumber \\
B_\zeta & = & 0. \nonumber
\end{eqnarray}
Following the same procedure as for the quadrupole and the sextupole, the constraints on the coefficients $c_j$ and $b_j$ are:
\begin{eqnarray}
- ic_1b_1\left(\frac{u}{b_1} + b_1v + iz\right)^3 - ic_1b_1\left(\frac{u}{b_1} + b_1v - iz\right)^3
- ic_2b_2\left(\frac{u}{b_2} + b_2v + iz\right)^3 - ic_2b_2\left(\frac{u}{b_2} + b_2v - iz\right)^3 \quad & & \nonumber \\
- ic_3b_3\left(\frac{u}{b_3} + b_3v + iz\right)^3 - ic_3b_3\left(\frac{u}{b_3} + b_3v - iz\right)^3
- ic_4b_4\left(\frac{u}{b_4} + b_4v + iz\right)^3 - ic_4b_4\left(\frac{u}{b_4} + b_4v - iz\right)^3 & = & iv^3, \nonumber \\
- i\frac{c_1}{b_1}\left(\frac{u}{b_1} + b_1v + iz\right)^3 - i\frac{c_1}{b_1}\left(\frac{u}{b_1} + b_1v - iz\right)^3
- i\frac{c_2}{b_2}\left(\frac{u}{b_2} + b_2v + iz\right)^3 - i\frac{c_2}{b_2}\left(\frac{u}{b_2} + b_2v - iz\right)^3 \quad & & \nonumber \\
- i\frac{c_3}{b_3}\left(\frac{u}{b_3} + b_3v + iz\right)^3 - i\frac{c_3}{b_3}\left(\frac{u}{b_3} + b_3v - iz\right)^3
- i\frac{c_4}{b_4}\left(\frac{u}{b_4} + b_4v + iz\right)^3 - i\frac{c_4}{b_4}\left(\frac{u}{b_4} + b_4v - iz\right)^3 & = & - iu^3, \nonumber \\
- ic_1\left(\frac{u}{b_1} + b_1v + iz\right)^3 - ic_1\left(\frac{u}{b_1} + b_1v - iz\right)^3
- ic_2\left(\frac{u}{b_2} + b_2v + iz\right)^3 - ic_2\left(\frac{u}{b_2} + b_2v - iz\right)^3 \quad & & \nonumber \\
- ic_3\left(\frac{u}{b_3} + b_3v + iz\right)^3 - ic_3\left(\frac{u}{b_3} + b_3v - iz\right)^3
- ic_4\left(\frac{u}{b_4} + b_4v + iz\right)^3 - ic_4\left(\frac{u}{b_4} + b_4v - iz\right)^3 & = & 0. \nonumber
\end{eqnarray}
There are five equations to satisfy, namely:
\begin{eqnarray}
c_1 + c_2 + c_3 + c_4 & = & 0, \nonumber \\
c_1b_1^2 + c_2b_2^2 + c_3b_3^2 + c_4b_4^2 & = & 0, \nonumber \\
\frac{c_1}{b_1^2} + \frac{c_2}{b_2^2} + \frac{c_3}{b_3^2} + \frac{c_4}{b_4^2} & = & 0, \nonumber \\
2c_1b_1^4 + 2c_2b_2^4 + 2c_3b_3^4 + 2c_4b_4^4 & = & - 1, \nonumber \\
2\frac{c_1}{b_1^4} + 2\frac{c_2}{b_2^4} + 2\frac{c_3}{b_3^4} + 2\frac{c_4}{b_4^4} & = & 1. \nonumber
\end{eqnarray}
The constraints on the coefficients $b_j$ and $c_j$ can be written:
\begin{eqnarray}
c_1 & = & - \frac{b_1^2}{2\left(b_1^2 - b_2^2\right)\left(b_1^2 - b_3^2\right)\left(b_1^2 - b_4^2\right)}, \nonumber \\
c_2 & = & - \frac{b_2^2}{2\left(b_2^2 - b_1^2\right)\left(b_2^2 - b_3^2\right)\left(b_2^2 - b_4^2\right)}, \nonumber \\
c_3 & = & - \frac{b_3^2}{2\left(b_3^2 - b_1^2\right)\left(b_3^2 - b_2^2\right)\left(b_3^2 - b_4^2\right)}, \nonumber \\
c_4 & = & - \frac{b_4^2}{2\left(b_4^2 - b_1^2\right)\left(b_4^2 - b_2^2\right)\left(b_4^2 - b_3^2\right)}, \nonumber \\
b_1 & = & \pm \frac{1}{b_2b_3b_4}. \nonumber
\end{eqnarray}

\subsection{General multipole magnets}

The derivation of the constraints on the coefficients $c_j$ and $b_j$ in the case of quadrupoles, sextupoles and octupoles can
be extended to any order of multipole.  In the body of the multipole, the components of the field are:
\begin{eqnarray}
B_u & = & iv^n, \nonumber \\
B_v & = & -iu^n, \nonumber \\
B_\zeta & = & 0. \nonumber
\end{eqnarray}
Substituting from (\ref{busummation}), (\ref{bvsummation}) and (\ref{bzetasummation}) gives:
\begin{eqnarray}
- ic_1b_1\left(\frac{u}{b_1} + b_1v + iz\right)^n - ic_1b_1\left(\frac{u}{b_1} + b_1v - iz\right)^n
- ic_2b_2\left(\frac{u}{b_2} + b_2v + iz\right)^n - ic_2b_2\left(\frac{u}{b_2} + b_2v - iz\right)^n \quad & & \nonumber \\
+ \cdots - ic_{n+1}b_{n+1}\left(\frac{u}{b_{n+1}} + b_{n+1}v + iz\right)^n - ic_nb_{n+1}\left(\frac{u}{b_{n+1}} + b_{n+1}v - iz\right)^n & = & iv^n, \nonumber \\
- i\frac{c_1}{b_1}\left(\frac{u}{b_1} + b_1v + iz\right)^n - i\frac{c_1}{b_1}\left(\frac{u}{b_1} + b_1v - iz\right)^n
- i\frac{c_2}{b_2}\left(\frac{u}{b_2} + b_2v + iz\right)^n - i\frac{c_2}{b_2}\left(\frac{u}{b_2} + b_2v - iz\right)^n \quad & & \nonumber \\
+ \cdots - i\frac{c_{n+1}}{b_{n+1}}\left(\frac{u}{b_{n+1}} + b_{n+1}v + iz\right)^n
- i\frac{c_{n+1}}{b_{n+1}}\left(\frac{u}{b_{n+1}} + b_{n+1}v - iz\right)^n & = & - iu^n, \nonumber \\
- ic_1\left(\frac{u}{b_1} + b_1v + iz\right)^n - ic_1\left(\frac{u}{b_1} + b_1v - iz\right)^n
- ic_2\left(\frac{u}{b_2} + b_2v + iz\right)^n - ic_2\left(\frac{u}{b_2} + b_2v - iz\right)^n \quad & & \nonumber \\
+ \cdots - ic_{n+1}\left(\frac{u}{b_{n+1}} + b_{n+1}v + iz\right)^n - ic_{n+1}\left(\frac{u}{b_{n+1}} + b_{n+1}v - iz\right)^n & = & 0. \nonumber
\end{eqnarray}
There are $n + 2$ equations to satisfy.  Two of the equations can be written:
\begin{eqnarray}
2c_1b_1^n + 2c_2b_2^n + \cdots + 2c_{n+1}b_{n+1}^n & = & - 1, \nonumber \\
2\frac{c_1}{b_1^n} + 2\frac{c_2}{b_2^n} + \cdots + 2\frac{c_{n+1}}{b_{n+1}^n} & = & 1. \nonumber
\end{eqnarray}
The remaining equations have the form:
\begin{eqnarray}
c_1b_1^\alpha + c_2b_2^\alpha + \cdots + c_{n+1}b_{n+1}^\alpha & = & 0, \nonumber \\
\frac{c_1}{b_1^\alpha} + \frac{c_2}{b_2^\alpha} + \cdots + \frac{c_{n+1}}{b_{n+1}^\alpha} & = & 0, \nonumber
\end{eqnarray}
where, for convenience, we have defined $\alpha = n - 2p$ with $p$ a positive integer. The number of
equations is determined by the order of the multipole and the restriction that $\alpha \ge 0$. Solving the
equations leads to the following constraints on the coefficients $b_j$ and $c_j$:
\begin{eqnarray}
b_1 & = & \pm \frac{i^{n-2}}{\prod_{j = 2}^{n+1}b_j}, \nonumber \\
c_1 & = & - \frac{b_1^{n-2}}{2\prod_{j = 2}^{n+1}\left(b_1^2 - b_j^2\right)}, \nonumber \\
c_2 & = & - \frac{b_2^{n-2}}{2\prod_{j = 1,j\neq2}^{n+1}\left(b_2^2 - b_j^2\right)}, \nonumber \\
 & \vdots & \nonumber \\
c_n & = & - \frac{b_n^{n-2}}{2\prod_{j = 1,j\neq n}^{n+1}\left(b_n^2 - b_j^2\right)}. \nonumber
\end{eqnarray}

\subsection{Significance of parameters and scaling laws}

\begin{figure}[htb]
\centering
\includegraphics[trim=40 60 40 50 ,clip,width=0.55\textwidth]{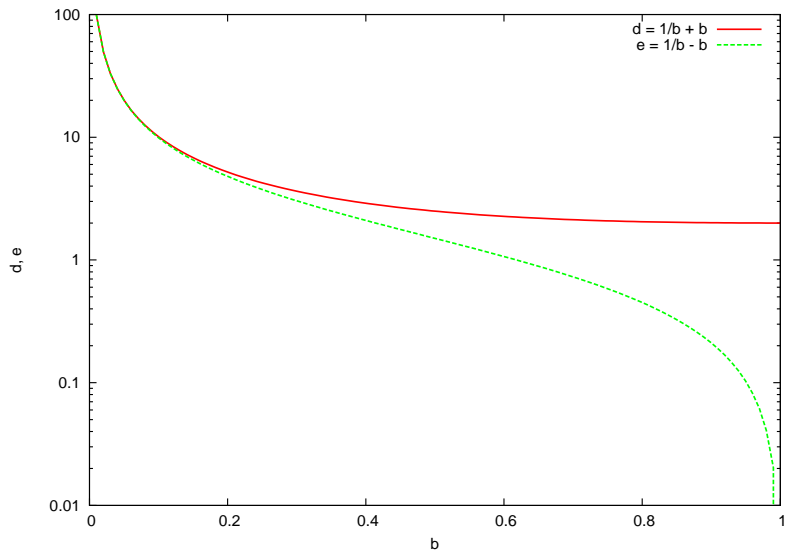}
\caption{Behaviour of the factors $d$ and $e$.}
\label{d+e}
\end{figure}

From equations (\ref{resbu}), (\ref{resbv}) and (\ref{resbzp}), it is clear that in the case of the elementary solution, the constant
$b$ acts as a scaling parameter. This remains the case when several elementary solutions are added together to represent the fringe
field of a multipole magnet. We shall see later that the constant $b$ affects the rate of fall-off of the field strength with distance from
the axis (keeping $z$ constant). This can be seen by inspecting the singularities present in the components of the field, as illustrated
in the case of quadrupoles in the next section. The singularities are an inevitable consequence of the way in which the magnetostatic Maxwell
equations in three dimensions were solved, by strict analogy with the two dimensional case. For quadrupole magnets, the locations of
the singularities depend on $b$; it is expected that the same is true for higher order multipoles. The singularities depend on the inverse
of $d$ or $e$; therefore, if $b$ is increased from zero towards one, the singularities move (transversely) away from the axis of the
magnet. Looking at the individual components of the magnetic field in Cartesian co-ordinates (equations (\ref{resbx}), (\ref{resby}) and
(\ref{resbz})), it can be seen from the way that the scaling constant $b$ enters through the constants $d$, $e$ and $c$ (with $c = 4/de$),
that the smaller the value of $b$, the larger the values of $d$ and $e$ and vice versa.  These relationships are illustrated in Fig.~\ref{d+e}.

Suppose that, for simplicity, we scale all the field components by a factor of $c$.  Then, $B_x \sim d$, $B_y \sim e$ and $B_z$ is
independent of $d$ or $e$.  For small values of $b$, the transverse components $B_x$ and $B_y$ of the magnetic field dominate
over the longitudinal component $B_z$. This agrees with the fact that, for a small value of $b$, the transverse fall-off of the field
is very rapid, so the appearance in the fringe region of the longitudinal component of the magnetic field $B_z$ does not need to be
as prolonged as for larger values of $b$. Ultimately, we shall see that the constant $b$, for quadrupoles, can be used as a parameter
to fit the field behaviour with distance from the axis, in the same way that the Enge coefficients can be used as parameters to fit
the field behaviour with distance along the axis.  The constant $b$ and the Enge coefficients are ultimately dependent on the
geometry of the magnet.

The number of constants increases as $(n - 2)/2$ with increasing order $n$ of the multipole. For a quadrupole, there are
effectively two transverse directions in which the rate of the fall-off of the field in the fringe region can vary. The rate of fall-off
in one direction can be adjusted by an overall re-scaling; to vary the rate of fall-off in the second direction (independently of the
first direction) requires a separate parameter. Similarly, in a sextupole, there are three independent transverse directions in which
the rate of fall-off of the field can vary, which means that there are two independent parameters to control the transverse behaviour.
For octupoles, with four independent transverse directions, three parameters are required, and so on.

\section{Multipole magnets with Enge-type fringe fields}

Having constructed general expressions for fringe fields in multipole magnets, it is worth investigating these expressions further,
to show that the field has the appropriate behaviour.  For a multipole magnet of arbitrary order, the expressions for the field can be
rather complicated: therefore, we consider in detail only the case of a quadrupole.  In this section, we shall discuss the behaviour of the
fields in a quadrupole, first for the elementary solution, and then for the full solution (in which the fields are given by real numbers)
obtained by adding two versions of the elementary solution.  In order to plot the field, we need to make some assumption for the form of the
roll-off of the field along the axis of the magnet: we shall consider the case that the roll-off is described by an Enge function.  At the
end of the section, we shall generalise the expressions for a full quadrupole solution with Enge roll-off of the gradient to higher
order multipole magnets.

Of particular interest is the appearance of singularities in the magnetic field.  Singularities are expected from the properties of
Laplace's equation: in two dimensions, solutions to Laplace's equation are either constant everywhere, or have singularities somewhere.
Our expressions for multipole fringe fields have been obtained by extending the two dimensional case to three dimensions.
Mathematically, it is no surprise that singularities appear, but if the expressions we have derived for the fields are to be applied
in physical situations, we should understand the position and nature of the singularities.

\subsection{Elementary solution for a quadrupole with Enge-type fringe field}\label{elementaryquadsolution}

We take $n = 1$ for the elementary solution and examine its behaviour.  For simplicity we assume that the fall-off functions
are related by $F_1 = - G_1$. To simplify the notation we drop the subscript $j = 1$, so the field components
(\ref{busummation})--(\ref{bzetasummation}) become:
\begin{eqnarray}
B_u & = & b c \left( (\zeta + ih) G(\zeta + ih) - (\zeta - ih) G(\zeta - ih) \right), \nonumber \\
B_v & = & \frac{c}{b} \left( (\zeta + ih) G(\zeta + ih) - (\zeta - ih) G(\zeta - ih) \right), \nonumber \\
B_\zeta & = & - i c \left( (\zeta + ih) G(\zeta + ih) + (\zeta - ih) G(\zeta - ih) \right). \nonumber
\end{eqnarray}
We shall see when we consider the full quadrupole solution in Section \ref{sectionfullquadrupoleenge} that an Enge-type
fall-off in the quadrupole gradient occurs when we choose the function $G(\zeta)$ as follows:
\[
G(\zeta) = \frac{\ln(1 + e^\zeta)}{\zeta} - \frac{\ln 2}{\zeta} - 1.
\]
%The function $G(\zeta)$ is plotted as a function of distance along the axis of the quadrupole magnet. Also shown in this figure (for
%comparison) is the Enge function:
%\[
%g(\zeta) = \frac{1}{1 + e^\zeta}.
%\]

The components of the field (in a Cartesian basis) can be written in this case:
\begin{eqnarray}
B_x & = & \frac{b}{2\sqrt{2}(1 - b^2)} \left( \ln(1 + e^{\zeta + i h}) - \ln(1 + e^{\zeta - i h}) -  2i h \right), \nonumber \\
B_y & = & \frac{i b}{2\sqrt{2}(1 + b^2)} \left( \ln(1 + e^{\zeta + i h}) - \ln(1 + e^{\zeta - i h}) -  2i h \right), \nonumber \\
B_z & = & \frac{i b^2}{\sqrt{2}(1 - b^4)} \left( 2\zeta + 2\ln 2 - \ln(1 + e^{\zeta + i h}) - \ln(1 + e^{\zeta - i h}) \right). \nonumber
\end{eqnarray}
Along the axis of the magnet ($h = 0$), the transverse components of the field vanish and the longitudinal component of the field
is purely imaginary.  If we identify the real parts of the field with the physical magnetic field, then the longitudinal component
of the field also vanishes along the axis.
Singularities in the field occur when:
\begin{equation}
\zeta \pm i h = i \ell \pi, \label{singularitycondition}
\end{equation}
where $\ell$ is any odd integer.  In terms of the Cartesian co-ordinates, the singularities occur at:
\[
(x, y) = \left( \frac{\sqrt{2} \ell \pi}{\frac{1}{b} + b} , \pm \frac{2 z}{\frac{1}{b} - b}\right).
\]
Note, however, that the singularities in the different terms in the transverse components of the field cancel out when $z = 0$.

\begin{figure}[t!]
\centering
\begin{tabular}{cc}
\includegraphics[trim=0 0 0 8.5cm,clip,width=0.28\textwidth]{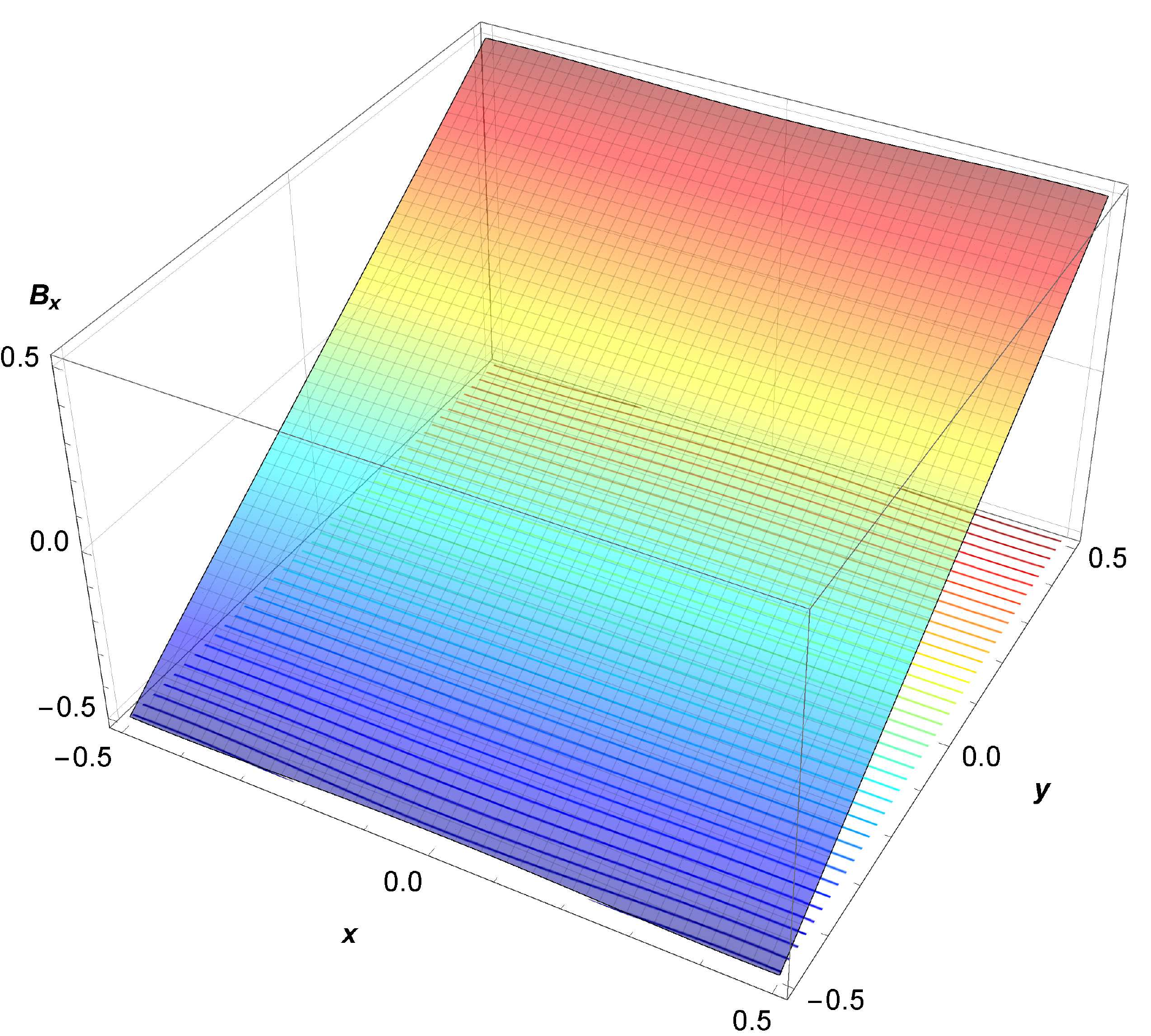} &
\includegraphics[trim=0 0 0 8.5cm,clip,width=0.28\textwidth]{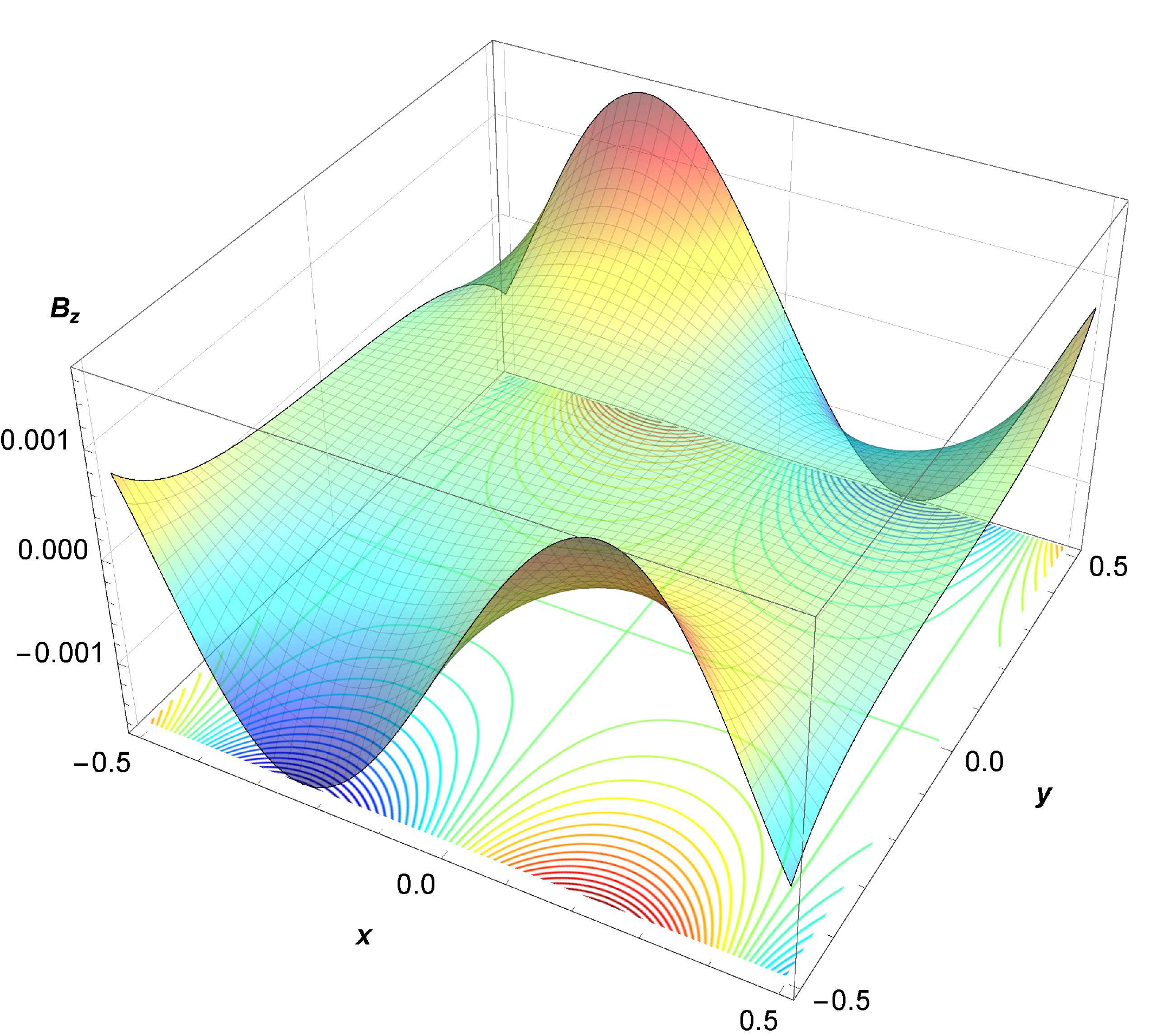} \\
\includegraphics[trim=0 0 0 8.5cm,clip,width=0.28\textwidth]{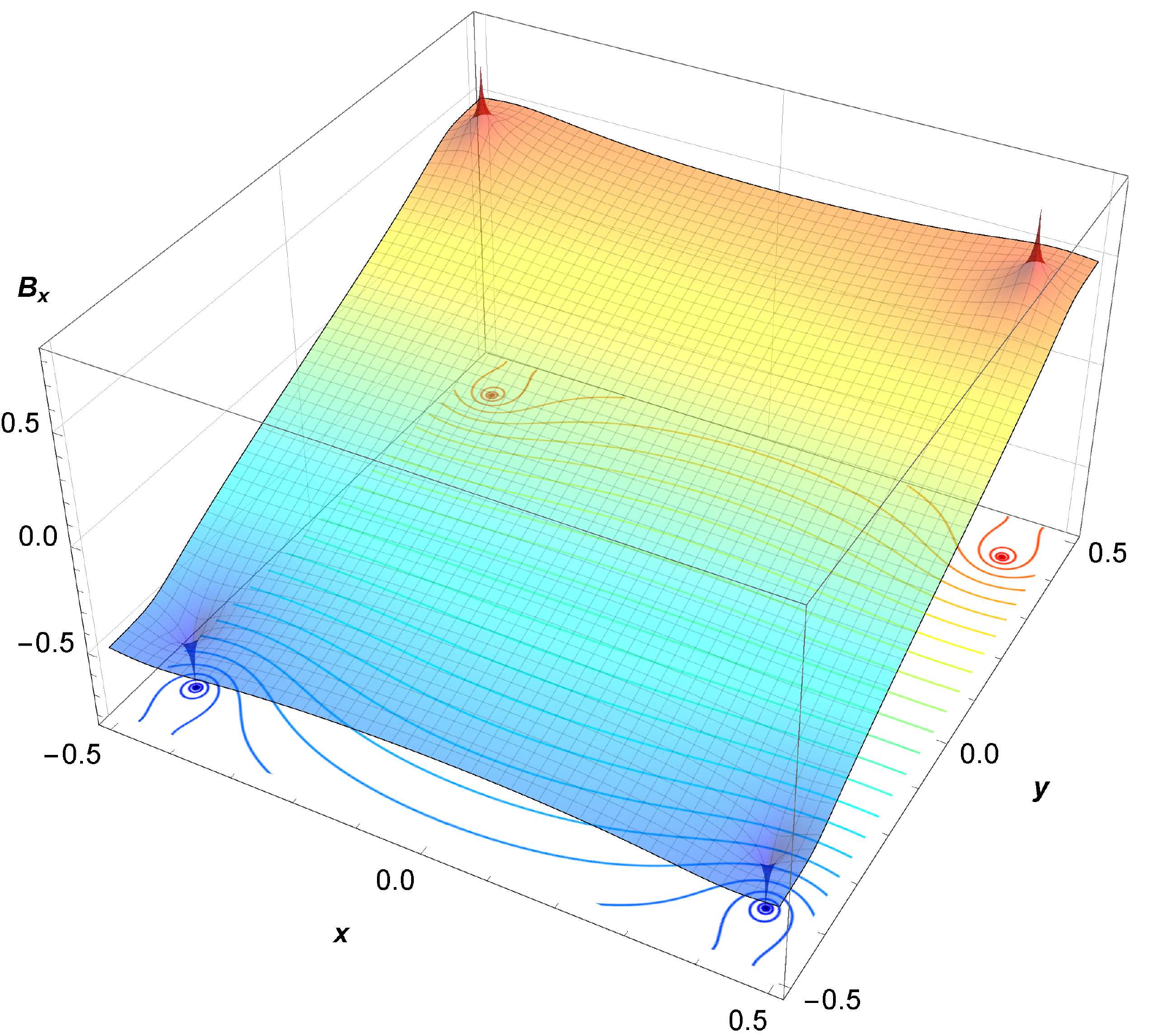} &
\includegraphics[trim=0 0 0 8.5cm,clip,width=0.28\textwidth]{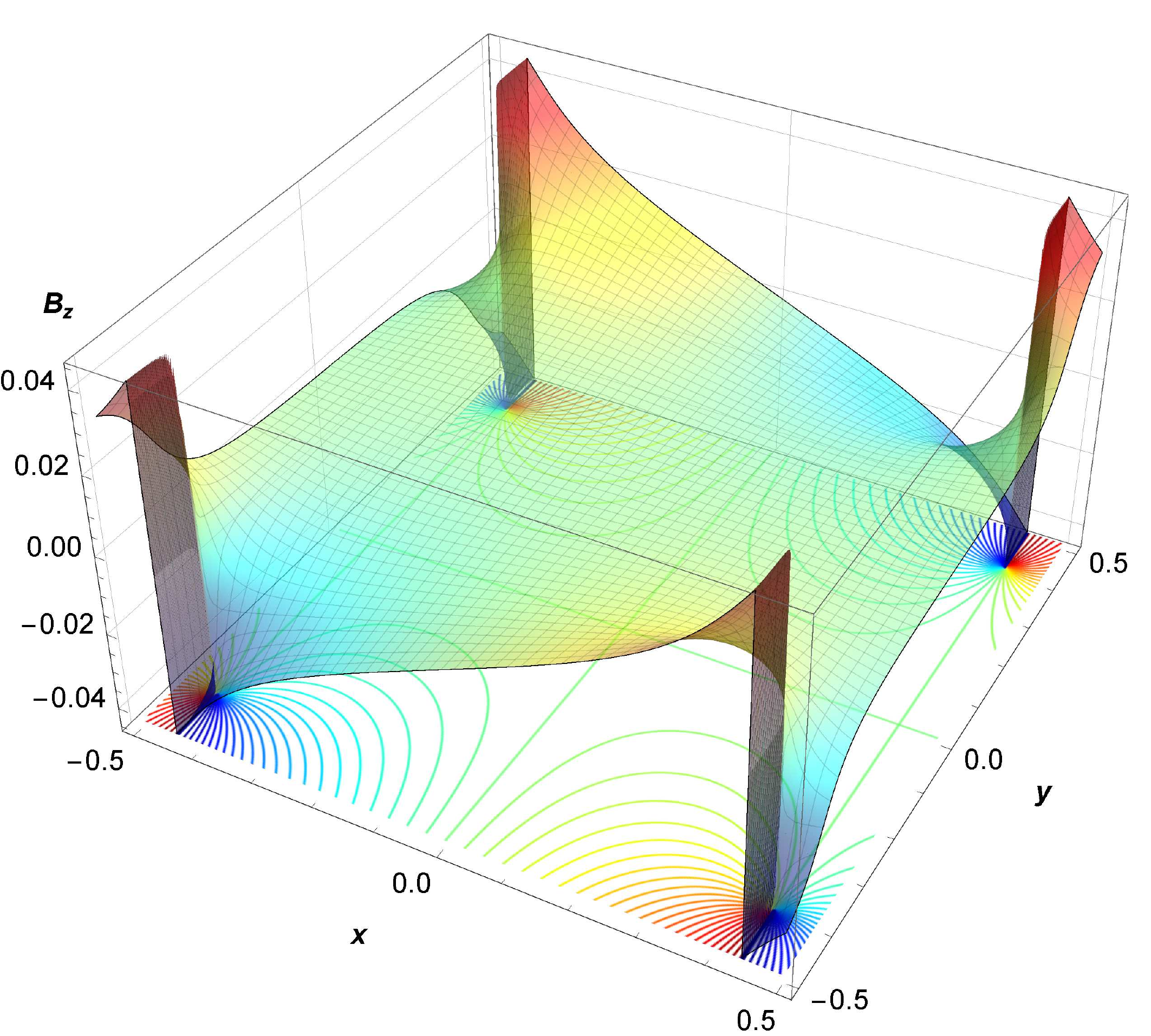} \\
\includegraphics[trim=0 0 0 8.5cm,clip,width=0.28\textwidth]{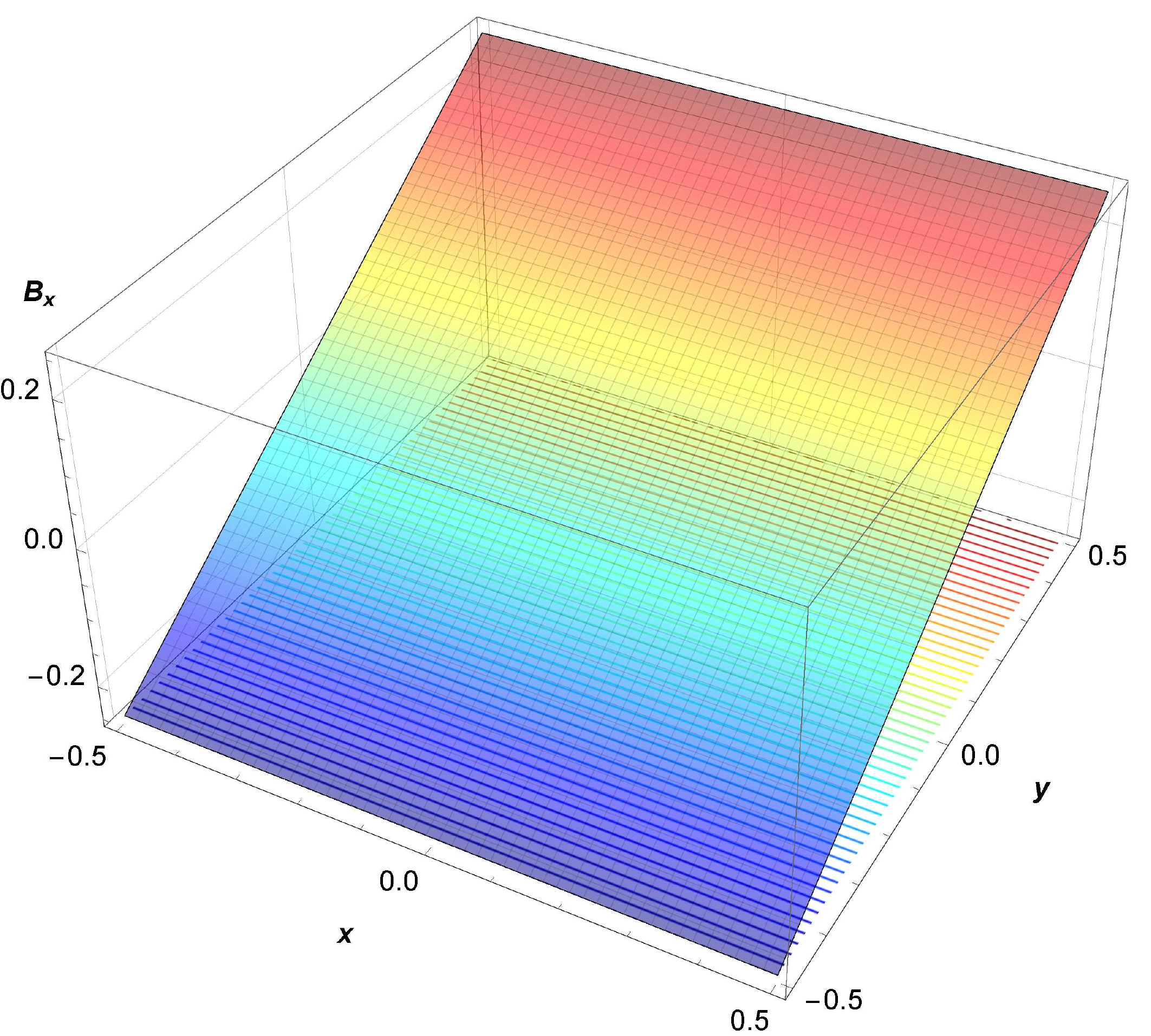} &
\includegraphics[trim=0 0 0 8.5cm,clip,width=0.28\textwidth]{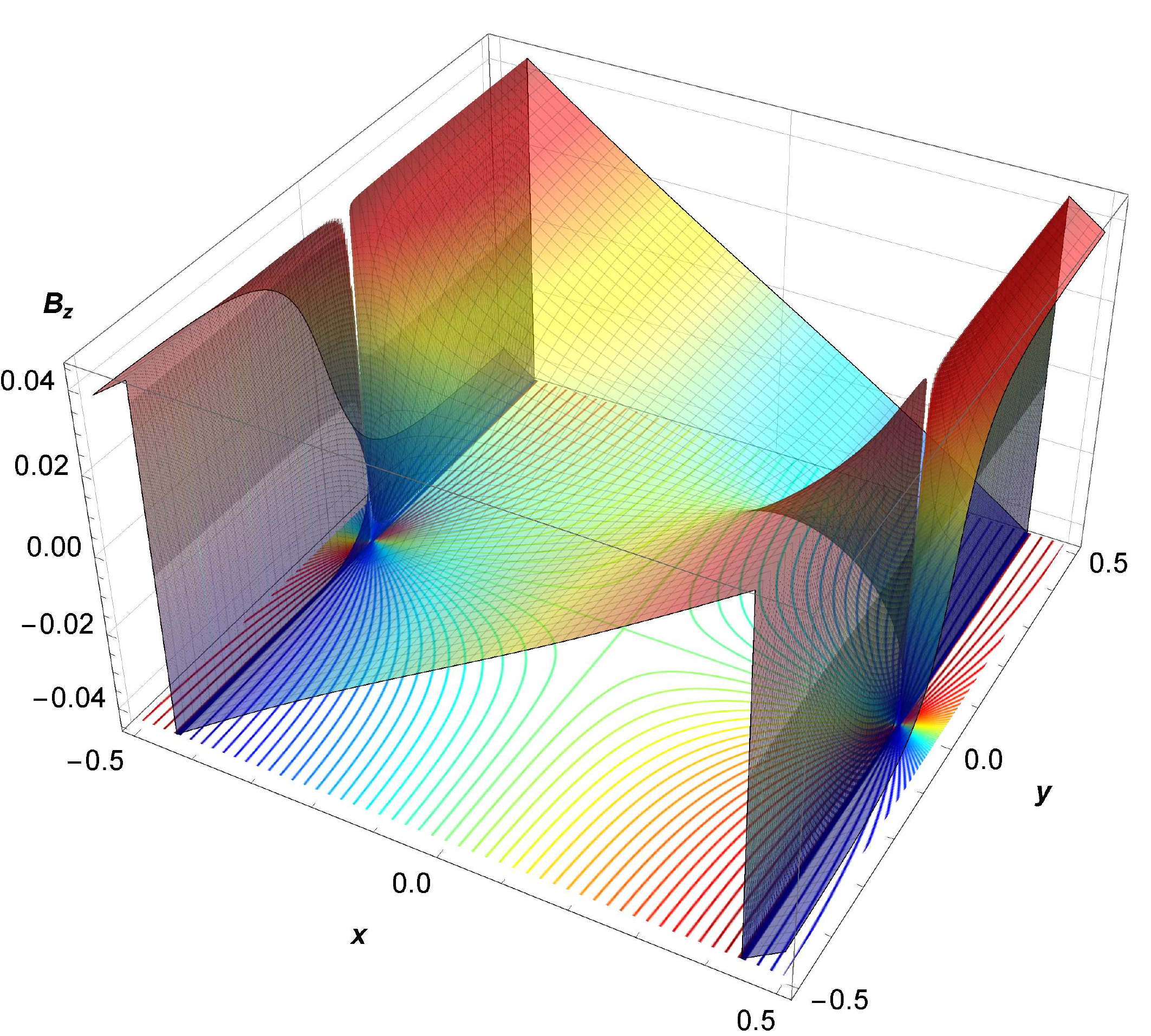} \\
\includegraphics[trim=0 0 0 8.5cm,clip,width=0.28\textwidth]{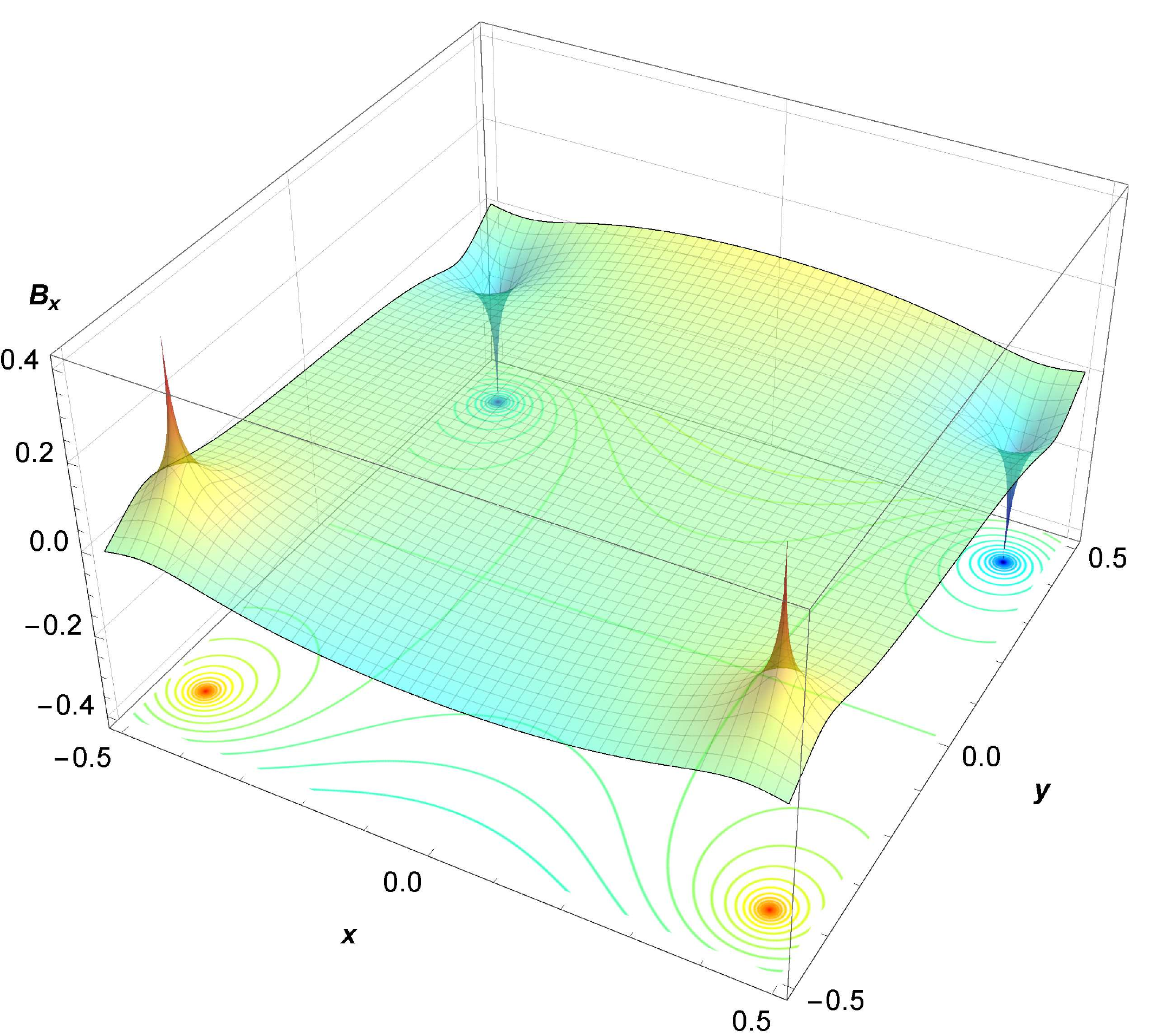} &
\includegraphics[trim=0 0 0 8.5cm,clip,width=0.28\textwidth]{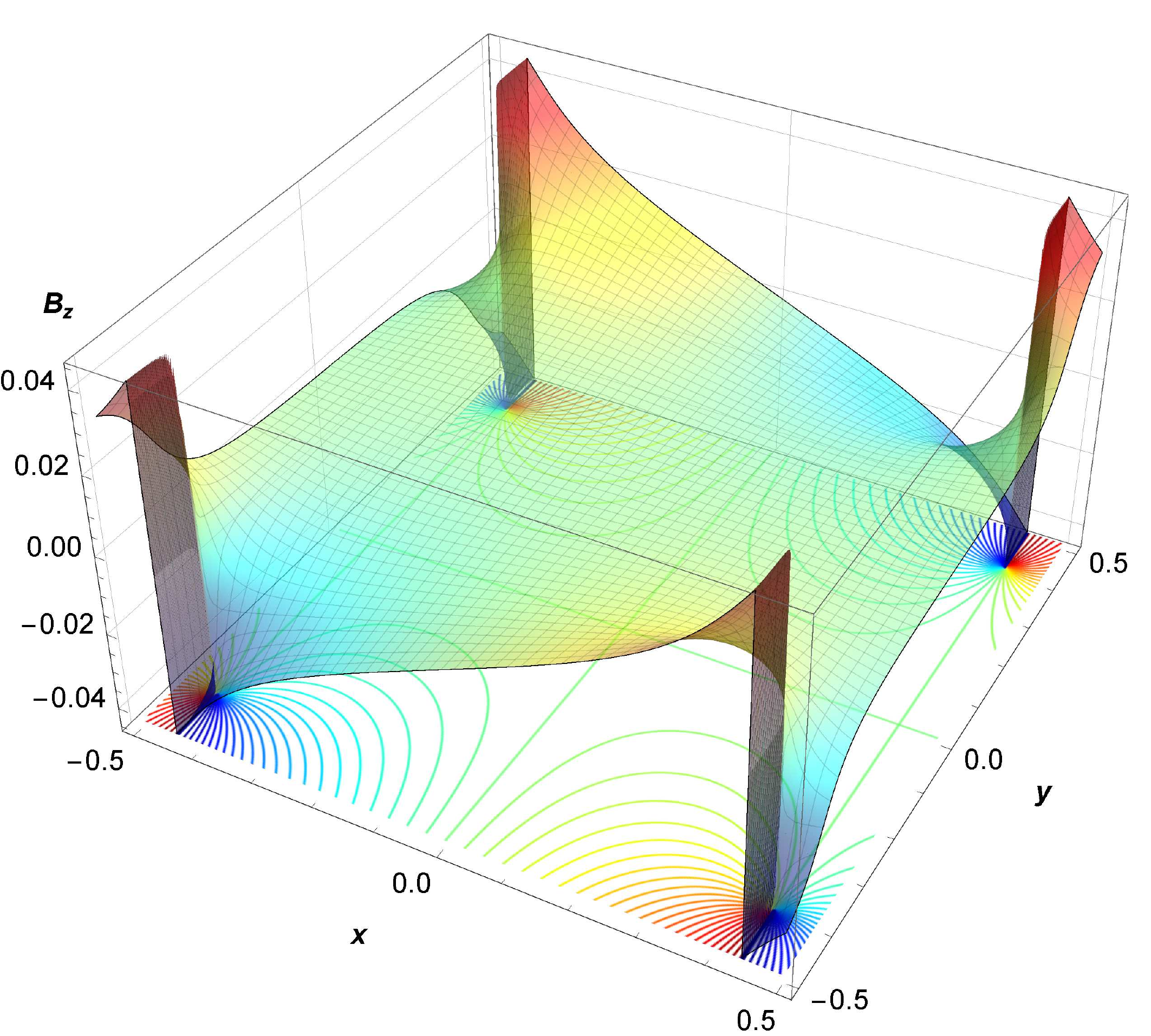} \\
\includegraphics[trim=0 0 0 8.5cm,clip,width=0.28\textwidth]{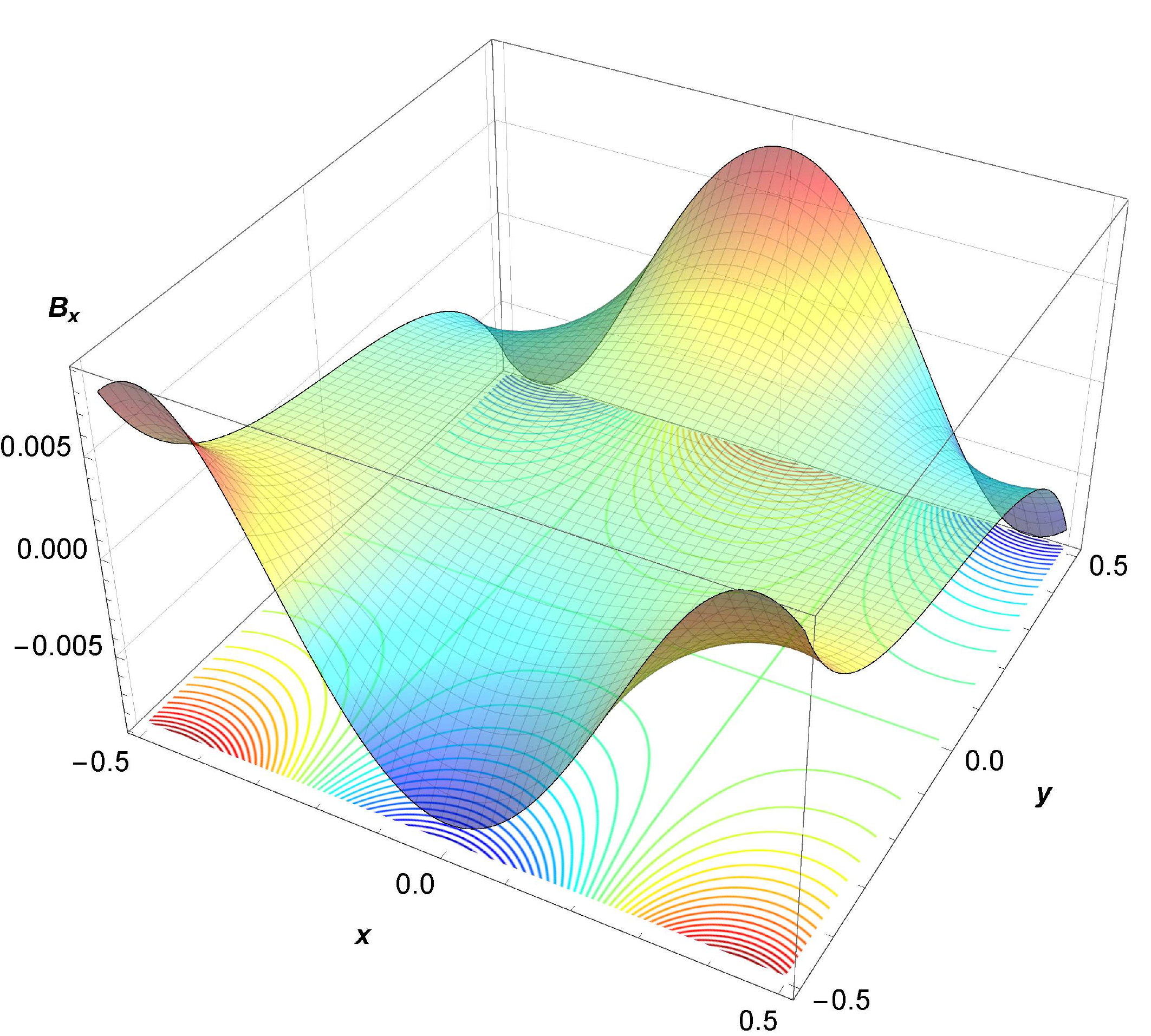} &
\includegraphics[trim=0 0 0 8.5cm,clip,width=0.28\textwidth]{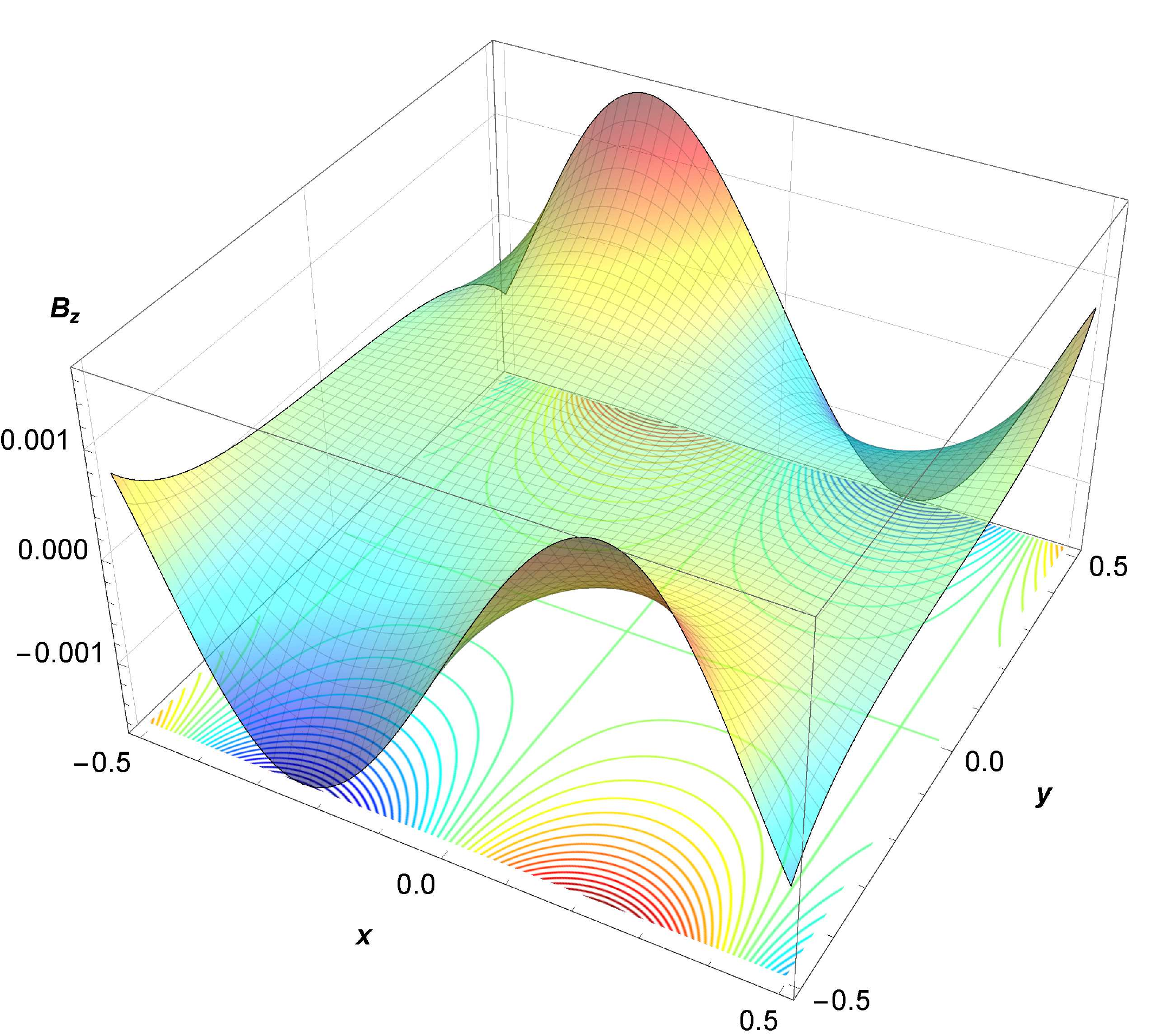}
\end{tabular}
\caption{Elementary quadrupole-like fringe field components $B_x$ (left) and $B_z$ (right) at  (top to bottom) $z = - 4.0$, $z = - 2.0$,
$z = 0$, $z = 2.0$ and $z = 4.0$.}
\label{engebxm4m2}
\end{figure}

The behaviour of the field can be seen in Fig.~\ref{engebxm4m2}, which shows the real parts of the $B_x$ and $B_z$ components of the
magnetic field as functions of the transverse co-ordinates at various $z$ locations. For the elementary quadrupole-like solution,
the constant $b$ amounts to a re-scaling in the coordinates and the field; therefore, we show plots for only a single value $b = 0.1$.
Also, since no new features occur in the behaviour of $B_y$ compared to the behaviour of $B_x$, we only show plots for $B_x$ and $B_z$.
The plots in Fig.~\ref{engebxm4m2} show that the field has the expected behaviour for a quadrupole: in particular, within the body of the
magnet, $B_x$ is linear in the co-ordinate $y$, and independent of $x$.  At increasing values of $z$, the slope of $B_x$ versus $y$ decreases
(the gradient falls off); at $z = 0$, the quadrupole gradient is half the value at large negative $z$. As $z$ increases further, the gradient
(and the field itself) falls to zero. The singularities in the field have the behaviour expected from (\ref{singularitycondition}). In
particular, we see that at $z = 0$, the singularities in the transverse field disappear completely.

Note that it was not necessary to demand that $F = - G$ in the fall-off functions above. This was only done for simplicity and it could
well be the case that a non-symmetric fall-off is desired and this would require $F$ and $G$ to be completely different. As they are just
functions that dictate the behaviour the fringe fall-off, nothing changes inside the quadrupole and the magneto-static Maxwell equations are
still satisfied.

\subsection{Full solution for a quadrupole with Enge-type fringe field}\label{sectionfullquadrupoleenge}

As shown above, we can construct physical (real) fields in a quadrupole by adding two versions of the elementary solution with $n = 1$:
\begin{eqnarray}
B_u & = & \sum_{j = 1}^{2} b_j c_j \left( (\zeta + ih_j) G(\zeta + ih_j) - (\zeta - ih_j) G(\zeta - ih_j) \right), \label{fullsolutionquadbu} \\
B_v & = & \sum_{j = 1}^{2} \frac{c_j}{b_j} \left( (\zeta + ih_j) G(\zeta + ih_j) - (\zeta - ih_j) G(\zeta - ih_j) \right),  \\
B_\zeta & = & - i \sum_{j = 1}^{2} c_j \left( (\zeta + ih_j) G(\zeta + ih_j) + (\zeta - ih_j) G(\zeta - ih_j) \right),\label{fullsolutionquadbzeta}
\end{eqnarray}
where:
\[
h_j = d_jx + ie_jy,
\]
and:
\begin{eqnarray}
d_j & = & \frac{b_j}{\sqrt{2}} + \frac{1}{\sqrt{2}b_j}, \nonumber \\
e_j & = & \frac{1}{\sqrt{2}b_j} - \frac{b_j}{\sqrt{2}}, \nonumber
\end{eqnarray}
together with the definitions of $b_j$ and $c_j$ given in (\ref{bq}) and (\ref{cq}), keeping in mind we took the negative solution in equation
(\ref{bq}).
%If we plot the two constants $d_j$ and $e_j$ as functions of $b_j$ (as shown in Fig.~\ref{d+e}) we see that the field components $B_x$
%and $B_y$ will decay at different rates. The transverse field components only show the same behaviour when $b_2 = 0$; this value of $b_2$ is
%not allowed, since as $b_2 \to 0$, $e_2 \to \pm \infty$. Therefore, the symmetry $B_x \leftrightarrow B_y$ under $x \leftrightarrow y$ has to
%be imposed. This can be done by adding equivalent expressions to $B_x$, $B_y$ and $B_z$, with $x$ and $y$ interchanged but keeping $b_2$ the
%same, and dividing the result by $2$. Therefore, in what follows, the \emph{full quadrupole solution} refers to the fully symmetric quadrupole
%solution. There may be cases where exact quadrupole symmetry is not required; however, we do not consider such cases here.

Working in cylindrical polar co-ordinates $r$, $\theta$, $z$, with the $x$ axis corresponding to the line $z = \theta = 0$, the
radial field component can be expressed as a Taylor series in $r$:
\begin{equation}
B_r = -r \sin(2\theta) \left( G(\zeta) + \zeta G^\prime(\zeta) \right) + O(r^2), \label{quadrupolegradient}
\end{equation}
where $G^\prime$ is the derivative of $G$ with respect to its argument.  Equation (\ref{quadrupolegradient}) describes the
behaviour we would expect from a magnet in which the quadrupole gradient $g(z)$ varies along the axis as:
\begin{equation}
g(\zeta) =  - G(\zeta) - \zeta G^\prime(\zeta). \label{gradientbigg}
\end{equation}
A conventional model for the fringe field in a quadrupole magnet describes the gradient as an Enge function \cite{Brown}.
Again to keep the analysis as simple as possible, we consider the case that the gradient varies as an Enge function with a
single parameter:
\[
g(\zeta) = \frac{1}{1 + e^\zeta}. 
\]
Writing the gradient in this way determines $z = 0$ as the location along the axis at which the quadrupole gradient falls to
half its nominal value within the body of the magnet.
Integrating (\ref{gradientbigg}) gives:
\[
G(\zeta) = \frac{\ln(1 + e^\zeta)}{\zeta} - \frac{\ln 2}{\zeta} - 1,
\]
where the constant of integration has been chosen so that $G(\zeta)$ remains finite as $\zeta \to 0$.

\begin{figure}[htb]
\centering
\includegraphics[width=0.5\textwidth]{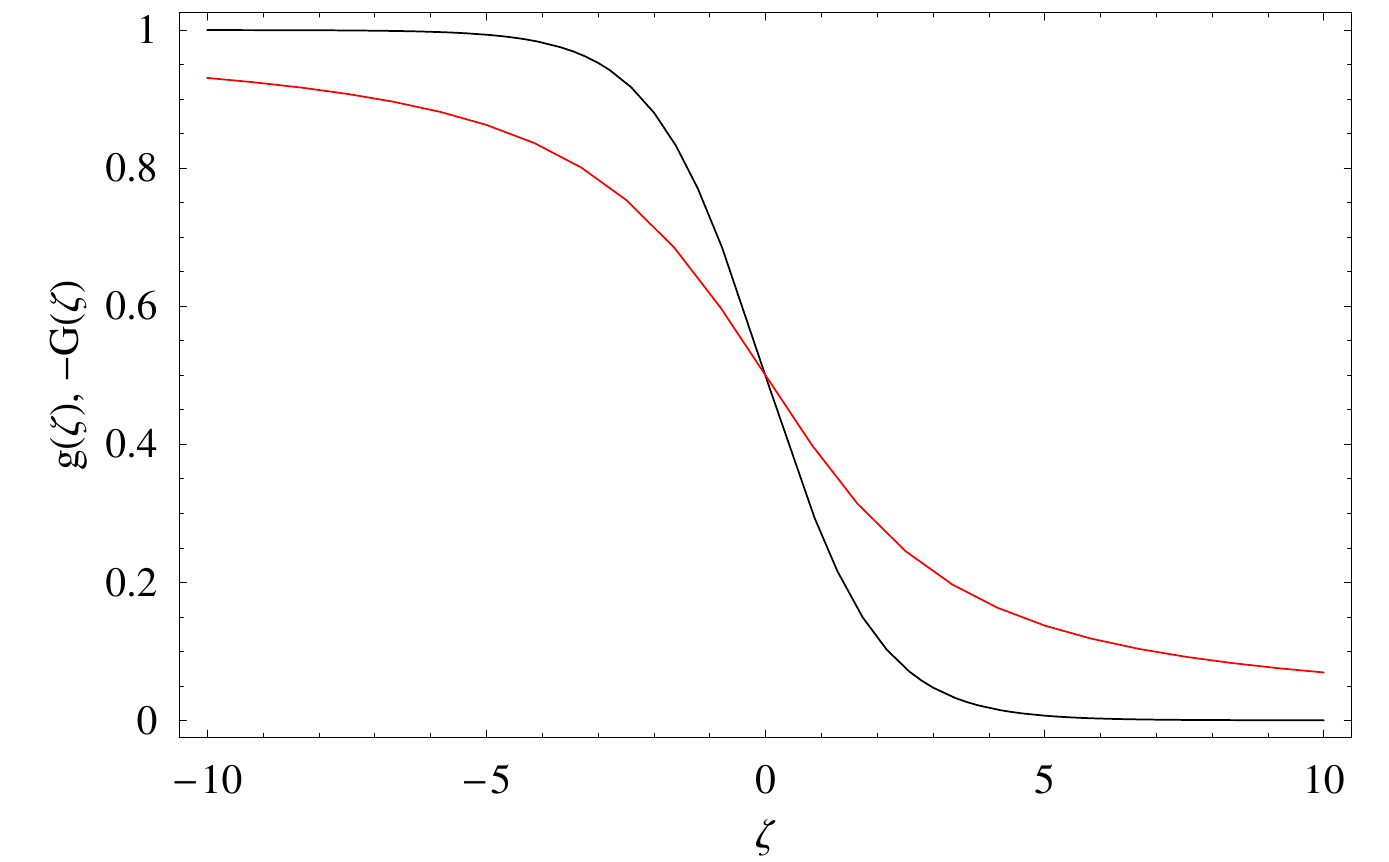}
\caption{Quadrupole gradient $g(\zeta)$ given by an Enge function (black line) and the corresponding function $-G(\zeta)$ (red line),
as a function of distance along the axis of a quadrupole magnet.}
\label{fall-offs}
\end{figure}

The components of the field can now be written (in a Cartesian basis):
\begin{eqnarray}
B_x & = & \sum_{j = 1}^{2}\frac{c_jd_j}{2}\left( - 2ih_j - \ln(1 + e^{\sqrt{2}z - ih_j}) + \ln(1 + e^{\sqrt{2}z + ih_j}) \right), \nonumber \\
B_y & = & i\sum_{j = 1}^{2}\frac{c_je_j}{2}\left( - 2ih_j - \ln(1 + e^{\sqrt{2}z - ih_j}) + \ln(1 + e^{\sqrt{2}z + ih_j}) \right), \nonumber \\
B_z & = & i\sum_{j = 1}^{2}\frac{c_j}{\sqrt{2}}\left( 2\ln(2) + 2\sqrt{2}z - \ln(1 + e^{\sqrt{2}z - ih_j}) - \ln(1 + e^{\sqrt{2}z + ih_j}) \right).
\nonumber
\end{eqnarray}
The quadrupole gradient $g(\zeta)$ and the corresponding function $G(\zeta)$ are plotted as a function of distance along the axis of the
quadrupole magnet in Fig.~\ref{fall-offs}.

If we plot the two constants $d_j$ and $e_j$ as functions of $b_j$ (as shown in Fig.~\ref{d+e}) we see that the field components $B_x$
and $B_y$ will decay at different rates. The transverse field components only show the same behaviour when $b_2 = 0$; this value of $b_2$ is
not allowed, since as $b_2 \to 0$, $e_2 \to \pm \infty$. Therefore, the symmetry $B_x \leftrightarrow B_y$ under $x \leftrightarrow y$ has to
be imposed. This can be done by adding equivalent expressions to $B_x$, $B_y$ and $B_z$, with $x$ and $y$ interchanged but keeping $b_2$ the
same, and dividing the result by $2$. Therefore, in what follows, the \emph{full quadrupole solution} refers to the fully symmetric quadrupole
solution. There may be cases where exact quadrupole symmetry is not required; however, we do not consider such cases here.

\begin{figure}[t!]
\centering
\begin{tabular}{ccc}
\includegraphics[trim=0 0 0 8.5cm,clip,width=0.28\textwidth]{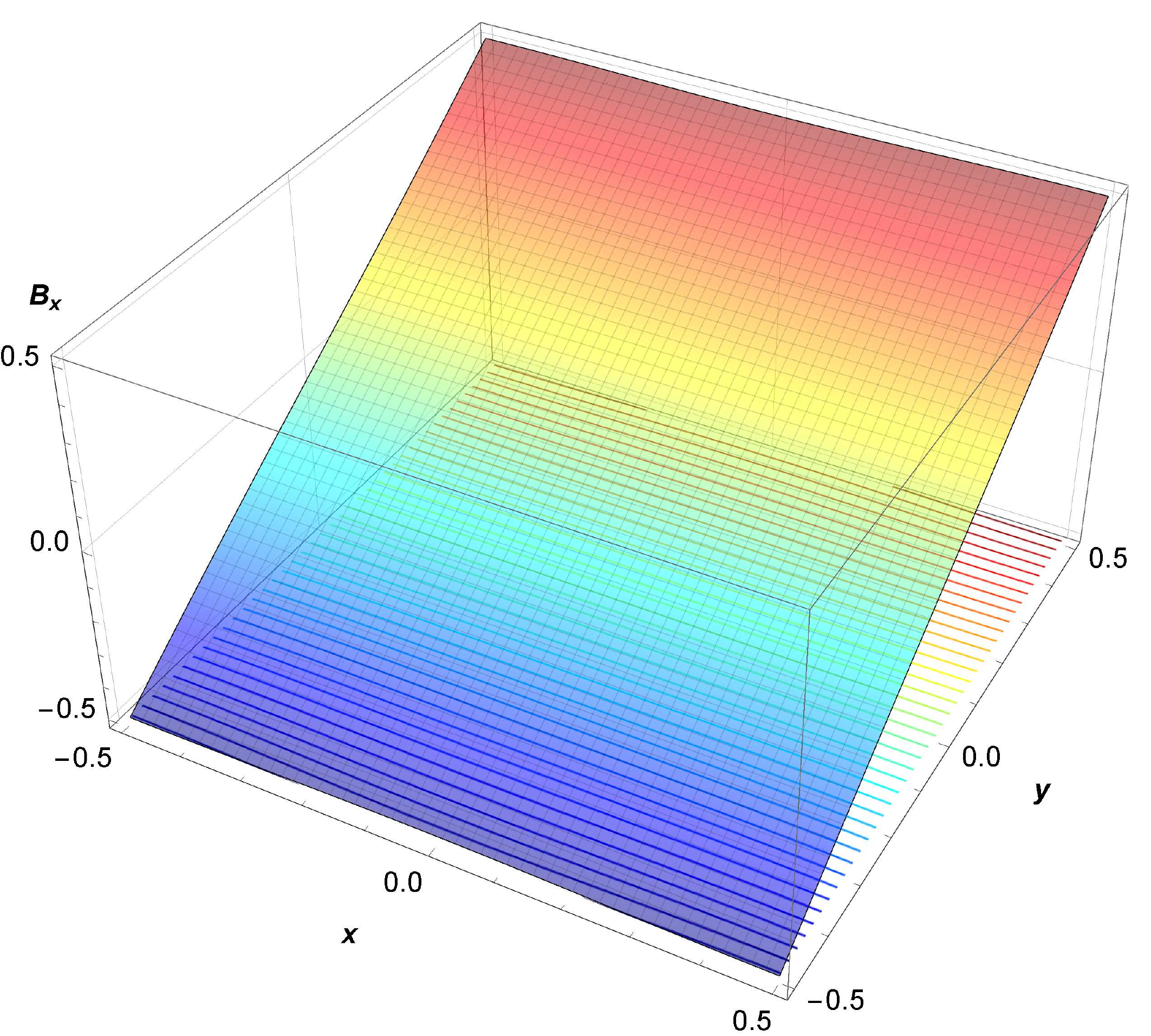} &
\includegraphics[trim=0 0 0 8.5cm,clip,width=0.28\textwidth]{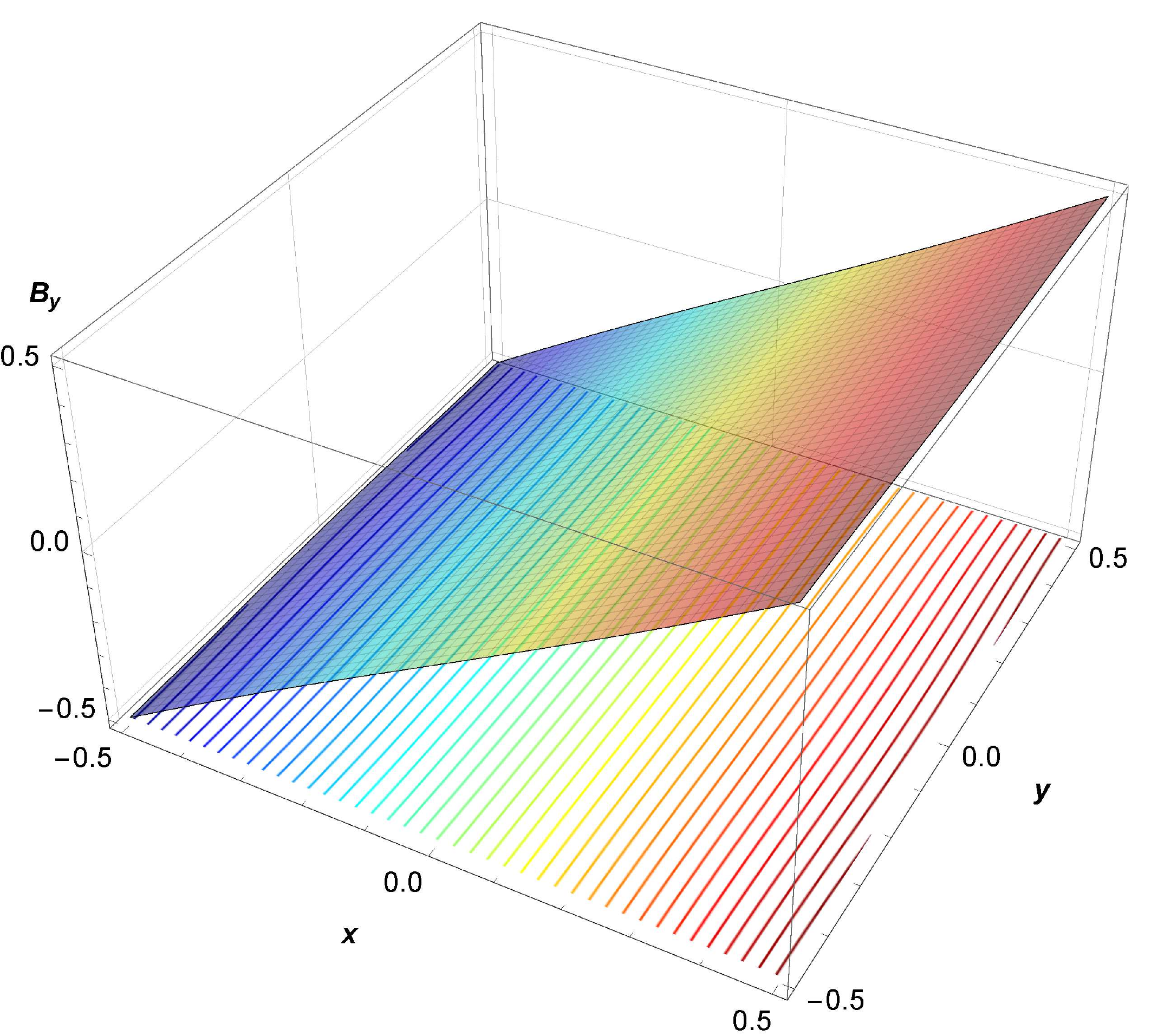} &
\includegraphics[trim=0 0 0 8.5cm,clip,width=0.28\textwidth]{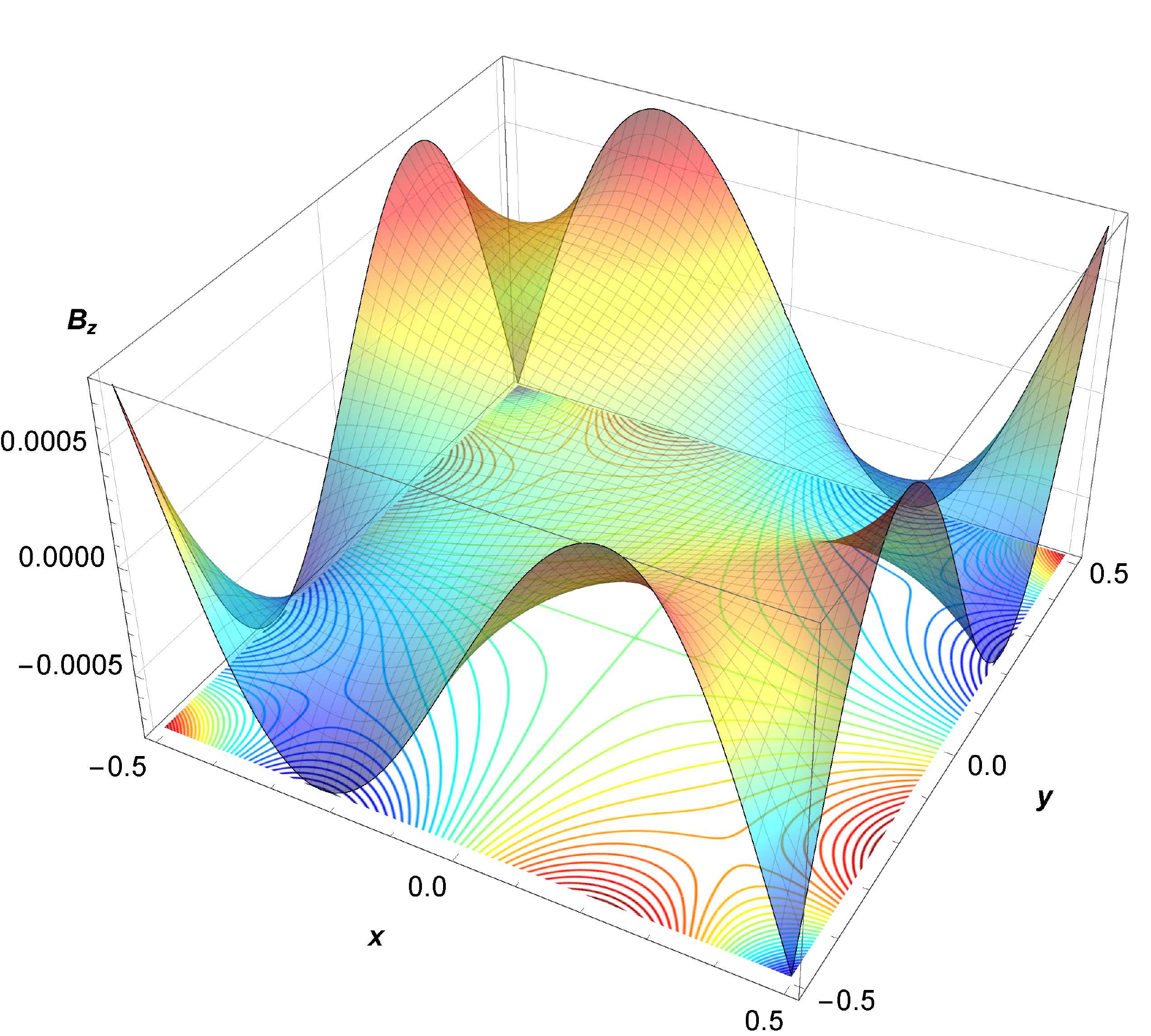} \\
\includegraphics[trim=0 0 0 8.5cm,clip,width=0.28\textwidth]{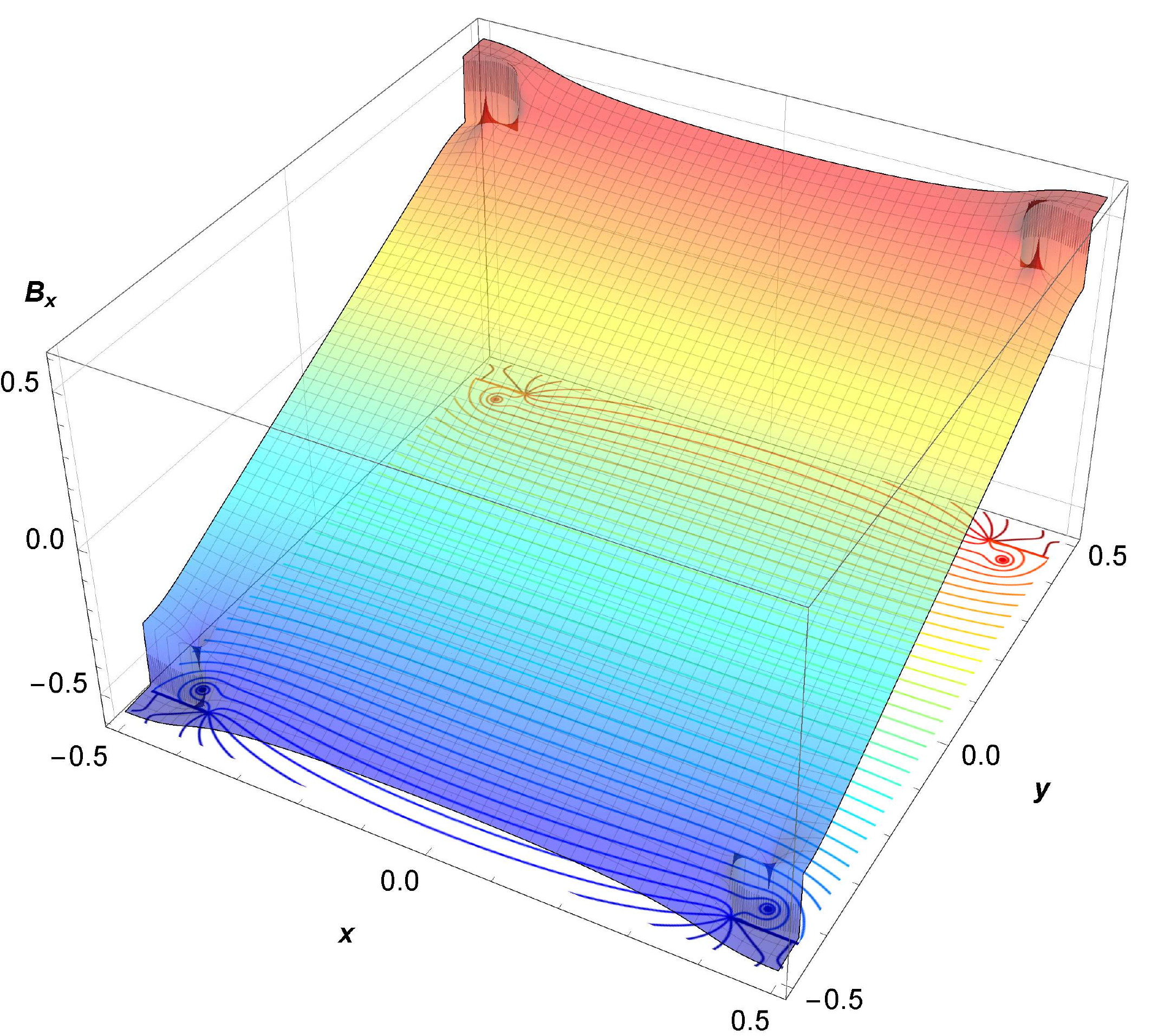} &
\includegraphics[trim=0 0 0 8.5cm,clip,width=0.28\textwidth]{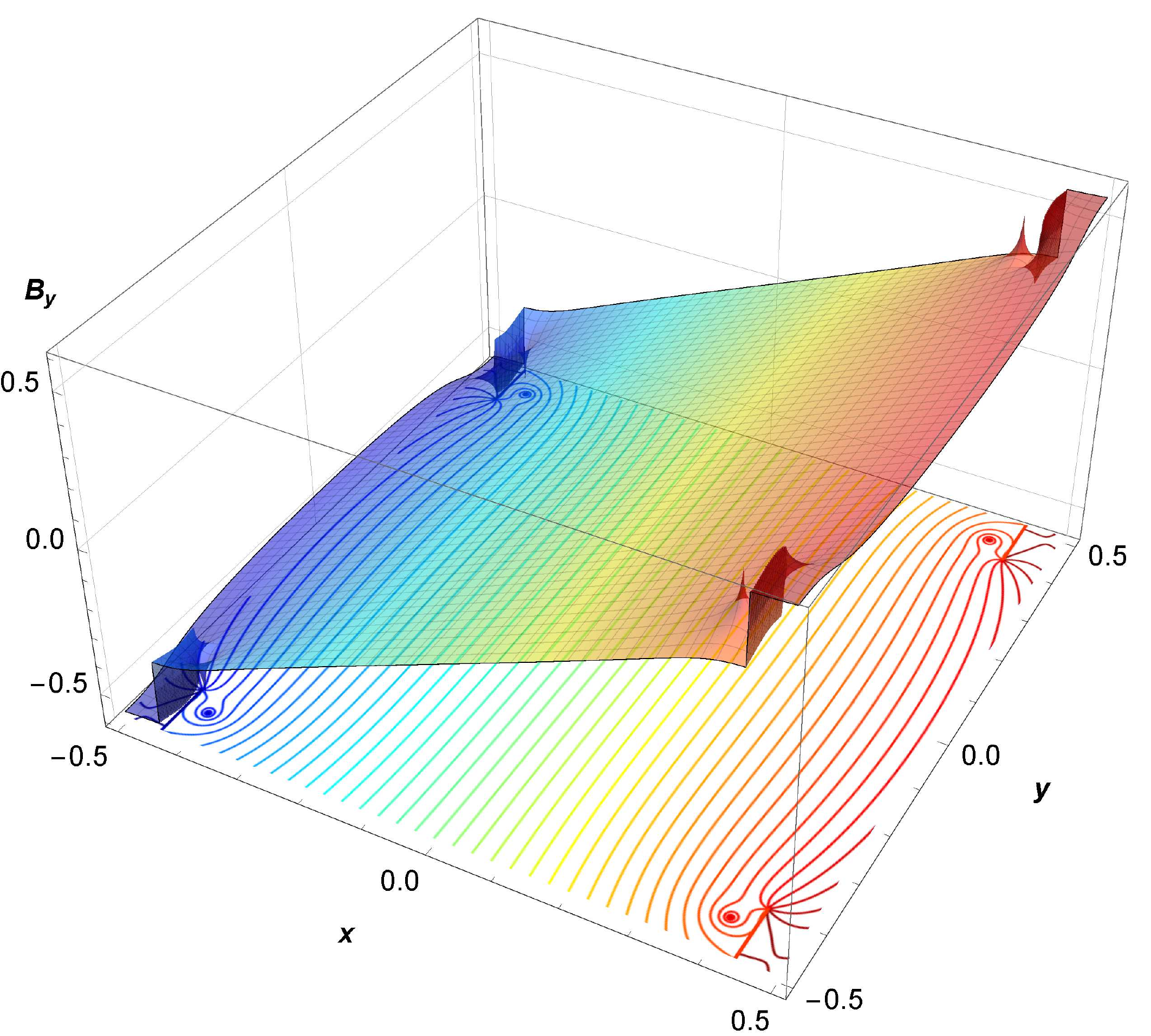} &
\includegraphics[trim=0 0 0 8.5cm,clip,width=0.28\textwidth]{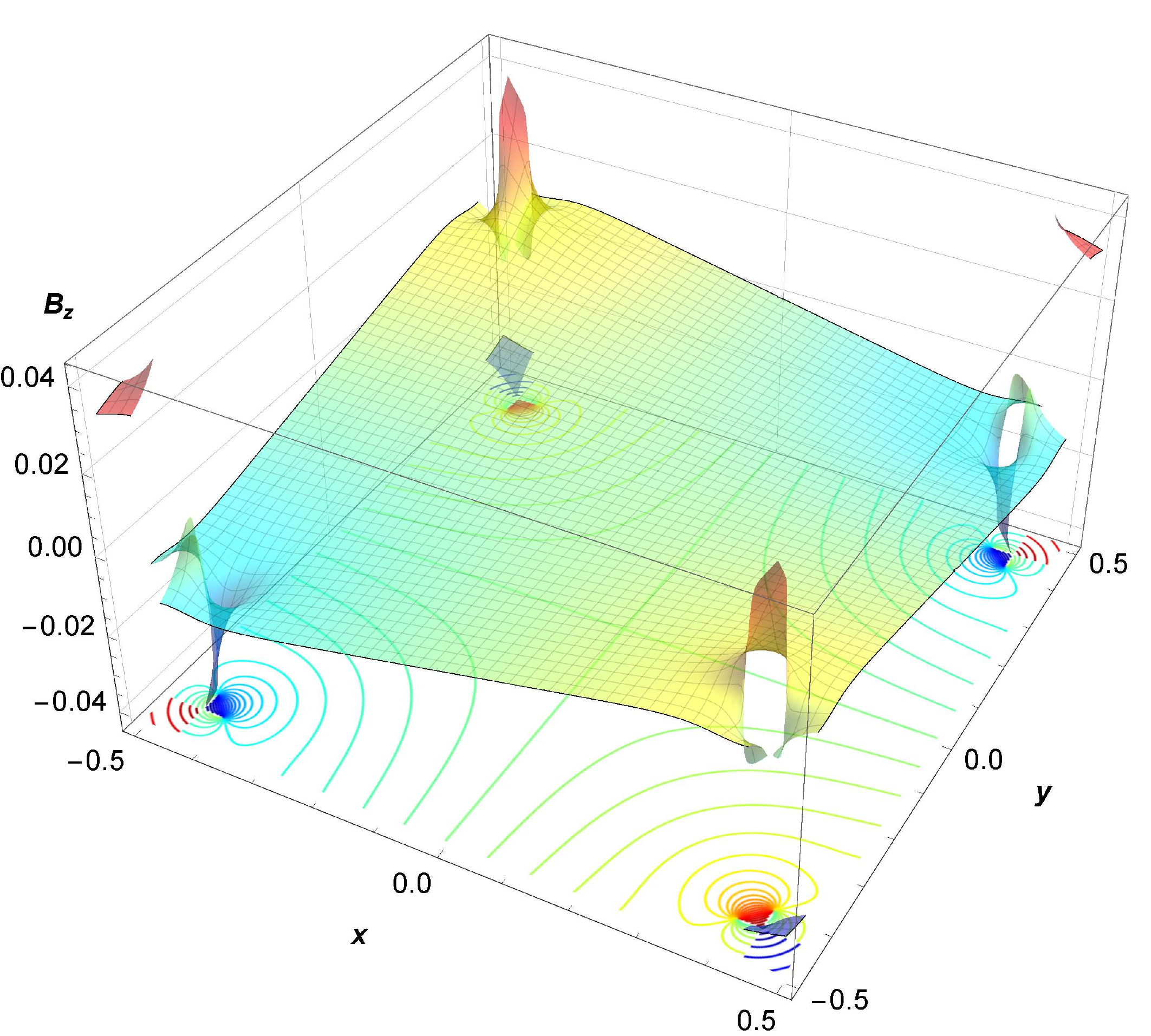} \\
\includegraphics[trim=0 0 0 8.5cm,clip,width=0.28\textwidth]{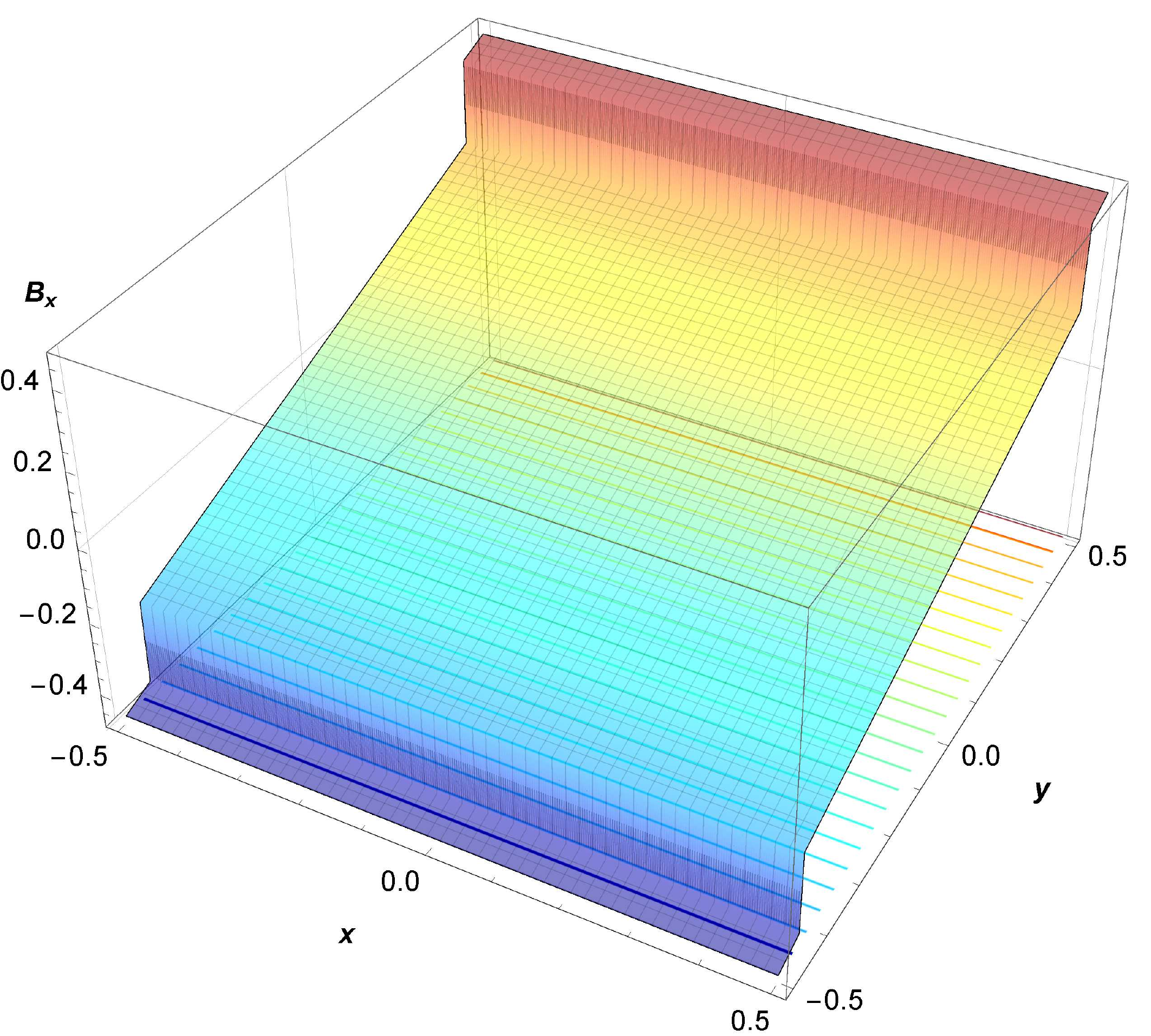} &
\includegraphics[trim=0 0 0 8.5cm,clip,width=0.28\textwidth]{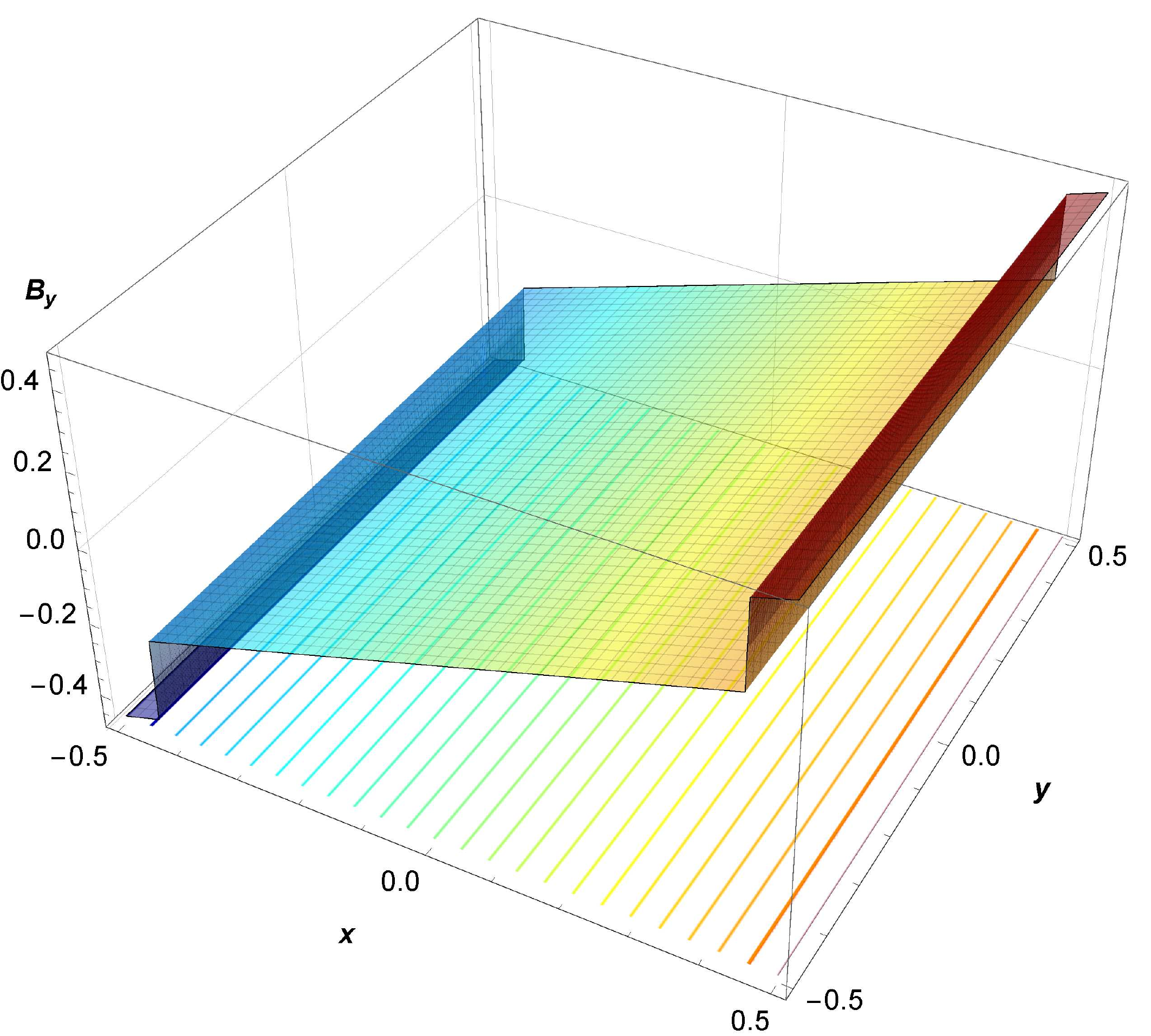} &
\includegraphics[trim=0 0 0 8.5cm,clip,width=0.28\textwidth]{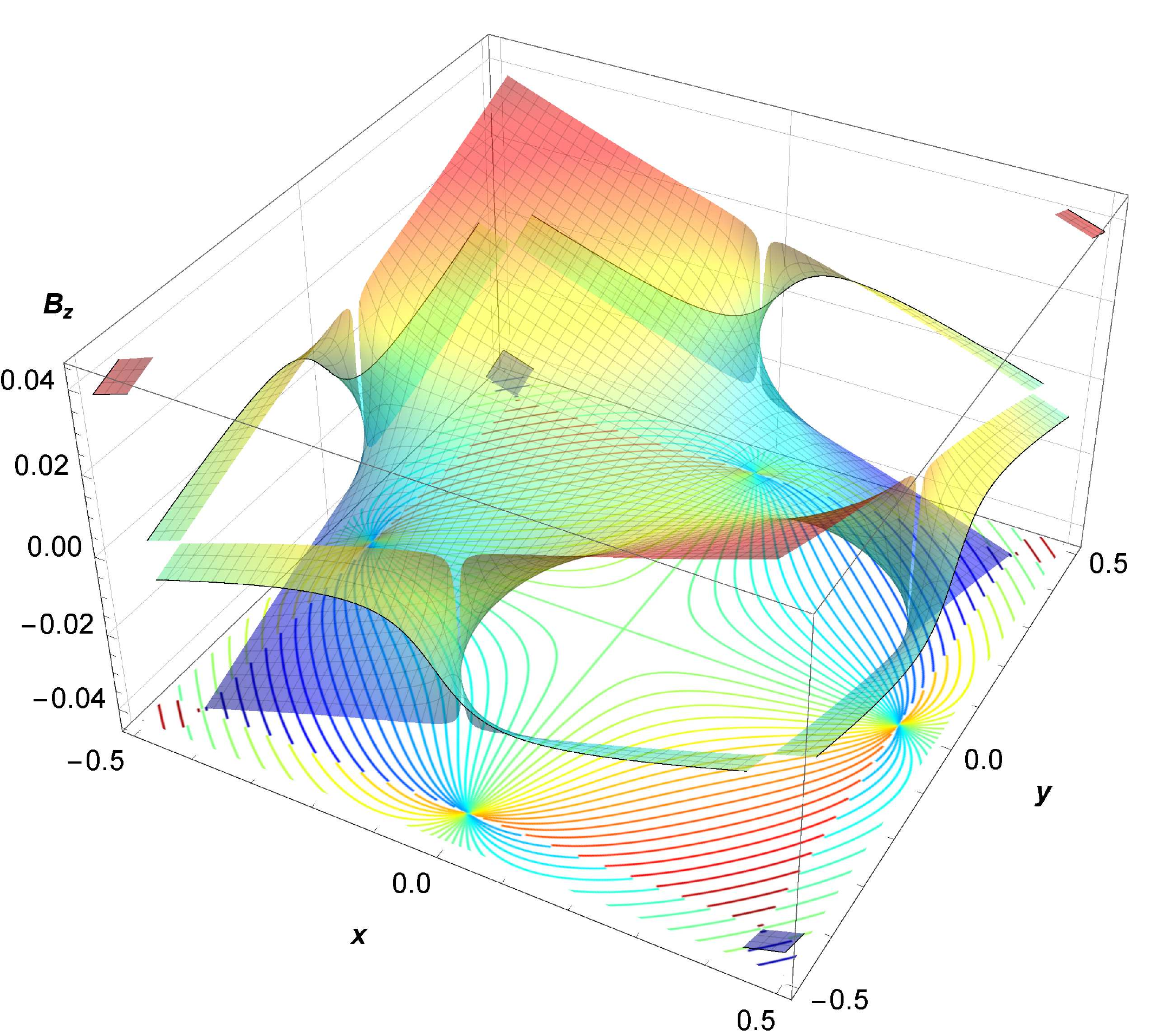} \\
\includegraphics[trim=0 0 0 8.5cm,clip,width=0.28\textwidth]{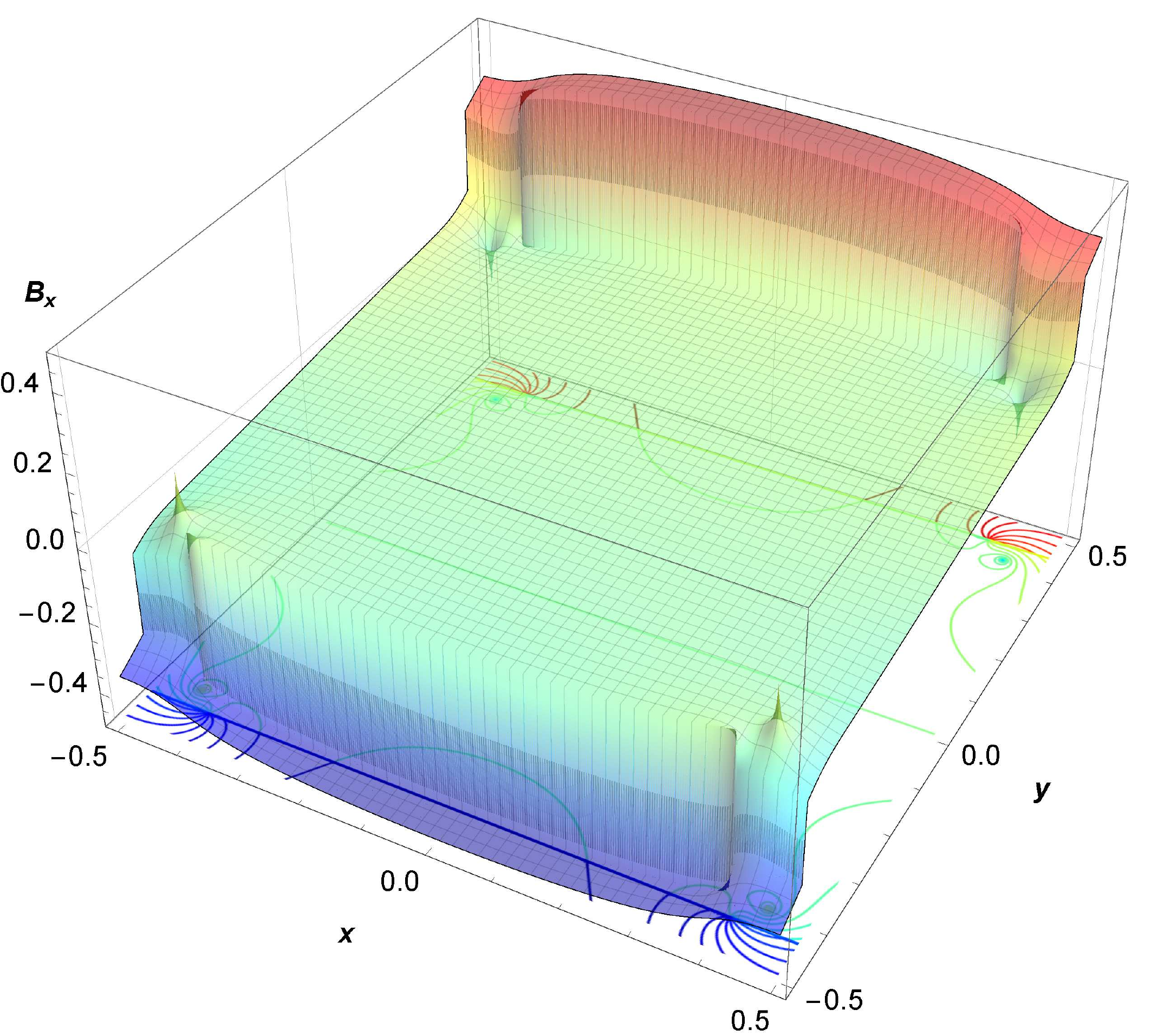} &
\includegraphics[trim=0 0 0 8.5cm,clip,width=0.28\textwidth]{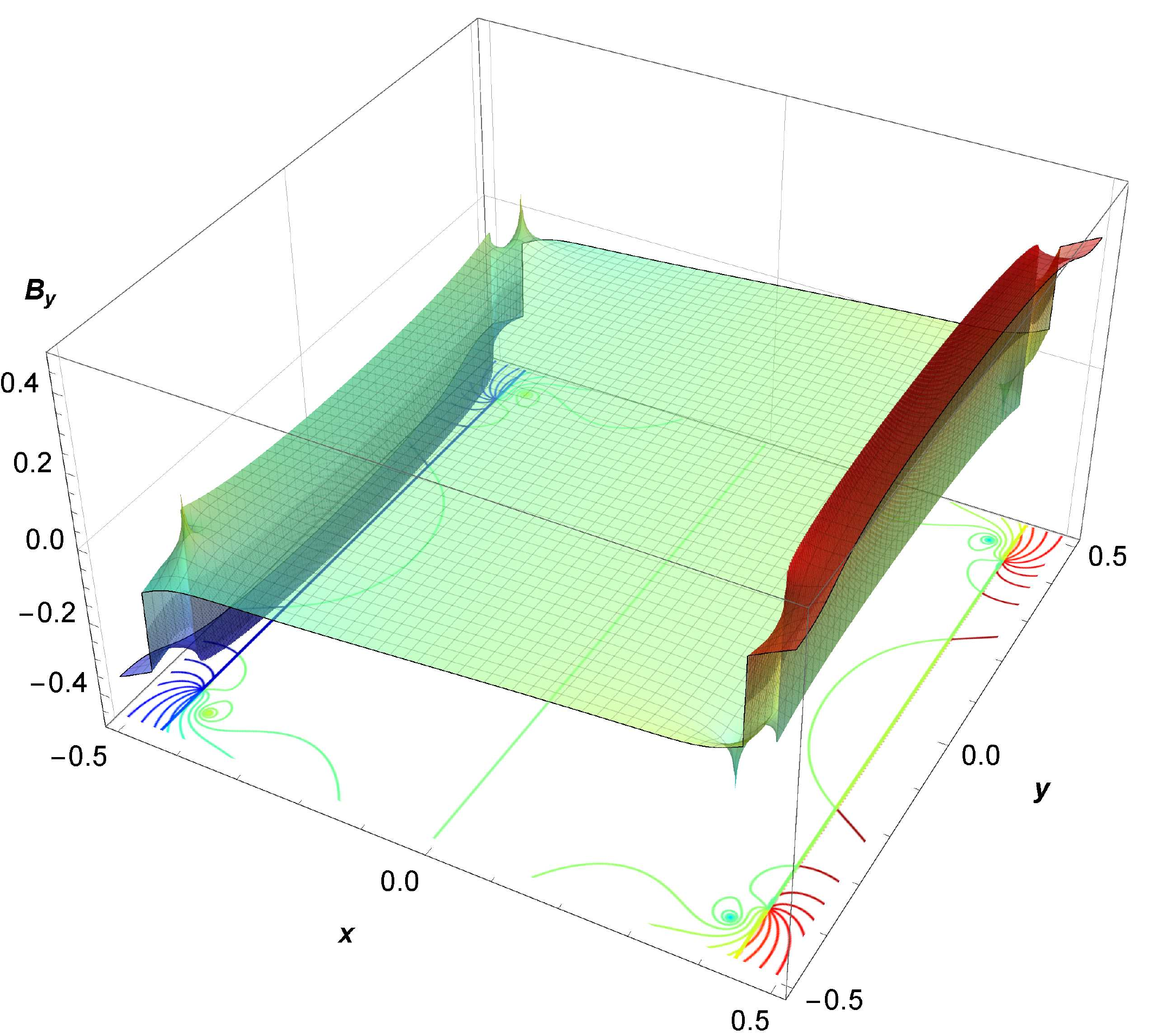} &
\includegraphics[trim=0 0 0 8.5cm,clip,width=0.28\textwidth]{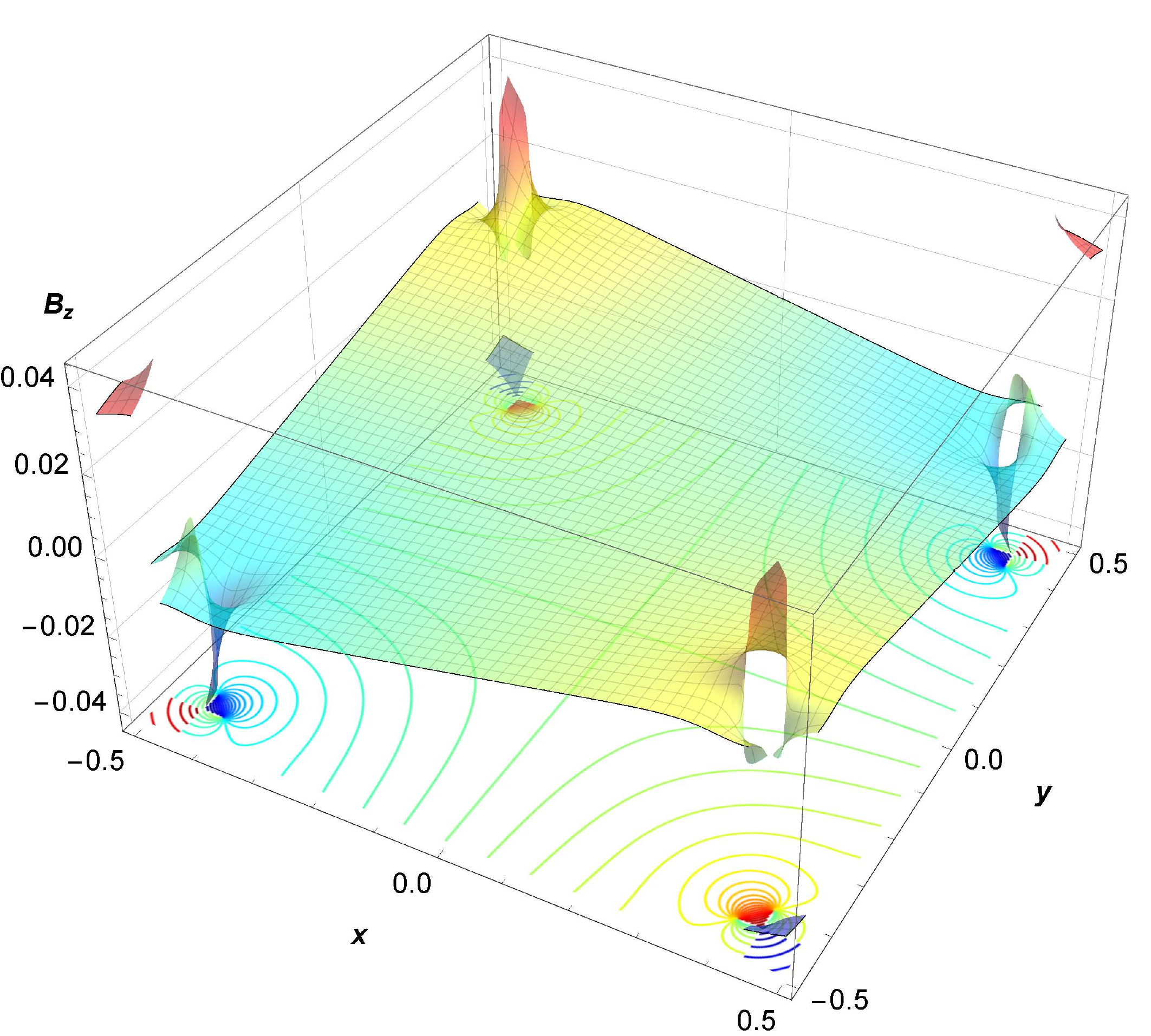} \\
\includegraphics[trim=0 0 0 8.5cm,clip,width=0.28\textwidth]{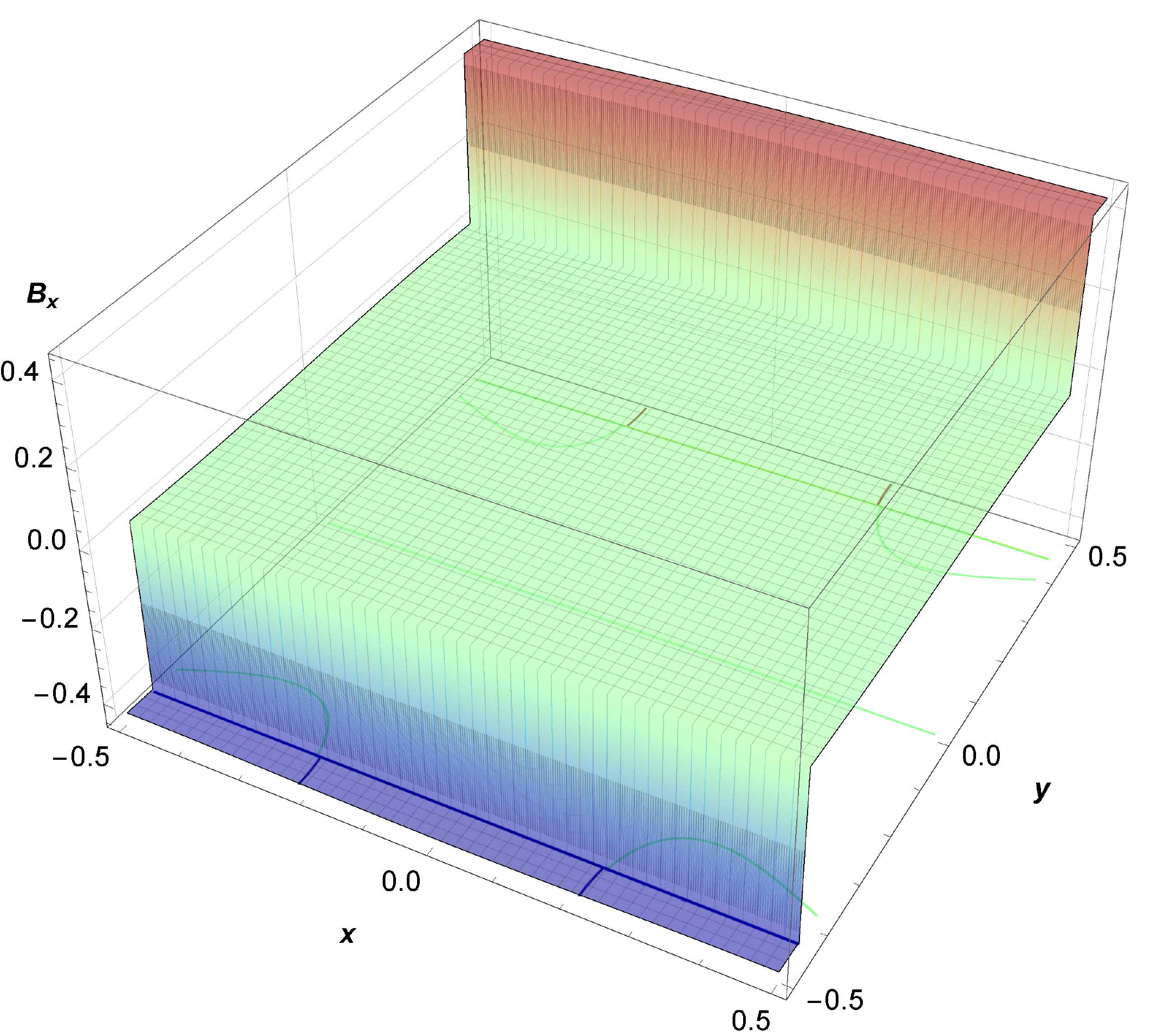} &
\includegraphics[trim=0 0 0 8.5cm,clip,width=0.28\textwidth]{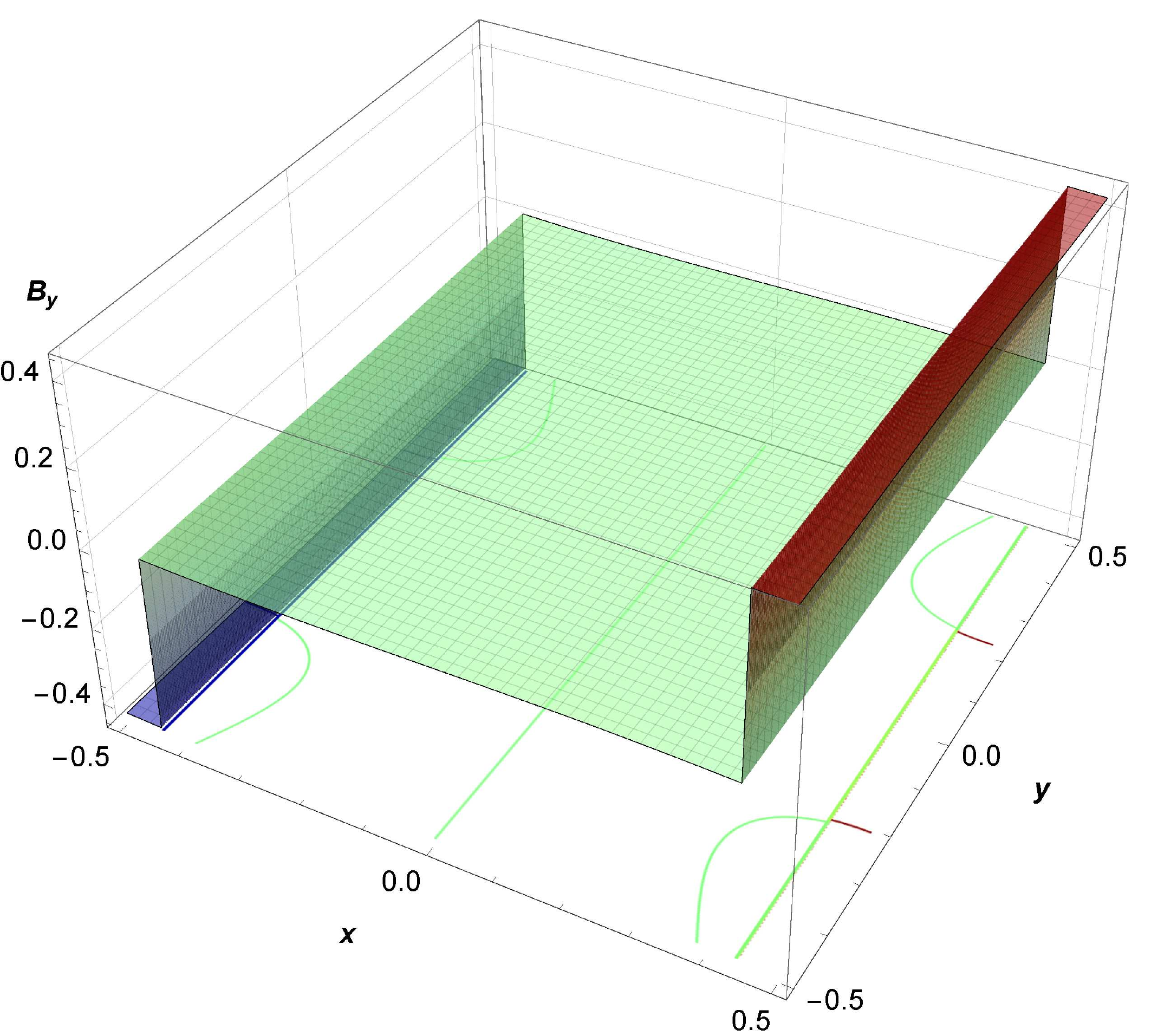} &
\includegraphics[trim=0 0 0 8.5cm,clip,width=0.28\textwidth]{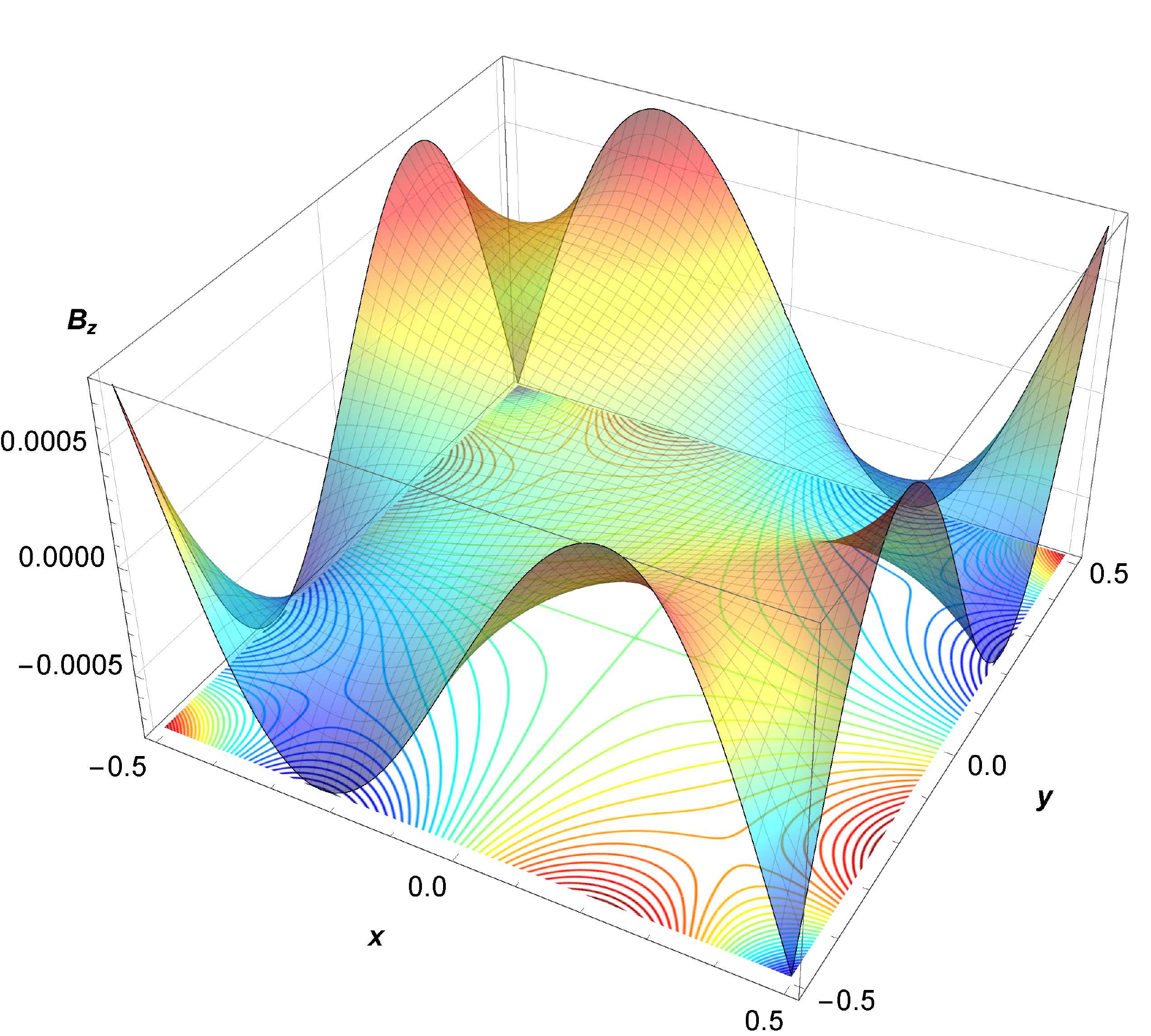}
\end{tabular}
\caption{Quadrupole fringe field components $B_x$ (left), $B_y$ (centre) and $B_z$ (right) at (top to bottom) $z = - 4.0$, $z = - 2.0$,
$z = 0$, $z = 2.0$ and $z = 4.0$.}
\label{engebxfm4m2}
\end{figure}

Figure \ref{engebxfm4m2} shows the behaviour of the field components $B_x$, $B_y$ and $B_z$ for the full quadrupole solution, with
the fixed value of the parameter $b_2 = 0.1$, and with the imposition of the symmetry $B_x \leftrightarrow B_y$ under $x \leftrightarrow y$. 
Within the body of the quadrupole (large negative $z$) we see that, as expected, $B_x \propto y$, $B_y \propto x$ and $B_z \approx 0$. 
At $z = 0$ the quadrupole gradient (the constant of proportionality between $B_x$ and $y$, or between $B_y$ and $x$) falls to half of its
value within the body of the magnet. At a large distance from the quadrupole (large positive $z$), the field approaches zero.

\subsection{Full solution for a multipole magnet with Enge-type fringe field}\label{sectionfullmultipoleenge}

The full solution for a multipole of order $n$ can be written:
\begin{eqnarray}
B_u & = & - \sum_{j = 1}^{n+1} b_j c_j \left( (-i)^{n+1} (\zeta + ih_j)^n G_n(\zeta + ih_j) - i^{n+1} (\zeta - ih_j)^n G_n(\zeta - ih_j) \right),
\nonumber \\
B_v & = & - \sum_{j = 1}^{n+1} \frac{c_j}{b_j} \left( (-i)^{n+1} (\zeta + ih_j)^n G_n(\zeta + ih_j) - i^{n+1} (\zeta - ih_j)^n G_n(\zeta - ih_j)
\right), \nonumber \\
B_\zeta & = &  i \sum_{j = 1}^{n+1} c_j \left( (-i)^{n+1} (\zeta + ih_j)^n G_n(\zeta + ih_j) + i^{n+1} (\zeta - ih_j)^n G_n(\zeta - ih_j) \right).
\nonumber
\end{eqnarray}
The function $G_0(\zeta)$ is an Enge function (with one coefficient):
\begin{equation}
G_0(\zeta) = \frac{1}{1 + e^\zeta}. \label{engefunctionG0}
\end{equation}
Functions $G_n(\zeta)$ for $n > 0$ are obtained by induction:
\begin{equation}
\zeta^n \, G_n(\zeta) = \int_0^\zeta \zeta^{\prime n-1}\, G_{n-1}(\zeta^\prime) \, \mathrm{d}\zeta^\prime. \label{engefunctionGn}
\end{equation}
It is then found that expanding the radial field component $B_r$ as a Taylor series in $r$ gives:
\[
B_r = \frac{r^n \sin ((n+1)\theta)}{2^\frac{n-1}{2} \, n! \, (1 + e^{\sqrt{2} z})} + O(r^{n+1}),
\]
where $O(r^{n+1})$ represents terms in $r$ of order $n+1$.  Thus, the field has the behaviour expected of a multipole magnet,
with gradient that decays as an Enge function in the fringe field region.

The functions $G_n(\zeta)$ can also be written in the form:
\[
G_n(\zeta) = \frac{1}{n!} + \frac{\polylog_n(-e^\zeta)}{\zeta^n} - \sum_{j = 1}^n \frac{\polylog_j(-1)}{\zeta^j (n-j)!},
\]
where $\polylog_n(\zeta)$ is the polylogarithm (or Jonqui\`ere function \cite{Jonquiere}) of order $n$.

\section{Potentials for multipole magnet fringe fields}

For some applications it may be useful to have analytical expressions for scalar and vector potentials from which the fringe fields
in multipole magnets can be derived.  For example, in an iron dominated magnet the pole faces correspond to surfaces of constant
scalar potential; this makes it possible to inspect the geometry of a magnet that would have a fringe field of a given form.
The vector potential can be useful for symplectic tracking of particles through the field (see, for example, \cite{WFR}).
General expressions for the scalar and vector potentials for the elementary solution were given in Section \ref{sectiongeneralthreedsolution}.
In this section, we consider in particular the case that the multipole gradient $g_n(\zeta)$ has a roll-off in the fringe field that can be
represented by an Enge function.

\subsection{Scalar potential and pole-face geometry}

Consider a fringe field of the form:
\begin{eqnarray}
B_u & = & - \sum_{j = 1}^{n+1} b_j c_j \left( (-i)^{n+1} g_n(\zeta + ih_j) - i^{n+1} g_n(\zeta - ih_j) \right), \label{generalfringebu} \\
B_v & = & - \sum_{j = 1}^{n+1} \frac{c_j}{b_j} \left( (-i)^{n+1} g_n(\zeta + ih_j) - i^{n+1} g_n(\zeta - ih_j) \right), \label{generalfringebv} \\
B_\zeta & = &  i \sum_{j = 1}^{n+1} c_j \left( (-i)^{n+1} g_n(\zeta + ih_j) + i^{n+1} g_n(\zeta - ih_j) \right). \label{generalfringebzeta}
\end{eqnarray}
In the case that:
\[
g_n(\zeta) = \zeta^n G_n(\zeta),
\]
where the functions $G_n(\zeta)$ are defined in (\ref{engefunctionG0}) and (\ref{engefunctionGn}),
this field represents the full solution for the fringe field in a multipole magnet of order $n$.  For the moment, we do not assume any
particular form for the function $g_n(\zeta)$.  The field (\ref{generalfringebu})--(\ref{generalfringebzeta}) may be derived from a scalar
potential given by:
\[
\varphi = \sum_{j = 1}^{n+1} (-i)^n c_j \tilde{g}_n(\zeta + i h_j) - i^n c_j \tilde{g}_n(\zeta - i h_j),
\]
where:
\[
\tilde{g}_n(\zeta) = \int_0^\zeta g_n(\zeta^\prime)\, \mathrm{d}\zeta^\prime.
\]
This expression for the scalar potential may be verified by substituting into $\vec{B} = \nabla \varphi$.
If it is possible to integrate the function $g_n(\zeta)$ describing the fringe field, then it is possible to write down a scalar potential for
the field. In the case of the fringe field of a multipole magnet of order $n$, with roll-off given by an Enge function, $\tilde{g}_n(\zeta)$
is given by:
\[
\tilde{g}_n(\zeta) = \int_0^\zeta g_n(\zeta^\prime)\, \mathrm{d}\zeta^\prime = \int_0^\zeta \zeta^{\prime n} G_n(\zeta^\prime)\,
\mathrm{d}\zeta^\prime = \zeta^{n+1}G_{n+1}(\zeta).
\]
Hence, in a multipole magnet of order $n$ with fringe field falling off as an Enge function, the scalar potential can be written:
\begin{equation}
\varphi = \sum_{j = 1}^{n+1} (-i)^n c_j (\zeta + i h_j)^{n+1}G_{n+1}(\zeta + i h_j) - i^n c_j (\zeta - i h_j)^{n+1}G_{n+1}(\zeta - i h_j).
\label{scalarpotentialmultipolefullsolution}
\end{equation}

In an iron dominated magnet, the pole faces can be identified with surfaces of constant scalar potential: an example of such a surface (for
an inner triplet quadrupole for the HL-LHC) is shown in Fig.~\ref{hllhcitpolefaceshape}.  The surface has the geometry expected of the poles
in an iron dominated quadrupole.  Within the body of the magnet, the intersection of the pole face with a plane $z = $constant forms a
hyperbola. As a function of position along the axis, the pole face has little dependence on $z$ for large negative $z$; but in the fringe field
region the pole face abruptly ``flattens'' off to lie close to the plane $z = 0$.

\subsection{Vector potential}

It can be useful to know the vector potential for a given field for symplectic integration of the equations of motion for a charged particle
moving through the field: see, for example, \cite{WFR}.
It is possible to write down an expression for a vector potential $\vec{A}$ from which the field (\ref{generalfringebu})--(\ref{generalfringebzeta})
may be derived.  In a gauge with $A_x = 0$, the components $A_y$ and $A_z$ are given by:
\begin{eqnarray}
A_y & = & - \sqrt{2} i \sum_{j = 1}^{n+1} \frac{c_j}{d_j} \left( (-i)^n \tilde{g}_n(\zeta + ih_j) + i^n \tilde{g}_n(\zeta - ih_j) \right), \nonumber \\
A_z & = &  - i \sum_{j = 1}^{n+1} \frac{c_j e_j}{d_j} \left( (-i)^n \tilde{g}_n(\zeta + ih_j) - i^n \tilde{g}_n(\zeta - ih_j) \right). \nonumber
\end{eqnarray}
This form for the vector potential may be verified by substitution\footnote{The result $d_j^2 = e_j^2 + 2$ may be useful for verifying
the expressions for the vector potential.} into $\vec{B} = \nabla \times \vec{A}$.  

In the case that the field (\ref{generalfringebu})--(\ref{generalfringebzeta}) represents the fringe field in a multipole magnet:
\[
g_n(\zeta) = \zeta^n G_n(\zeta),
\]
with $G_0(\zeta)$ an Enge function, it is possible to express the vector potential in terms of the functions $G_n(\zeta)$
(defined in (\ref{engefunctionG0}) and (\ref{engefunctionGn})) using:
\[
\tilde{g}_n(\zeta) = \zeta^{n+1}G_{n+1}(\zeta).
\]

\section{Examples}

To illustrate the application of the methods and results described in the previous sections, we consider two examples: a quadrupole in a
final focus or ``inner triplet'' region of the high luminosity upgrade of the Large Hadron Collider (HL-LHC), and a quadrupole in the EMMA
non-scaling fixed-field alternating gradient accelerator.  These magnets are chosen to provide two contrasting cases in terms of magnet
technology and parameter regime.  The HL-LHC inner triplet quadrupole is a superconducting magnet with large aperture and gradient
(respectively, 150\,mm diameter and 140\,T/m).  Design studies are still in progress; some field maps are available, but the impact of the
fringe fields on the beam dynamics are still under investigation.  The EMMA quadrupoles are normal-conducting electromagnets with more
conventional gradient.  Here, we consider one of the two types of EMMA magnet, namely the F quadrupoles \cite{emmaquads}.  These
magnets have large apertures (74\,mm diameter) given the length of the iron poles (73\,mm); as a result, the fringe fields make a
dominant contribution to the focusing effects. The specified integrated gradient is 0.387\,T.  Although it is possible to represent an EMMA
quadrupole by a hard-edge model, beam dynamics studies (supported by experimental results) \cite{yoelthesis} indicate the need for a
more realistic representation in order to give an accurate description of the longitudinal and transverse dynamics in the machine.  

We should emphasise that the purpose of considering these illustrative cases is not to demonstrate close agreement between the numerical
and analytical field models, but simply to show that it is possible, using analytical formulae from the previous sections, to construct a model
of the fringe field in each case that satisfies Maxwell's equations, and is closer to reality than a simple hard-edge model.

For the HL-LHC inner triplet quadrupole and for the EMMA F quadrupole we use the full solution for a quadrupole with Enge-type
fringe field, presented in Section \ref{sectionfullquadrupoleenge}.  The numerical field data in each case are fitted using a roll-off
function such that the quadrupole gradient has the form of an Enge function:
\begin{equation}
g(z) = \frac{a_0}{1 + e^{a_1 + \sqrt{2} a_2 z}}.
\label{examplegradientenge}
\end{equation}
The fit is optimised by varying the parameters $a_0$, $a_1$ and $a_2$.  Following the same procedure as in Section \ref{sectionfullquadrupoleenge},
we find that the radial component of the magnetic field can be written to first order in the radial co-ordinate $r$:
\begin{equation}
B_r = \frac{a_0 \sin(2\theta) r}{1 + e^{a_1 + \sqrt{2} a_2 z}} + O(r^3).
\end{equation}
If the radial component of the field is known as a function of longitudinal position for some $\theta$ and (small) $r$,
then $a_0$, $a_1$ and $a_2$ can be obtained from a nonlinear fit to the data.  With the gradient described by an
Enge function as in (\ref{examplegradientenge}), the function $G(\zeta)$ in (\ref{fullsolutionquadbu})--(\ref{fullsolutionquadbzeta})
is given by:
\begin{equation}
G(\zeta) = a_0 \left( 1 - \frac{\ln(1 + e^{a_1 + a_2 \zeta}) - \ln(1 + e^{a_1})}{a_2 \zeta} \right).
\label{exampleGfunction}
\end{equation}
All field components at any point in the magnet (including the fringe field) can then be obtained using
(\ref{fullsolutionquadbu})--(\ref{fullsolutionquadbzeta}).

\subsection{HL-LHC inner triplet quadrupoles}\label{sectionhllhcitquads}

As a first example, we show the results of a fit to the field in an HL-LHC inner triplet quadrupole.
Field data are obtained from a magnetic model, and a fit to the field based on a function of the form (\ref{exampleGfunction})
is then obtained using the method described above.  The parameter values obtained by fitting the radial field component along a line
at $\theta = \pi/4$ and various values of $r$ are shown in Table \ref{hllhcfitparams}.  If the chosen function provides a good
description of the field, then we expect little variation in the parameters obtained by fitting along lines with different $\theta$
and $r$, at least for small $r$ (since the fit is based on the linear term in a series expansion of the field).  For the HL-LHC inner
triplet quadrupole, inspection of the values shown in Table \ref{hllhcfitparams} shows changes in the fit parameters of less than
0.7\%, comparing values obtained by fitting the field along lines with $r = r_\mathrm{max}/10$ and $r = r_\mathrm{max}/6$
(and fixed $\theta = \pi/4$).

\begin{table}
\centering
\caption{Parameters for a fit of an Enge function of the form (\ref{examplegradientenge}), to the quadrupole
gradient in an HL-LHC inner triplet quadrupole.  The maximum radius $r_\mathrm{max}$ in the field data is 75\,mm.}
\label{hllhcfitparams}
\begin{tabular}{cccc}
\hline
radius of fit & $a_0$ & $a_1$ & $a_2$ \\
\hline
$r_\mathrm{max}/10$ & -55.9503 & -0.520120 & 8.98913 \\
$r_\mathrm{max}/8$ & -55.9504 & -0.521262 & 9.00549 \\
$r_\mathrm{max}/6$ & -55.9505 & -0.523743 & 9.04082 \\
$r_\mathrm{max}/4$ & -55.9510 & -0.530875 & 9.14139 \\
$r_\mathrm{max}/3$ & -55.9517 & -0.540984 & 9.28116 \\
$r_\mathrm{max}/2$ & -55.9539 & -0.570142 & 9.66955 \\
\hline
\end{tabular}
\end{table}

\begin{figure}
\centering
\includegraphics[width=0.5\textwidth]{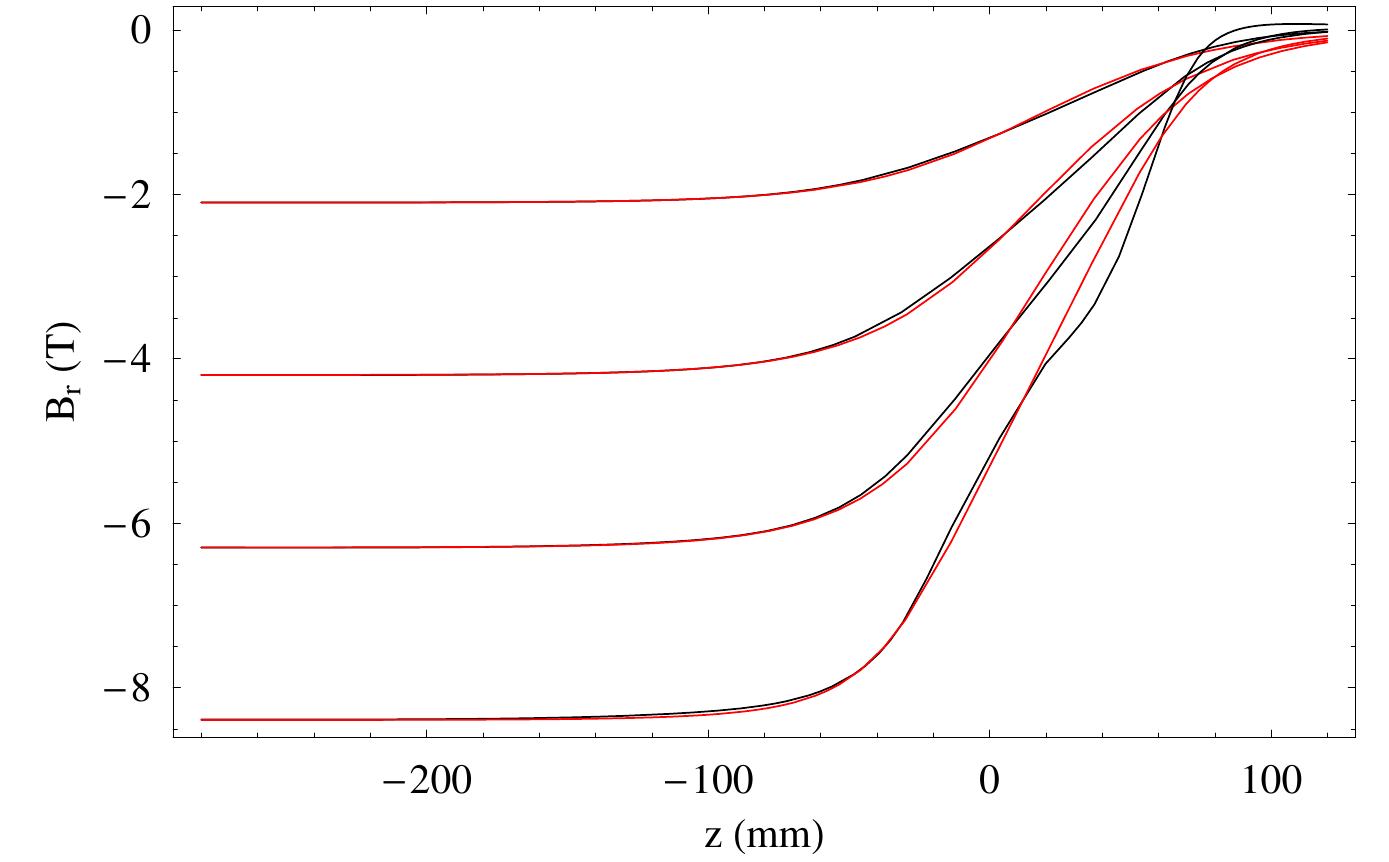}
\caption{Radial component of the magnetic field in an HL-LHC inner triplet quadrupole
as a function of position along lines of given radius and with cylindrical polar co-ordinate $\theta = \pi/4$.
Each pair of black and red lines shows the field at a different radius, from $r_\mathrm{max}/5$ to
$4r_\mathrm{max}/5$, with $r_\mathrm{max} = 75$\,mm.
The black lines show the field obtained from the (numerical) magnetic model; the red lines show the
analytical model (\ref{fullsolutionquadbu})--(\ref{fullsolutionquadbzeta}), with fit parameters given
in Table \ref{hllhcfitparams} with radius of fit $r_\mathrm{max}/10$, and $b_1 = 2.5$.}
\label{examplefigbrvszvariousr}
\end{figure}

Having obtained values for the fit parameters, we can make direct comparisons of the field given by equations
(\ref{fullsolutionquadbu})--(\ref{fullsolutionquadbzeta}) with the field obtained by the numerical magnetic model.
An example of such a comparison, for the radial field component, is shown in Fig.~\ref{examplefigbrvszvariousr}.
Although the analytical model does not match the numerical model exactly, the general behaviour of the field
is reproduced quite closely.  The impact of the residuals from the fit on the beam dynamics still needs to be studied;
but the analytical model does include features of the field that would be completely omitted in a hard-edged magnet model.

\begin{figure}
\centering
\includegraphics[width=0.5\textwidth]{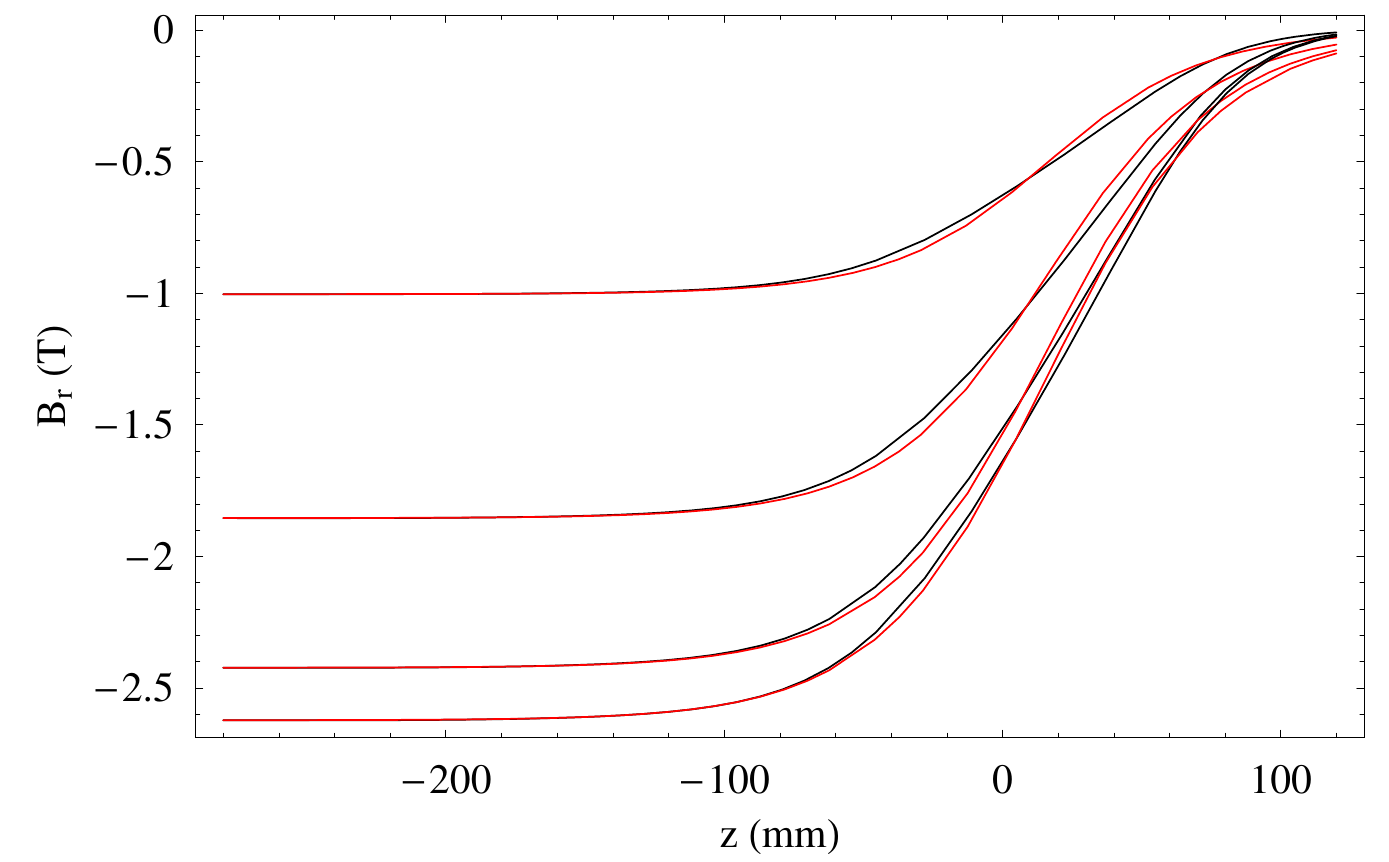}
\caption{Radial component of the magnetic field in an HL-LHC inner triplet quadrupole as a function of
position along lines of fixed distance $r = r_\mathrm{max}/4$.
Each pair of black and red lines shows the field at a different value of the 
cylindrical polar co-ordinate $\theta$, from $\pi/16$ to $\pi/4$ (in steps of $\pi/16$).
The black lines show the field obtained from the (numerical) magnetic model; the red lines show the
analytical model (\ref{fullsolutionquadbu})--(\ref{fullsolutionquadbzeta}), with fit parameters given
in Table \ref{hllhcfitparams} with radius of fit $r_\mathrm{max}/10$, and $b_1 = 2.5$.}
\label{examplefigbrvszvarioustheta}
\end{figure}

\begin{figure}
\centering
\includegraphics[width=0.5\textwidth]{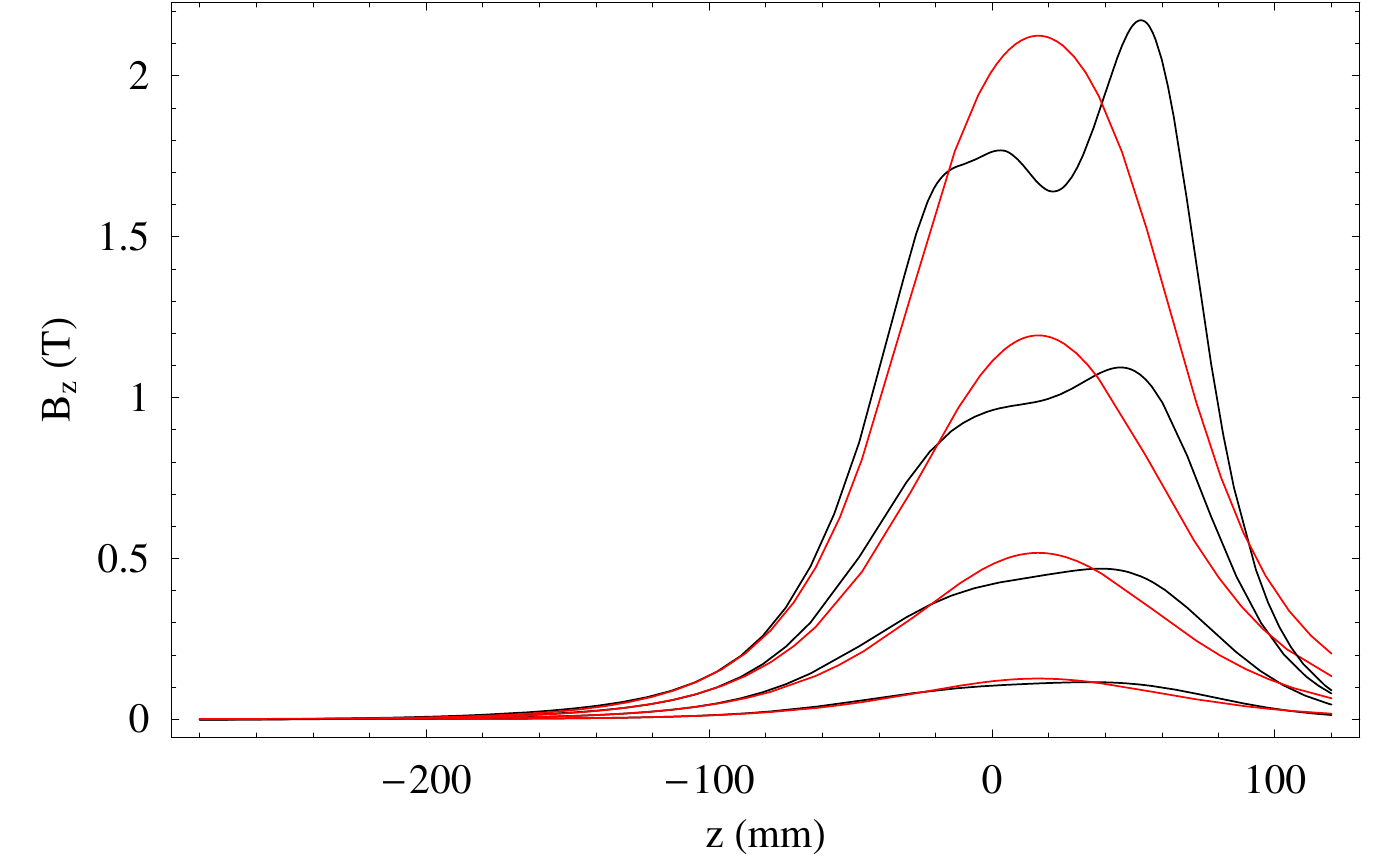}
\caption{Longitudinal component of the magnetic field in an HL-LHC inner triplet quadrupole
as a function of position along lines of given radius and with cylindrical polar co-ordinate $\theta = \pi/4$.
Each pair of black and red lines shows the field at an increasing radius, from $r_\mathrm{max}/5$ to
$4r_\mathrm{max}/5$, with $r_\mathrm{max} = 75$\,mm.
The black lines show the field obtained from the (numerical) magnetic model; the red lines show the
analytical model (\ref{fullsolutionquadbu})--(\ref{fullsolutionquadbzeta}), with fit parameters given
in Table \ref{hllhcfitparams} with radius of fit $r_\mathrm{max}/10$, and $b_1 = 2.5$.}
\label{examplefigbzvszvariousr}
\end{figure}

Figure \ref{examplefigbrvszvarioustheta} shows the radial field component as a function of position along lines of
fixed radius and for different values of the polar co-ordinate $\theta$.  Again, there is reasonable agreement between
the numerical field data (black lines) and the analytical fit (red line).  Finally, Fig.~\ref{examplefigbzvszvariousr}
shows the longitudinal field component as a function of position along lines of given radius and with fixed $\theta = \pi/4$.
Here, it appears that there are more significant discrepancies between the numerical field data and the analytical model;
however, there is general agreement in the main features, especially for the region close to the axis of the magnet.
The more detailed structure that appears at larger distances from the axis cannot be reproduced by the relatively simple
(Enge) function that is used to describe the fall-off of the quadrupole gradient in the fringe field.

\begin{figure}
\centering
\includegraphics[width=0.45\textwidth]{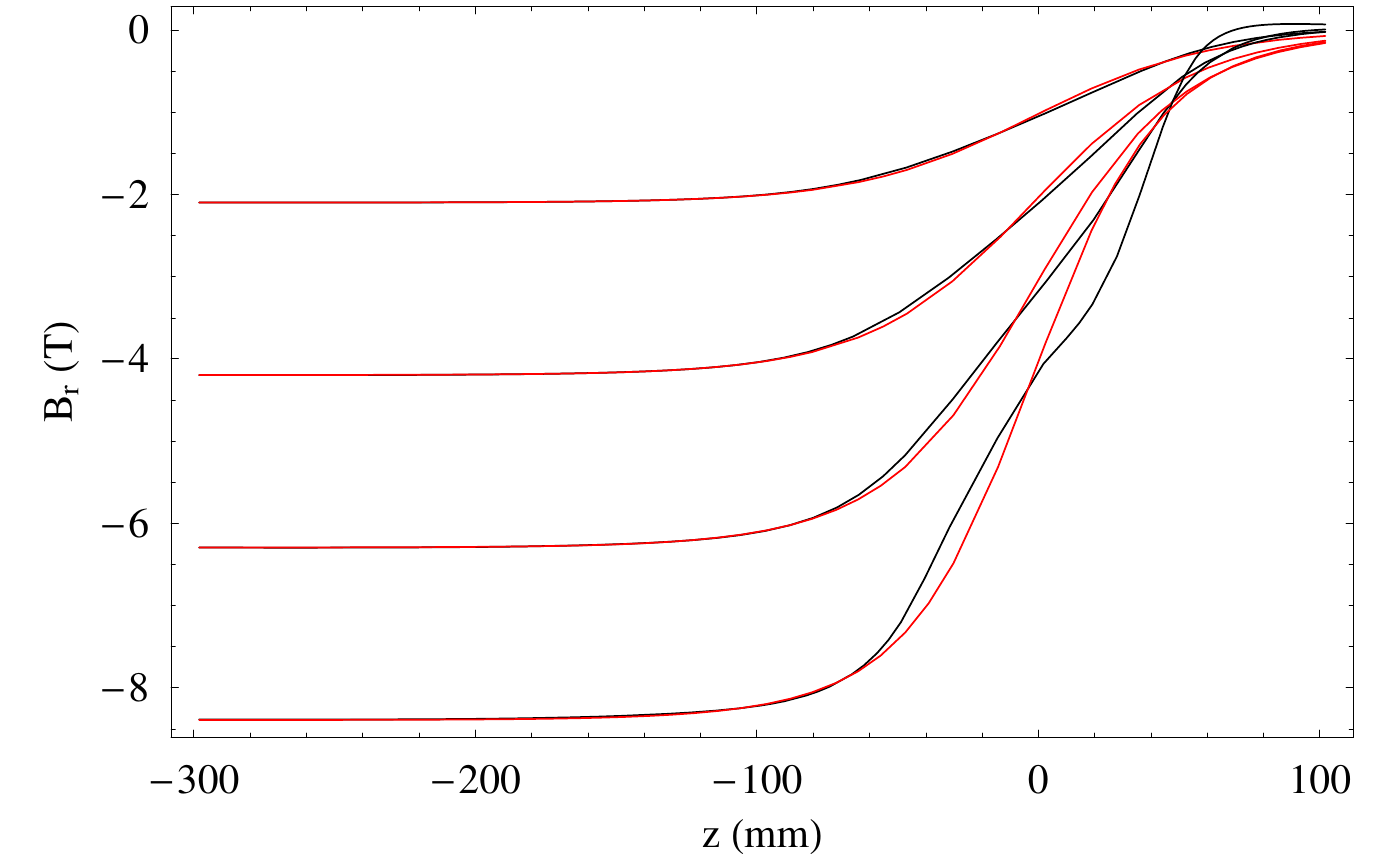}
\includegraphics[width=0.45\textwidth]{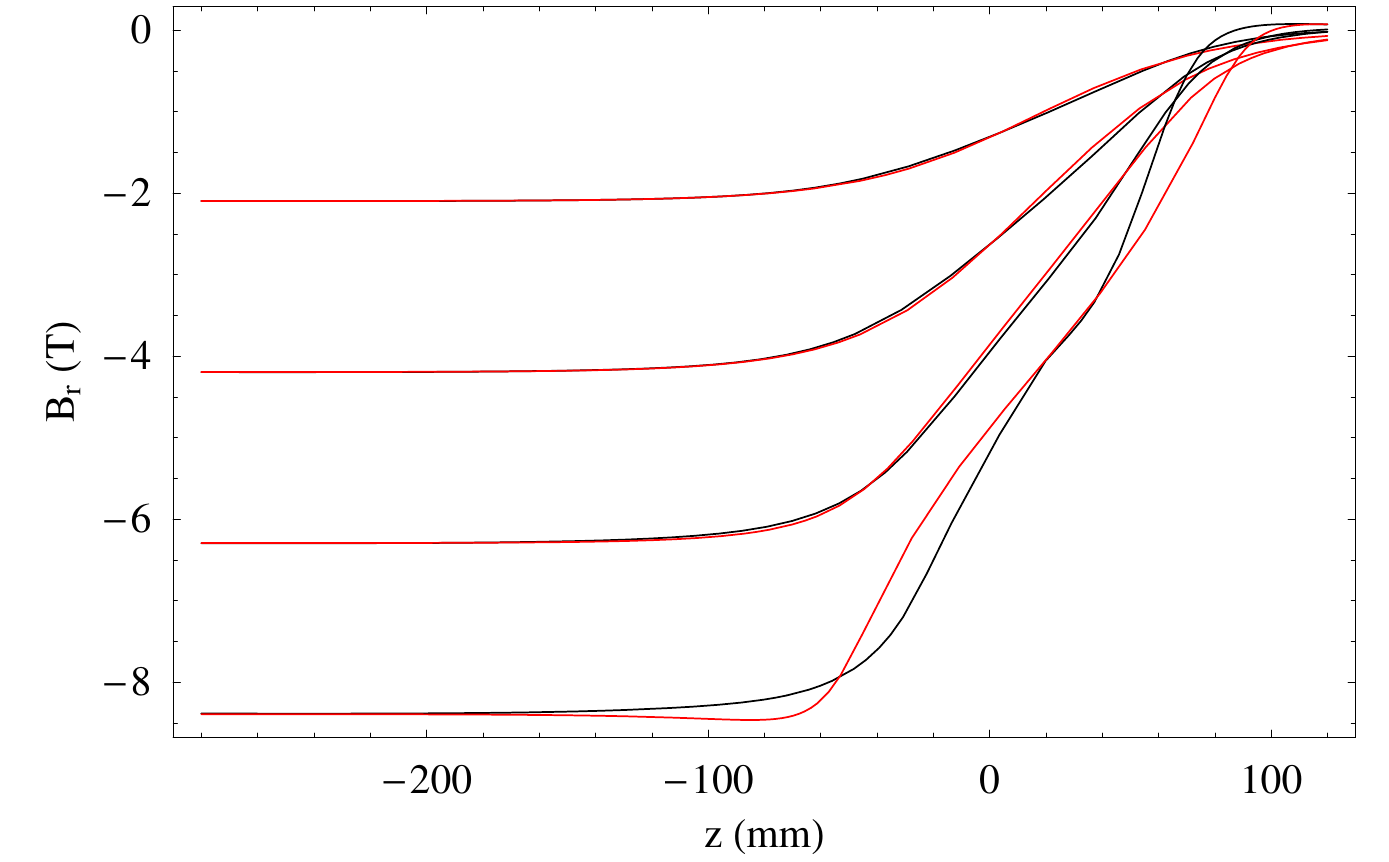}
\caption{As Fig.~\ref{examplefigbrvszvariousr}, but with $b_1 = 1.5$ (left) and $b_1 = 3.5$ (right).}
\label{examplefigbrvszvariousrb}
\end{figure}

It is worth considering the dependence of the field on the parameter $b_1 (= 1/b_2)$.  Changing the value of 
$b_1$ has an effect on the way that the fringe field varies with distance from the magnetic axis.  This
can be seen by comparing the plots in  Fig.~\ref{examplefigbrvszvariousrb}, which show the
radial field component as a function of position along lines with fixed polar co-ordinate $\theta = \pi/4$
and different distance $r$ from the axis of the magnet.  The plot on the left in Fig.~\ref{examplefigbrvszvariousrb}
shows a comparison between the numerical field map and the analytical model with $b_1 = 1.5$; the plot on
the right compares the numerical field map and the analytical model with $b_1 = 3.5$.  In regions close to 
the axis of the magnet, changes in the value of $b_1$ in the range 1.5 to 3.5 have little effect on the field;
but differences are apparent at larger distances from the axis.  The parameter $b_1$ therefore allows control
over the field behaviour at large $r$, after the parameters $a_0$, $a_1$ and $a_2$ have been chosen to
fit the field behaviour close to the axis of the magnet.

\begin{figure}
\centering
\includegraphics[width=0.6\textwidth]{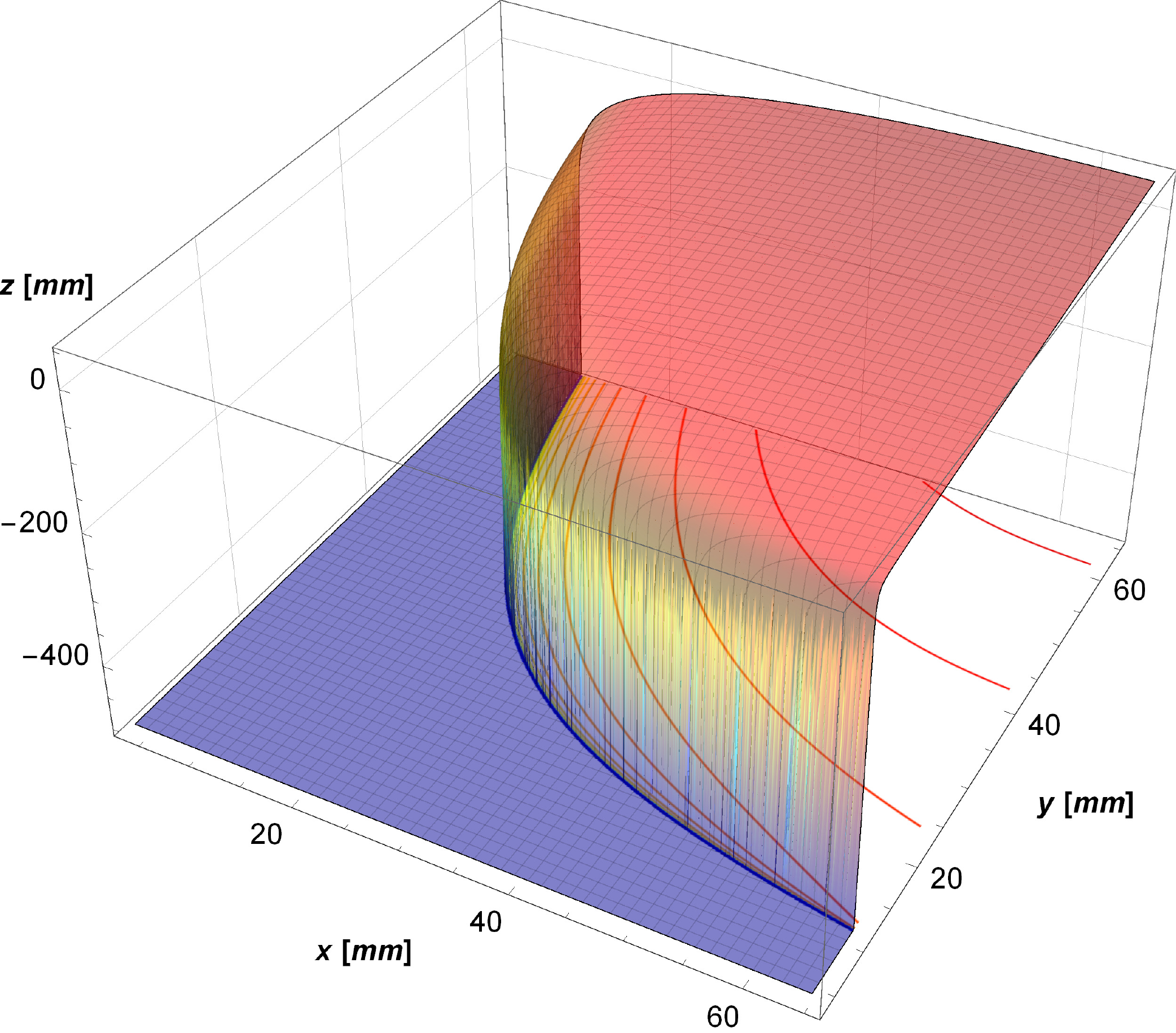}
\caption{Surface of constant scalar potential $\varphi = 0.25$\,T\,m in a representation of the HL-LHC inner triplet quadrupole,
with gradient falling as an Enge function in the fringe field.  The parameters of the Enge function (\ref{examplegradientenge})
correspond to the first line in Table \ref{hllhcfitparams}.}
\label{hllhcitpolefaceshape}
\end{figure}

Although the HL-LHC inner triplet quadrupoles are superconducting magnets, we can nevertheless inspect the shape
of the pole face that would be needed to give the same field.  Assuming poles with infinite magnetic permeability,
the shape of a pole face corresponds to a surface of constant magnetic scalar potential $\varphi$; in the case of a
quadrupole with gradient falling off as an Enge function, the scalar potential is given by
(\ref{scalarpotentialmultipolefullsolution}).  Figure \ref{hllhcitpolefaceshape} shows an equipotential surface for
the analytical field model fitted to the numerical HL-LHC inner triplet quadrupole field data.  The surface has been
chosen so that the magnetic scalar potential has the (arbitrary) value $\varphi = 0.25$\,T\,m.  We see that the
equipotential surface has the shape that might be expected
of an iron-dominated quadrupole, with the curved surace following an hyperbola in a plane of constant negative $z$, and
the end of the pole being reasonably flat (close to a plane of constant $z \approx 0$).

\subsection{EMMA quadrupoles}

The quadrupole magnets in EMMA are iron-dominated, normal conducting magnets.  The unusual feature of the EMMA
quadrupoles is that the diameter of the aperture is comparable to the length of the magnet; this allows the accelerator to have
a large transverse acceptance in a lattice consisting of magnets packed very close together.  However, the relatively large
aperture of the quadrupoles means that the gradient falls off rapidly from the centre of the magnet.  There is no appreciable
distance along the axis for which the gradient is constant, and a realistic model of the field must include some representation
of the fringe fields.

\begin{figure}
\centering
\includegraphics[width=0.5\textwidth]{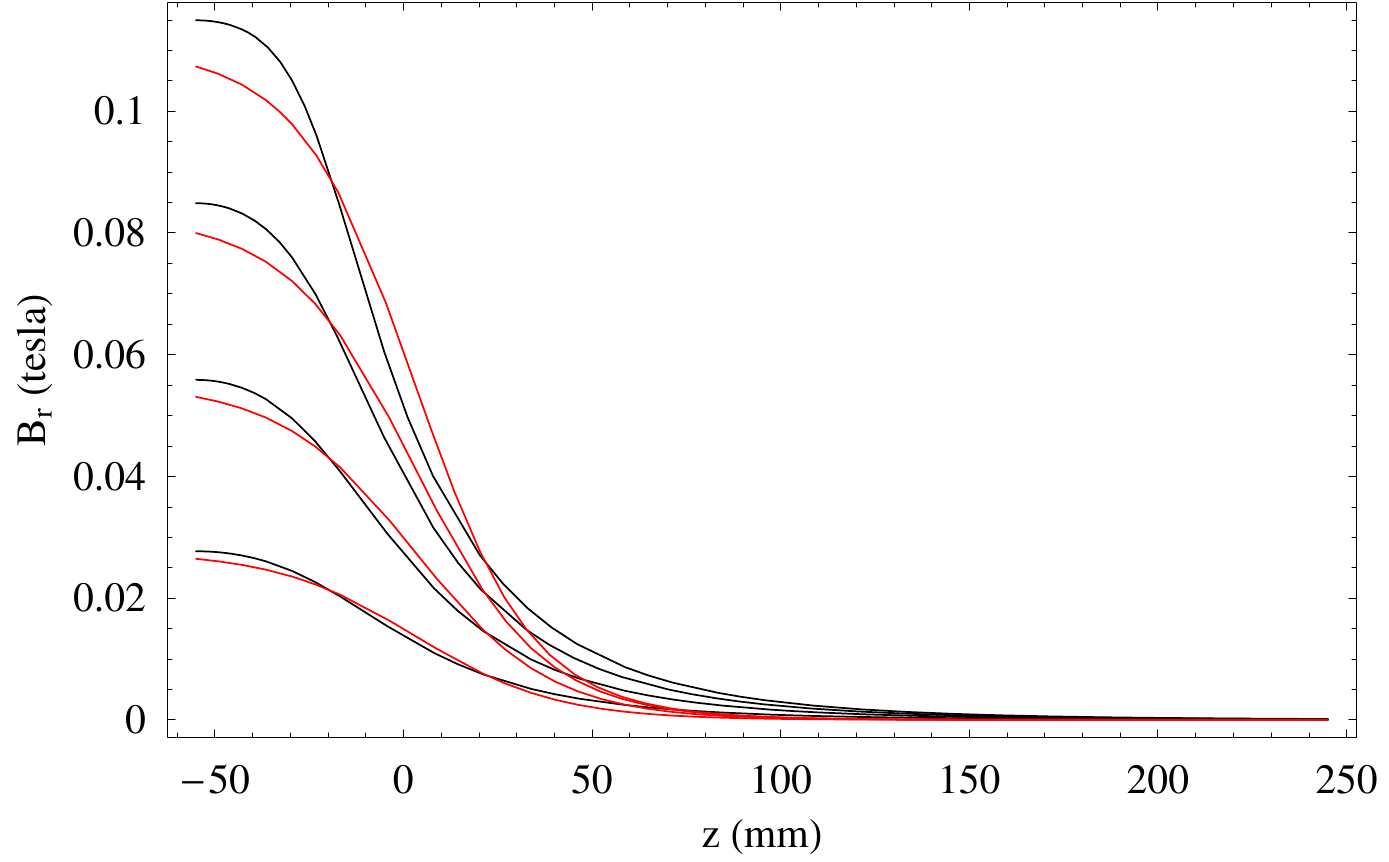}
\caption{Radial component of the magnetic field in an EMMA F quadrupole
as a function of position along lines of given radius and with cylindrical polar co-ordinate $\theta = \pi/4$.
Each pair of black and red lines shows the field at a different radius, from $r_\mathrm{max}/5$ to
$4r_\mathrm{max}/5$, with $r_\mathrm{max} = 36$\,mm.
The black lines show the field obtained from the (numerical) magnetic model; the red lines show the
analytical model (\ref{fullsolutionquadbu})--(\ref{fullsolutionquadbzeta}), with fit parameters given
in Table \ref{emmaquadfitparams} with radius of fit $r_\mathrm{max}/10$, and $b_1 = 1.8$.
The centre of the quadrupole is at the far left of the plot, $z = -53.6$\,mm.}
\label{examplefigbrvszvariousremma}
\end{figure}

\begin{figure}
\centering
\includegraphics[width=0.5\textwidth]{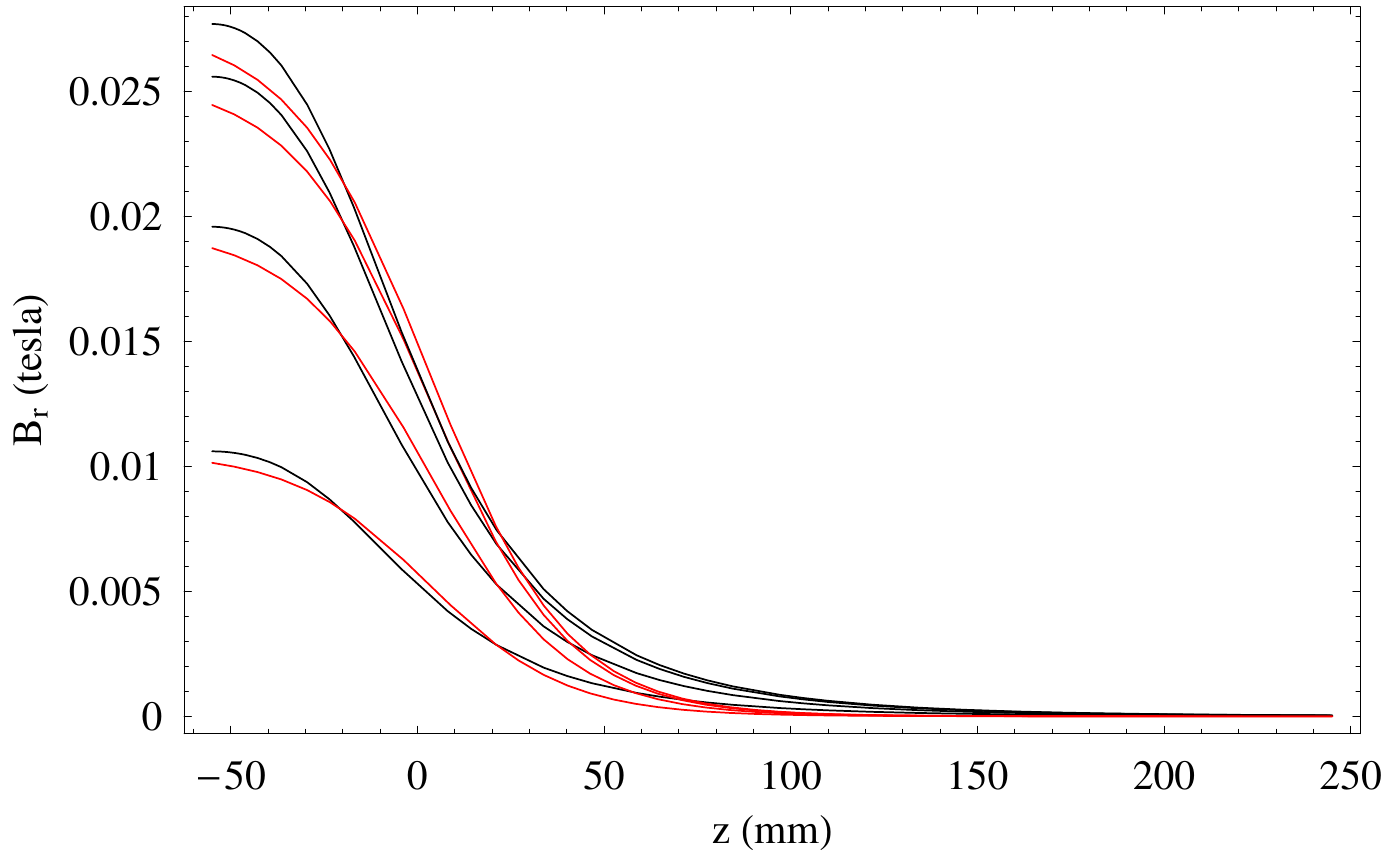}
\caption{Radial component of the magnetic field in an EMMA F quadrupole
as a function of position along lines of fixed distance $r = r_\mathrm{max}/4$.
Each pair of black and red lines shows the field at a different value of the 
cylindrical polar co-ordinate $\theta$, from $\pi/16$ to $\pi/4$ (in steps of $\pi/16$).
The black lines show the field obtained from the (numerical) magnetic model; the red lines show the
analytical model (\ref{fullsolutionquadbu})--(\ref{fullsolutionquadbzeta}), with fit parameters given
in Table \ref{emmaquadfitparams} with radius of fit $r_\mathrm{max}/10$, and $b_1 = 1.8$.
The centre of the quadrupole is at the far left of the plot, $z = -53.6$\,mm.}
\label{examplefigbrvszvariousthetaemma}
\end{figure}

\begin{figure}
\centering
\includegraphics[width=0.5\textwidth]{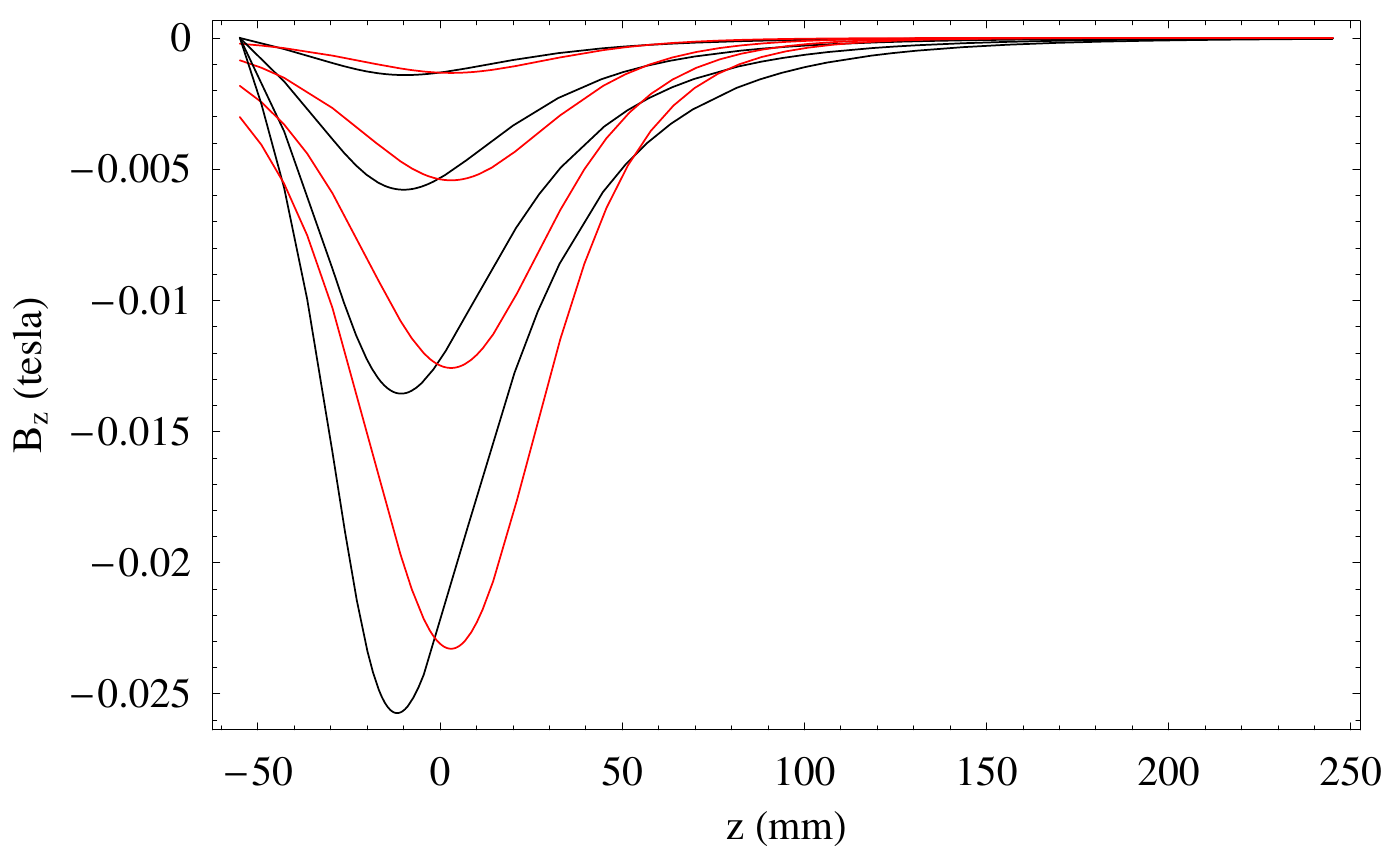}
\caption{Longitudinal component of the magnetic field in an EMMA F quadrupole
as a function of position along lines of given radius and with cylindrical polar co-ordinate $\theta = \pi/4$.
Each pair of black and red lines shows the field at an increasing radius, from $r_\mathrm{max}/5$ to
$4r_\mathrm{max}/5$, with $r_\mathrm{max} = 36$\,mm.
The black lines show the field obtained from the (numerical) magnetic model; the red lines show the
analytical model (\ref{fullsolutionquadbu})--(\ref{fullsolutionquadbzeta}), with fit parameters given
in Table \ref{emmaquadfitparams} with radius of fit $r_\mathrm{max}/10$, and $b_1 = 1.8$.
The centre of the quadrupole is at the far left of the plot, $z = -53.6$\,mm.}
\label{examplefigbzvszvariousremma}
\end{figure}

Given numerical field data from a magnetic model of an EMMA quadrupole, we can repeat the analysis used for the HL-LHC inner
triplet quadrupole in Section \ref{sectionhllhcitquads}.  We again use an Enge function of the form (\ref{examplegradientenge})
to represent the fall-off of the gradient along the axis of the magnet.  The field data cover an area closely approaching the pole
tip (the field values are given on a rectangular grid, with transverse co-ordinates extending to 36\,mm).  The results of fitting
the parameters in the Enge function to field data along a line parallel to the axis (at different distances from the axis and fixed
polar angle $\theta = \pi/4$) are shown in Table \ref{emmaquadfitparams}.  We see that there is much larger variation in the
parameters if the fit is performed at different distances from the axis, compared to the case of the HL-LHC inner triplet quadrupole.
This suggests that the quality of the fit will not be as good.  Making a direct comparison of the field based on the analytical formula
with the numerical field data confirms that this is the case: see Figs.~\ref{examplefigbrvszvariousremma},
\ref{examplefigbrvszvariousthetaemma} and \ref{examplefigbzvszvariousremma}.  Inspecting Fig.~\ref{examplefigbrvszvariousremma}
suggests the reason for the poor quality of the fit.  The gradient initially falls off quite rapidly along the axis from the centre of the
magnet, but there is a long ``tail'' as the gradient approaches zero: this asymmetric behaviour cannot be represented accurately using
an Enge function with a small number of coefficients.  Using a larger number of Enge coefficients improves the quality of the fit for the
field at a given radius (and polar angle $\theta$), but then performing the integral of the quadrupole gradient $g(z)$ to find the function
$G(\zeta)$ becomes difficult.

It is of course possible to plot an equipotential surface to represent the shape of the pole in an EMMA quadrupole, in the same
way that we did for a ``normal conducting equivalent'' HL-LHC inner triplet quadrupole.  However, we find that the shape of the
pole is much as expected for a normal conducting quadrupole, i.e. the plot is qualitatively very similar to that shown in
Fig.~\ref{hllhcitpolefaceshape}.

\begin{table}
\centering
\caption{Parameters for a fit of an Enge function of the form (\ref{examplegradientenge}), to the quadrupole
gradient in an EMMA F quadrupole.  The maximum radius $r_\mathrm{max}$ in the field data is 36\,mm.}
\label{emmaquadfitparams}
\begin{tabular}{cccc}
\hline
radius of fit & $a_0$ & $a_1$ & $a_2$ \\
\hline
$r_\mathrm{max}/10$ & 0.0138149 & -0.162670 & 11.2914 \\
$r_\mathrm{max}/8$ & 0.0172756 & -0.163023 & 11.3031 \\
$r_\mathrm{max}/6$ & 0.0230543 & -0.163794 & 11.3286 \\
$r_\mathrm{max}/4$ & 0.0346675 & -0.166075 & 11.4036 \\
$r_\mathrm{max}/3$ & 0.0463818 & -0.169480 & 11.5140 \\
$r_\mathrm{max}/2$ & 0.0702287 & -0.180796 & 11.8694 \\
\hline
\end{tabular}
\end{table}

\section{Conclusions}

A closed form analytic expression was presented for fringe fields in multipole magnets.  For quadrupoles, the field described by the
analytic expression was shown to have the expected properties.  The expression can be extended to describe multipoles of any order.
For ease of explanation and illustration, we looked in particular at fringe fields in which the multipole gradient had a roll-off
along the axis of the magnet described by an Enge function with only a single parameter in the exponent.  However, the technique
can be applied to any function with the appropriate dependence on the co-ordinates (i.e.~any function that depends on the co-ordinates
combined in the form $\sqrt{2}z \pm i h$).  Examples of other (non-Enge) functions that may be suitable for describing fringe
fields may be found in \cite{Kato1,Kato2}.

Expressions were also given for scalar and vector potentials from which the multipole fringe fields presented here could be derived.
Again, the expressions for the potentials can be extended to apply to multipole magnets of any order.  The scalar potential is of
interest since, in iron-dominated magnets, the pole faces form surfaces of constant scalar potential.  This provides a connection
between studies of the dynamics of particles moving through the fringe fields a particular magnet, and design studies of the magnet
geometry.  It is hoped that by having access to realistic analytical descriptions of fringe fields at an early stage in the design of an
accelerator beamline, the design process (typically involving many iterations between beam dynamics studies and magnet design work)
may be made more efficient.

The vector potential is of interest for particle tracking.  In particular, some techniques for symplectic integration of the equations
of motion for particles moving in magnetic fields are based on analytical expressions for the vector potential (see, for example,
\cite{WFR}).  Again, it is hoped that there will be benefits in being able to perform symplectic tracking through
realistic fringe field models at an early stage in the design of an accelerator.

In some types of magnet, such as those used in non-scaling FFAGs, fringe fields dominate the effects of the magnet.  In such cases,
being able to study the impact of fringe fields at an early stage of the accelerator design is essential for making efficient progress
with the design.  It should be possible to implement the methods presented here in standard accelerator tracking codes; this will allow
accurate modelling of fringe field effects in multipole magnets of arbitrary order, and enhance the range of tools available for
accelerator design and simulation.

\section{Acknowledgements}

This work has been ongoing for many years and BDM thanks Deepa Angal-Kalinin for giving him the opportunity and required support to finish it.
BDM is also pleased to acknowledge useful discussions on both form and content, as well as general encouragement over the years from the
following: Jim Clarke, Chris Edmonds, Bas van der Geer, Fay Hannon, Werner Herr, David Holder, Marieke de Loos, Neil Marks, Giovanni Muratori
Snr., Hywel Owen and Peter Williams.  The authors would like to thank Susana Izquierdo Bermudez for the LHC IT quadrupole field data, and
Chris Edmonds for the EMMA F quadrupole field data.

This work was partially supported by the FP7 HiLumi LHC Design Study http://hilumilhc.web.cern.ch.

%\end{document}

%\include{conclusion}
%\include{additions_5.4_5.5_6}

\end{document}